\documentclass[a4paper,11pt]{article}
\pdfoutput=1
\usepackage{jheppub}
\usepackage[T1]{fontenc}
\usepackage{tensor}
\usepackage{multirow, tabularx}
\usepackage[mathscr]{euscript}
\usepackage{float}
\usepackage{tikz}
\usepackage{soul}

\newcommand{\half}{\frac{\scriptstyle 1}{\scriptstyle 2}}
\newcommand{\C}{\mathbb{C}}

\newcommand{\R}{\mathbb{R}}
\renewcommand{\P}{\mathbb{P}}

\newcommand{\p}{\partial}

\newcommand{\D}{\mathrm{D}}

\renewcommand{\P}{\mathbb{P}}

\newcommand{\be}{\begin{equation}\label}
\newcommand{\ee}{\end{equation}}
\newcommand{\bea}{\begin{eqnarray}\label}
\newcommand{\eea}{\end{eqnarray}}

\title{Symmetries and Covering Maps for the Minimal Tension String on $\mathbf{AdS_3\times S^3\times T^4}$}

\author{N. M. McStay$^*$ \& R. A. Reid-Edwards$^{\dagger}$}

\affiliation{Department of Applied Mathematics and Theoretical Physics,\\University of Cambridge, CB3 0WA, United Kingdom}

\emailAdd{$^*$nm646@cam.ac.uk, $^{\dagger}$rar31@cam.ac.uk}

\abstract{This paper considers a recently-proposed string theory on $AdS_3\times S^3\times T^4$ with one unit of NS-NS flux ($k=1$). We discuss interpretations of the target space, including connections to twistor geometry and a more conventional spacetime interpretation via the Wakimoto representation. We propose an alternative perspective on the role of the Wakimoto formalism in the $k=1$ string, for which no large radius limit is required by the inclusion of extra operator insertions in the path integral. This provides an exact Wakimoto description of the worldsheet CFT. We also discuss an additional local worldsheet symmetry, $Q(z)$, that emerges when $k=1$ and show that this symmetry plays an important role in the localisation of the path integral to a sum over covering maps. We demonstrate the emergence of a rigid worldsheet translation symmetry in the radial direction of the $AdS_3$, for which again the presence of $Q(z)$ is crucial. We conjecture that this radial symmetry plays a key role in understanding, in the case of the $k=1$ string, the encoding of the bulk physics on the two-dimensional boundary.}

\begin{document} 
\maketitle
\flushbottom

\pagenumbering{arabic}

\section{Introduction}

In \cite{Maldacena:1997re}, Maldacena proposed that the large $N$ limit of certain conformal field theories (without gravity) appear to be equivalent to string theories in asymptotically Anti-de Sitter spaces. Each of these theories are defined perturbatively, meaning they are only well understood for small values of their perturbative parameters. Fascinatingly, the matching of the parameters in the AdS/CFT duality is a strong-weak correspondence, meaning that when the inverse string tension is taken to be small (the supergravity approximation), the corresponding CFT is strongly coupled and vice versa. This makes it difficult to meaningfully compare observables on each side of the picture and directly prove Maldacena's conjecture in perturbation theory. Nevertheless, in a remarkable series of papers \cite{Gaberdiel:2018rqv,Eberhardt:2018ouy,Eberhardt:2019ywk,Eberhardt:2019qcl,Eberhardt:2020akk,Dei:2020zui,Gaberdiel:2020ycd,Knighton:2020kuh,Gaberdiel:2021njm,Gaberdiel:2021kkp,Gaberdiel:2022bfk,Dei:2022pkr,Gaberdiel:2022oeu,Naderi:2022bus,Eberhardt:2019}\footnote{
See also \cite{Giribet:2018ada}.
},
Gaberdiel, Gopakumar, Eberhardt and collaborators propose a type IIB string theory on $AdS_3 \times S^3 \times T^4$ in the minimal tension limit\footnote{
We will use the term ``minimal tension'', rather than ``tensionless'' to describe this string theory.
}
of one unit of NS-NS flux wrapping the $S^3$, and argue that it is exactly dual to the symmetric product orbifold Sym$^N(T^4)$ CFT in the large $N$ limit. There is, of course, a natural application of these ideas also to $AdS_3 \times S^3 \times K3$.

The RNS string theory on $AdS_3 \times S^3 \times T^4$ \cite{Maldacena:2000hw,Maldacena:2000kv,Maldacena:2001km,Giveon:1998ns,deBoer:1998gyt} with $k$ units of NS-NS flux may be described by a supersymmetric WZW model of
$$ \mathfrak{sl}(2,\mathbb{R})_k^{(1)} \oplus \mathfrak{su}(2)_k^{(1)} \oplus \mathfrak{u}(1)^4 \, ,$$
where the superscript $(1)$ denotes that it is an $\mathcal{N} = 1$ superconformal affine algebra. The level $k \in \mathbb{N}$ is the amount of NS-NS flux present in the background and is related to the curvature of the spacetime by
$$R_{AdS}^2 = k\ell_s^2.$$
The minimal tension limit is at $k=1$ and for this reason, we refer to the minimal tension string on $AdS_3 \times S^3 \times T^4$ as the ``$k=1$ string'' in this paper. This is far from the supergravity regime and stringy effects cannot be neglected, suggesting that this theory may yield new insights into the nature of spacetime in string theory.

Unfortunately, defining the $k=1$ string is not a simple task; the RNS formulation of this theory is not well-defined at $k=1$ as unitarity is broken \cite{Eberhardt:2018ouy}. Yet the hybrid formalism of \cite{Berkovits:1999im} proffers a free worldsheet CFT that circumvents these unitarity issues. A key property of the theory is a shortening of the spectrum from the full continuous and discrete representations uncovered in \cite{Maldacena:2001km} for string theory on $AdS_3$ at generic $k$. The only highest weight states that survive at $k=1$ are those sitting at the bottom of the continuum. Further features include an apparently topological quality of the theory \cite{Eberhardt:2018ouy, Eberhardt:2021jvj}, intriguing connections to twistor theory \cite{Bhat:2021dez}, and a localisation of correlation functions to covering maps of the boundary \cite{Eberhardt:2019ywk,Dei:2020zui} (as foreshadowed in \cite{Pakman:2009zz}). The localisation to covering maps is of particular significance, since it provides a manifest realisation of the AdS/CFT duality, potentially providing a mechanism explicitly relating observables in the bulk to those in the boundary. 

The duality arises from considering a D1-D5 system \cite{Maldacena:1997re}, which has a twenty-dimensional parameter space. It is therefore a challenge to identify which CFT corresponds to which particular string theory. As mentioned above, it has been conjectured that the string with one unit ($k=1$) of pure NS-NS flux on $AdS_3\times S^3\times T^4$ is dual to the free symmetric product orbifold CFT $\text{Sym}^N(T^4)$ \cite{Gaberdiel:2018rqv,Eberhardt:2018ouy}. A number of important tests have been successfully checked, including the matching of the physical spectrum \cite{Eberhardt:2018ouy,Eberhardt:2020bgq,Naderi:2022bus}, the matching of correlation functions for the ground states \cite{Dei:2020zui,Eberhardt:2019ywk,Eberhardt:2020akk,Knighton:2020kuh} and the BPS sector \cite{Gaberdiel:2022oeu}, as well as progress in understanding deformations away from the orbifold point \cite{Fiset:2022erp}. Yet, the physical interpretation of this theory, what it calculates (and how) are still not well-understood. In particular, there have been hints that the $k=1$ theory describes a topological string \cite{Eberhardt:2018ouy,Eberhardt:2021jvj} --- whether this reflects a simplification of the physics at $k=1$ or that the theory only captures a subsector of the full physics (as with the more familiar topological strings \cite{Vonk:2005yv}) is an open question.

It is our aim in this paper to shed light on some of these issues. A particular focus will be the effect of a worldsheet gauge symmetry generated by a constraint we shall write as 
$$
Q(z)=0,
$$
where $Q(z)$ is expressed explicitly in terms of worldsheet fields by (\ref{eq:Q}), (\ref{eq:Q_bosonized}) and (\ref{eq:Q_WZW}). The fact that $Q(z)$ is nilpotent only when $k=1$\footnote{
This statement is proven in Appendix \ref{sec:Qappendix}.
}
and seems to play a special role in the theory demonstrates the simplification of AdS/CFT at $k=1$. We shall show how the requirement that physical vertex operators are invariant under this symmetry plays a key role in the connection between correlation functions and the covering map. We shall also see the emergence of a global radial symmetry in $AdS_3$, closely linked to the existence of the nilpotent worldsheet symmetry. Such a global symmetry suggests that all of the worldsheet degrees of freedom can be taken to effectively live at the boundary of spacetime, providing intuition for the holographic principle. Along the way, we will motivate the $k=1$ string as a rather natural construction, the efficacy of which does not rest solely on the hybrid construction.

One of our goals is to demystify aspects of the $k=1$ string. In order to keep the article relatively self-contained, various key results from the literature are reviewed. The novel results presented here are:
\begin{itemize}
\item The twistor geometry of the target space is elucidated and it is shown that, whilst some constraints can be written in terms of bulk twistors, the $Q$ constraint is naturally written in terms of the boundary twistors (rather than the full boundary ambitwistors).
\item Using only the spectral flow properties of the supercurrents and the $Q$ constraint, the bosonization of \cite{Naderi:2022bus} is used to provide an efficient and intuitive proof of the covering map localization result of \cite{Eberhardt:2019ywk,Dei:2020zui}.
\item The connection with the spacetime target space description is discussed and an alternative interpretation is proposed that does not require the worldsheet to be pinned on the boundary of the spacetime. This is in contrast to the perspective presented in \cite{Eberhardt:2019ywk,Bhat:2021dez}.
\item We show that, in this alternative perspective, the bulk theory has a rigid radial symmetry and a candidate for the ``secret representations'' of \cite{Eberhardt:2019ywk} naturally arises.
\end{itemize}

The structure of this paper is as follows: In Section \ref{sec:k=1_strings} we review and motivate aspects of the $k=1$ string from first principles and review the connection with the hybrid formalism. The interpretation of the target space is discussed, including connections to the bulk twistor theory and the ambitwistor space of the boundary theory. We introduce an alternative target space interpretation of the theory that makes the connection with the non-linear sigma model of $AdS_3$ more transparent. Section \ref{sec:spectrum} reviews the construction of correlation functions in the $k=1$ string. This section does not contain any novel material per se, but presents results that we will need in later sections and explains our interpretation of the $b$-ghost insertions. In Section \ref{sec:localisation} we show how the $Q$-invariance of correlation functions can be used to efficiently prove the localisation of the worldsheet to a covering map of the boundary. In Section \ref{sec:radial_profile} we show that there is a symmetry that removes the radial zero mode in $AdS_3$, suggesting that the physics can be naturally localised to the boundary. Finally, Section \ref{sec:discussion} discusses consequences of the results described here and open questions for the future. Various technical details are relegated to the Appendices.

\section{Strings on $\mathbf{AdS_3}$ at $\mathbf{k=1}$}
\label{sec:k=1_strings}

The hybrid formalism of \cite{Berkovits:1993xq,Berkovits:1994vy,Berkovits:1999im} can be applied to study the minimal tension $k=1$ string on $AdS_3 \times S^3 \times \mathcal{M}_4$, where ${\cal M}_4$ can be either $T^4$ or $K3$. The field redefinitions involved in deriving the hybrid formalism from the RNS string are complicated and, adding in the fact that the RNS string at $k=1$ appears to be ill-defined, the physical principles underlying the minimal tension hybrid string can be difficult to parse. We will begin this section by attempting to motivate some of the key mathematical structures required for the $k=1$ string, which ultimately lead to the hybrid formalism. The route from the RNS formalism to the hybrid formalism is long and convoluted and the end product cries out for a simpler interpretation. It would therefore be instructive to have an explicit construction of the $k=1$ string from first principles and without any reference to the hybrid formalism. We shall make some progress in this and hope to return to a more complete construction elsewhere. Since we would like to preserve the interpretation of this CFT as a sigma model, we will also discuss the target spaces of the theory. Though the theory seems to naturally live in the twistor space of $AdS_3$, a bosonization leads to a more conventional target space interpretation.

\subsection{The free field realisation}
\label{sec:FFR}

A natural starting point for studying string theory on $AdS_3 \times S^3$ is to consider the WZW model for the Kac-Moody current algebras associated to each of the group manifolds. Firstly, $S^3$ is described by $\mathfrak{su}(2)_1$, defined by the OPEs
$$K^+(z)K^-(w) \sim \frac{1}{(z-w)^2} + \frac{2K^3(w)}{z-w}, \quad K^3(z)K^{\pm}(w) \sim \frac{\pm K^{\pm}(w)}{z-w},$$
$$K^3(z)K^3(w) \sim \frac{1}{2(z-w)^2}.$$
The stress tensor for the CFT is given by the usual Sugawara construction \cite{DiFrancesco:1997nk}. It is well known that, at level $k=1$, the WZW model for this current algebra can be described by a free boson on a circle (at self-dual radius) \cite{Frenkel:1980rn,Segal:1981ap} and also as a free fermion theory \cite{DiFrancesco:1997nk}. In the latter case, we introduce weight $\left( \frac{1}{2},0 \right)$ worldsheet fermions $\psi^A$ and $\chi^A$ with OPEs
$$
\psi^A(z)\psi^B(w)\sim 0,	\qquad		\chi^A(z)\chi^B(w)\sim 0,	\qquad		\psi^A(z)\chi^B(w)\sim \frac{\epsilon^{AB}}{z-w},
$$
where $A,B \in \{\pm\}$ and $\epsilon^{+-} = 1 = -\epsilon_{+-}$. The relationship between the two descriptions is given by
$$
K^{\pm} =\pm\chi^{\pm}\psi^{\pm},	\qquad		K^3=-\frac{1}{2}:\left(\chi^+\psi^- + \chi^-\psi^+\right):.
$$
The map is not one-to-one due to the scaling redundancy
\begin{equation}
\label{eq:fermion_scaling}
    \psi^A\sim t\psi^A,	\qquad		\chi^A\sim t^{-1}\chi^A,
\end{equation}
which preserves the form of the $\mathfrak{su}(2)_1$ currents $K^a(z)$. It is less well known that the $\mathfrak{sl}(2,\R)_1$ WZW model for $AdS_3$ can also be written as a theory of free symplectic bosons \cite{Goddard:1987td,Gaberdiel:2018rqv}. These are weight $\left( \frac{1}{2},0 \right)$ worldsheet bosons $\xi^{\alpha}$ and $\eta^{\alpha}$ with OPEs
$$\xi^{\alpha}(z)\xi^{\beta}(w)\sim 0,	\qquad		\eta^{\alpha}(z)\eta^{\beta}(w)\sim 0,	\qquad		\eta^{\alpha}(z)\xi^{\beta}(w)\sim \frac{\epsilon^{\alpha\beta}}{z-w}.$$
where $\alpha,\beta \in \{\pm\}$. These free fields are related to the $\mathfrak{sl}(2,\R)_1$ generators satisfying
$$J^+(z)J^-(w) \sim \frac{1}{(z-w)^2} - \frac{2J^3(w)}{z-w}, \quad J^3(z)J^{\pm}(w) \sim \frac{\pm J^{\pm}(w)}{z-w},$$
$$J^3(z)J^3(w) \sim \frac{-1}{2(z-w)^2},$$
via
$$
J^{\pm}=\eta^{\pm}\xi^{\pm},	\qquad		J^3=-\frac{1}{2}:\left(\eta^+\xi^-+\eta^-\xi^+\right):.
$$
This is analogous to the free fermion description of the $\mathfrak{su}(2)_1$ theory, but there is no known analogue of the free boson construction for $\mathfrak{sl}(2,\R)_1$. Again, the free fields have a scaling symmetry
\begin{equation}
\label{eq:boson_scaling}
    \xi^{\alpha}\sim t\xi^{\alpha}, \qquad	\eta^{\alpha}\sim t^{-1}\eta^{\alpha},
\end{equation}
and live in $\C\P^1\times \C\P^1$.\footnote{There are two \emph{independent} scalings here; one given by the stress tensor, where $\xi^{\alpha}$ and $\eta^{\alpha}$ both scale as weight $1/2$ fields and one given by (\ref{eq:boson_scaling}) where these fields scale oppositely.} As indicated in \cite{Dei:2020zui}, the construction and the moduli space localisation is reminiscent of twistor constructions \cite{Berkovits:2004hg} and we will comment on this twistorial interpretation of the theory in \S\ref{sec:target_spaces} and in more depth elsewhere.

Thus, a starting point for a bosonic string at level $k=1$ on $AdS_3\times S^3$ is given by the action
\begin{equation}
\label{eq:free_field_action}
    S=\int d^2z\Big( \epsilon_{\alpha\beta}\eta^{\beta}\bar{D}\xi^{\alpha} + \epsilon_{AB}\chi^B\bar{D}\psi^A+b\bar{\p}c+c.c.\Big)
\end{equation}
where, as usual, one imagines this string as arising as a gauge-fixing of a theory with worldsheet gravity. This gauge-fixing introduces an integral over the moduli space of (bosonic) Riemann surfaces and left- and right-moving $(b,c)$ ghost systems which we have also included. The covariant derivative
$$\bar{D}\xi^{\alpha}=\bar{\p}\xi^{\alpha}+A\xi^{\alpha}$$
includes a weight $(0,1)$ field $A(z)$ that acts as a Lagrange multiplier for the constraint $Z:= \frac{1}{2}
:\left(\epsilon_{\alpha\beta}\eta^{\alpha}\xi^{\beta} +\epsilon_{AB} \chi^A\psi^B \right):=0$ that generates the scalings \eqref{eq:fermion_scaling} and \eqref{eq:boson_scaling}. The conformal invariance of the theory is described by the vanishing of the stress tensor, with the following contribution from the symplectic bosons and free fermions
$$T= -\frac{1}{2} \epsilon_{\alpha\beta} :\left( \xi^{\alpha}\partial\eta^{\beta} + \eta^{\alpha}\partial \xi^{\beta}\right):+\frac{1}{2}\epsilon_{AB}:\left(  \psi^A\partial\chi^B -\chi^A\partial\psi^B \right):.$$
We will denote the modes of this stress tensor by $L_n$. We could also add contributions from the $(b,c)$ ghost system and the ghosts required for the $Z=0$ gauge condition \cite{Gaberdiel:2022bfk}.\\

The above gives a free field description of the $\mathfrak{sl}(2,\R)_1 \oplus \mathfrak{su}(2)_1$ theory. It is a remarkable fact that this content is already rich enough to describe the full supersymmetric theory and provides an elegant way to avoid the unitarity issues that emerge in conventional attempts to give an RNS description of the $k=1$ theory \cite{Eberhardt:2018ouy,Dei:2020zui}. Indeed, it has been shown \cite{Eberhardt:2018ouy,Dei:2020zui} (see also Appendix A of \cite{Gaiotto:2017euk}) that the full $\mathfrak{psu}(1,1|2)_1$ algebra is generated by the free fields in the action \eqref{eq:free_field_action}, with the supercurrents $S^{\alpha A\pm}(z)$ realised by
\begin{equation}
\label{eq:supercurrents}
    S^{\alpha A+}=\xi^{\alpha}\chi^A,	\qquad		S^{\alpha A-}=-\eta^{\alpha}\psi^A,
\end{equation}
which combine with the bosonic currents $K^a(z)$ and $J^a(z)$ to give the superalgebra $\mathfrak{u}(1,1|2)_1$. Note that the currents $S^{\alpha A\pm}(z)$ are also invariant under the scaling symmetries \eqref{eq:fermion_scaling} and \eqref{eq:boson_scaling} when performed simultaneously, so that this scaling invariance is a property of the full supersymmetric theory. We could choose to view the scalings \eqref{eq:fermion_scaling} and \eqref{eq:boson_scaling} as independent symmetries, generated by the $\mathfrak{u}(1)$ currents
$$U= -\frac{1}{2}(\eta^+\xi^- - \eta^-\xi^+),	\qquad			V= -\frac{1}{2}(\chi^+\psi^- - \chi^-\psi^+).$$
It is often convenient to define the new basis $Z = U+V$ and $Y = U-V$ for which
$$Z(z)Z(w) \sim 0, \qquad Y(z)Y(w) \sim 0, \qquad Z(z)Y(w) \sim \frac{-1}{(z-w)^2}. $$
We can recover the desired $\mathfrak{psu}(1,1|2)_1$ theory via
\begin{equation}
\label{eq:psu_iso}
    \mathfrak{psu}(1,1|2)_1 \cong \frac{\mathfrak{u}(1,1|2)_1}{\mathfrak{u}(1)_U \oplus \mathfrak{u}(1)_V},
\end{equation}
by quotienting out the two $\mathfrak{u}(1)$ currents. Since these $\mathfrak{u}(1)$ currents have non-trivial OPEs with themselves, this quotient is slightly non-trivial. We first note that the $\mathfrak{u}(1)$ currents leave the (anti-)commutation relations of the $\mathfrak{psu}(1,1|2)_1$ generators invariant, except for the anticommutator of the supercharges
\begin{align*}
    \{S^{\alpha AI}_m,S^{\beta BJ}_n\} &= -\epsilon^{AB}\epsilon^{IJ}c_a(\sigma_a)^{\alpha\beta}J^a_{m+n} + \epsilon^{\alpha\beta}\epsilon^{IJ}(\sigma_a)^{AB}K^a_{m+n}\\
    &\quad + km\epsilon^{\alpha\beta}\epsilon^{AB}\epsilon^{IJ}\delta_{m+n,0} + \epsilon^{\alpha\beta}\epsilon^{AB}\delta^{I+J,0}Z_{m+n},
\end{align*}
where $c_a$ is a constant such that $c_- = -1$ and $c_+=c_3 = +1$ and conventions for the $\sigma_a$ can be found in \cite{Dei:2020zui}. It is then clear that we need to impose $Z_n = 0$ for all $n$.\footnote{It is sufficient to impose $Z_n = 0$ for $n \geq 0$, since $[Z_m,Z_n] = 0$.}
This is why we treated the two scalings symmetrically in \eqref{eq:free_field_action} and it ensures that the supercurrents \eqref{eq:supercurrents} are each preserved under the scaling.

The second $U(1)$ current $Y$ decouples from the theory because it is not generated by the commutators of the other generators. Instead, sectors with different $Y_0$ charge give different representations of the physics. It will play the role of a picture number: there is a copy of the full $\mathfrak{psu}(1,1|2)_1$ theory at each eigenvalue of $Y_0$. We must take account of this in the construction of physical correlators \cite{Dei:2020zui,Gaberdiel:2022bfk,Knighton:2022ipy}, identifying two correlators that differ only in their $Y_0$ charge as physically equivalent. To avoid such complications, we will only consider correlation functions where the overall $Y_0$ charge vanishes in \S\ref{sec:physical_correlators}.
\\

The action \eqref{eq:free_field_action} is therefore manifestly supersymmetric in the target space, and can be thought of as a Green-Schwarz \cite{Green:1987sp} treatment of the string. Surprisingly, it contains a, somewhat mysterious, additional local symmetry generated by the weight $(3,0)$ field\footnote{
Sometimes this gauge symmetry is written as $R = -\psi^+\psi^-(\eta^+\partial\eta^- - \eta^-\partial\eta^+)$ as in \cite{Gaberdiel:2022bfk}. This is just a matter of convention, which comes from exchanging $\xi^{\alpha}$ and $\chi^{A}$ with their conjugate variables. From a WZW perspective, it is equivalent to exchanging the roles of the supercharges of the $\mathfrak{psu}(1,1|2)_1$ algebra. From an ambitwistor perspective, it arises from the exchange of twistors and dual twistors, i.e. the choice of which $\C\P^1$ is identified with the boundary (see \S\ref{sec:target_spaces}).}
\begin{equation}
\label{eq:Q}
    Q = -\frac{1}{2} \epsilon_{AB} \chi^A\chi^BD\xi = \chi^+\chi^-(\xi^+\partial\xi^- - \xi^-\partial\xi^+),
\end{equation}
where $D\xi=-\epsilon_{\alpha\beta}\xi^{\alpha}\partial\xi^{\beta} $ is the pull-back of the projective measure on $\C\P^1$ to the worldsheet.
The $Q(z)$ transformation is given by $\delta_Q(\cdot) = [Q[\varepsilon] ,\cdot]$ where
$$
Q[\varepsilon]=\oint \mathrm{d}z \; \varepsilon(z)Q(z),
$$
for some weight $(-2,0)$ parameter field $\varepsilon(z)$. To impose the constraint $Q=0$, we introduce a weight (-2,1) Lagrange multiplier $\kappa$
\begin{equation}\label{SQ}
 S=\int d^2z\Big( \epsilon_{\alpha\beta}\eta^{\beta}\bar{D}\xi^{\alpha} + \epsilon_{AB}\chi^B\bar{D}\psi^A+b\bar{\p}c+\kappa Q+c.c.\Big).
\end{equation}
The gauge transformations generated by the constraint are
$$
    \delta_Q\eta^{\alpha} = \partial\varepsilon \, \chi^+ \chi^- \xi^{\alpha} + \varepsilon\left( \partial(\chi^+\chi^-)\xi^{\alpha} + 2\chi^+\chi^-\partial\xi^{\alpha} \right), \qquad
    \delta_Q\psi^A = \varepsilon \chi^A(\xi^+\partial\xi^- - \xi^-\partial\xi^+ ),
$$
$$
    \delta_Q\xi^{\alpha}= 0,   \qquad \delta_Q \chi^A = 0,   \qquad  \delta_Q\kappa=-\bar{\p}\epsilon,
$$
from which it can be checked that $Q$ generates a symmetry of (\ref{SQ}). We note that although $Q$ does not commute with the currents $J(z)$, $K(z)$ and $S(z)$, it does commute with all of the zero modes of $J^a(z)$ and $K^a(z)$ (the bosonic isometries) and \emph{half}\footnote{In this way we expect $Q$ to play a role similar to kappa symmetry in the Green-Schwarz superstring.} of the zero modes of $S^{\alpha A\pm}(z)$.

We now have an extended worldsheet symmetry algebra generated by $T$, $Q$ and $Z$. The $T$ and $Q$ part of this algebra looks somewhat like a simplified $W$-algebra,
\begin{align*}
    &[L_m,L_n] = (m-n)L_{m+n} + \frac{c}{12}(m^3-m)\delta_{m+n,0},\\
    &[L_m,Q_n] = (2m-n)Q_{m+n}, \quad [Q_m,Q_n] = 0,
\end{align*}
where $c$ is the central charge.\footnote{
The $\mathfrak{u}(1,1|2)_1$ free fields contribute $c=0$, whilst the $(b,c)$ ghost system contributes $c = -26$. Finally, the $Z$-gauging removes two bosonic degrees of freedom, contributing $c=-2$. This gives $c = -28$ overall.}
By ``simplified'' we simply mean that the weight 3 generator commutes with itself, which is not generically the case for $W$-algebras. The symmetry generated by $Q$ will play a starring role in what follows, whilst little attention will be given to the scaling symmetry of $Z$. This scaling symmetry is only present in the $k=1$ string for the free field realisation and not in the hybrid formalism. Its role is therefore not quite so fundamental to the theory, but is useful for interpreting the free fields as $AdS_3$ twistors \S\ref{sec:target_spaces}. For a physical state $\Phi$ that depends only on the free fields (i.e. is independent of the compact $\mathcal{M}_4$), the physical state conditions are
\begin{equation}
\label{eq:FFR_phys_state_conditions}
    L_n\Phi = Q_n\Phi = Z_n\Phi = 0, \quad \text{for all } n\geq 0.
\end{equation}

We have already accounted for the gauge-fixing of worldsheet gravity by including the $(b,c)$ ghost sector. We also need to gauge-fix $\kappa=0$. What are the Faddeev-Popov ghosts that do this? In the gauge-fixing of the worldsheet metric, we introduce the $c$-ghost which has the same weight as the gauge parameter (a worldsheet vector field) but opposite statistics. Similarly, we introduce an additional ghost $\tilde{c}$ of weight (-2,0) with fermionic statistics to gauge-fix $Q$. A putative BRST operator is then
\begin{equation}
\label{eq:putative_BRST}
    {\cal Q}=\oint dz\,\Big(cT+\tilde{c}Q+...\Big),
\end{equation}
where $T$ now includes ghost contributions and the $+...$ accounts for terms present due to the fact that $[T,\tilde{c}Q]\neq 0$, as well as terms from $\mathcal{M}_4$. It is difficult to understand what CFT the ghosts $\tilde{c}$ live in. A way forward comes from considering the bosonization of the $(b,c)$ ghost system,
$$b(z)=e^{i\sigma(z)},	\qquad		c(z)=e^{-i\sigma(z)},$$
where $\sigma$ is a linear dilaton theory (Appendix \ref{sec:linear_dilaton}) with stress tensor
$$T_{\sigma}=-\frac{1}{2}(\p\sigma)^2+\frac{3i}{2}\p^2\sigma,	\qquad		\sigma(z)\sigma(w)\sim -\ln(z-w).$$
Naively, we can deduce the form of $\tilde{c}$ by an analogous bosonization $\tilde{c} = e^{m\rho}$ for some real $m$, where $\rho$ is a linear dilaton with stress tensor
$$T_{\rho}= \epsilon \left(\frac{1}{2}(\p\rho)^2 - \frac{1}{2}q_{\rho}\p^2\rho \right),	\qquad		\rho(z)\rho(w)\sim \epsilon\ln(z-w).$$
In order to cancel the conformal anomaly, we would like $\tilde{c}$ to carry a central charge of $+28$. Combining this with the requirement that it is fermionic with weight $-2$, the only possibilities are
$$
\epsilon = -1, \qquad q_{\rho} = \pm 3, \qquad m= \mp 1.
$$
Hence, up to a transformation $\rho \mapsto -\rho$, we must have that $\tilde{c} = e^{-\rho}$, where $\rho$ carries a background charge of $q_{\rho} = +3$. We note that there is no obvious candidate for a $\tilde{b}$ ghost of weight $3$ and the $e^{-\rho}$ OPE is not what we would expect for a conventional $c$-ghost so this interpretation cannot be the full story. Nonetheless, the combination $e^{-\rho}Q$ acts as a conventional BRST current, as a consequence of the double zero in the OPE of $Q$ with itself.

With this field content, the $AdS_3 \times S^3$ part of the theory has a vanishing conformal anomaly and precisely matches the hybrid formalism of \cite{Berkovits:1999im} as we will outline in the following section. We expect that a formal treatment of the Faddeev-Popov method for gauge-fixing $Q$ will demonstrate that $c_{\rho} = +28$ is a necessary condition and thus fully motivate the field content of the hybrid string.\footnote{
It may appear that this six dimensional theory could stand in its own right, without the need for the four compact dimensions of $\mathcal{M}_4$. Unfortunately, the field content of the $\mathcal{M}_4$ sector is needed to construct a measure on the moduli space of Riemann Surfaces $\mathcal{M}_{g,n}$ that will cancel the anomaly from the $\rho$ ghost current, as we will see in \S\ref{sec:physical_correlators}.
}

\subsection{The hybrid formalism}
\label{sec:hybrid}
Traditionally, the hybrid formalism is constructed by rewriting the RNS string as an $\mathcal{N} = 4$ topological string \cite{Berkovits:1993xq,Berkovits:1994vy}, but we shall regard it as a completion of the string theory we were building up in the previous subsection. In particular, it describes $AdS_3 \times S^3$ by a $PSU(1,1|2)$ sigma model in terms of GS-like variables, in agreement with the free field realisation. We additionally require a CFT describing the embedding into the four-dimensional hyper-K\"ahler manifold $\mathcal{M}_4$, which we take to be $T^4$ but could equally be taken to be $K3$. Since the net central charge of the $AdS_3 \times S^3$ sector of the theory is zero, the four-dimensional CFT embedding into $T^4$ must also have central charge zero, suggesting that the four dimensional component is a topologically twisted CFT on the $T^4$. This resonates with hints we will discuss later that the theory in $AdS_3\times S^3$ described above is secretly a topological theory, as conjectured in \cite{Eberhardt:2018ouy,Eberhardt:2021jvj}. It is this combination of GS-type variables for $AdS_3 \times S^3$ and (topologically-twisted) RNS-like variables for $T^4$ from which the name ``hybrid'' is derived.

Following \cite{Gerigk:2012cq,Gaberdiel:2022bfk} and working at generic $k \in \mathbb{N}$, there is a (small) $\mathcal{N} = 4$ topological algebra in the string theory generated by
\begin{equation}
\label{eq:generators}
\begin{split}
    T &= T^{WZW} +T_{\sigma} + T_{\rho} + T_C,\\
    G^- &= e^{-i\sigma} + G^-_C,\\
    G^+ &= -e^{-2\rho-i\sigma}P + e^{-\rho}Q + e^{i\sigma}\mathcal{T} + G^+_C,\\
    \tilde{G}^+ &= e^{\rho+iH} + e^{\rho+i\sigma}\tilde{G}^+_C,\\
    \tilde{G}^- &= -e^{-3\rho-2i\sigma-iH}P + e^{-2\rho-i\sigma-iH}Q - e^{-\rho-iH}\mathcal{T} - e^{-\rho-i\sigma} \tilde{G}^-_C,\\
    J &= \frac{1}{2}\partial(\rho+i\sigma) + J_C,\\
    J^{\pm\pm} &= e^{\pm\rho \pm i\sigma \pm iH},
\end{split}
\end{equation}
where\footnote{Note that $:e^{i\sigma}({\cal T}-T^{WZW}-T_{\sigma}-T_{\rho}):=\p[:-e^{i\sigma}\p(\rho+i\sigma):]$ so that, as far as the BRST charge is concerned, the stress tensor appearing in $G^+$ is the natural stress tensor of the $AdS_3\times S^3$ part of the theory. Since the $T^4$ theory is topologically twisted, we expect to see $G^+_C$, rather than $T_C$ contributing to the BRST current $G^+$.
}
$$\mathcal{T} = T^{WZW} - \frac{1}{2}(\partial(\rho+i\sigma))^2 + \frac{1}{2}\partial^2(\rho+i\sigma).$$
The operator $T^{WZW}$ is the Sugawara energy momentum tensor associated to the WZW model for $\mathfrak{psu}(1,1|2)_k$ and similarly $P$ and $Q$ depend only on the WZW variables. $Q$ is weight three, depending on both the supercharges and the bosonic currents, whilst $P$ is quartic in the supercharges but independent of the bosonic currents. The operators containing a subscript $C$ correspond to the $\mathcal{N} = 4$ topological algebra of the compact $T^4$. We use the conventions of \cite{Gaberdiel:2021njm} for these operators,
\begin{equation}
\label{eq:compact_generators}
\begin{aligned}
    T_C &= \partial X^j \partial \bar{X}^j - \bar{\lambda}^j\partial \lambda^j, &
    G^-_C &= \bar{\lambda}^j\partial X^j , & \tilde{G}^-_C &= \epsilon_{ij}\bar{\lambda}^i \partial \bar{X}^j , & J^{--}_C = -\bar{\lambda}^1\bar{\lambda}^2,\\
    G^+_C &= \lambda^j\partial \bar{X}^j , & \tilde{G}^+_C &= \epsilon_{ij}\lambda^i\partial X^j , & J_C &= \frac{1}{2}\lambda^j \bar{\lambda}^j, &\\
    J^{++}_C &= \lambda^1\lambda^2,
\end{aligned}
\end{equation}
where $j \in \{1,2\}$ and $\epsilon_{12} = +1$. The topological twisting means that $(\partial X^j, \partial \bar{X}^j)$ are complex bosons, each of weight $(1,0)$, whilst $\lambda^j$ and $\bar{\lambda}^j$ are complex fermions of weights $(0,0)$ and $(1,0)$ respectively. Hence, the bilinears on the first, second and third lines are weight two, one and zero respectively. The topological twist is generated by including a factor of $\partial J_C$ in $T_C$. In what follows, we will often bosonize the fermions as $\lambda^j = e^{iH^j}$ and $\bar{\lambda}^j = e^{-iH^j}$, as well as defining the combination $H = H^1 + H^2$. There are non-trivial background charges associated to these bosonized fields as a consequence of the topological twisting, such that the fermions now play the role of ghosts.

In the minimal tension limit of $k=1$, the generator $P$ vanishes, whilst $T$ and $Q$ agree with \S\ref{sec:FFR}. It is then clear that \eqref{eq:putative_BRST} agrees with the form of the BRST operator $G^+_0$ in the hybrid string (see footnote 10) whilst $J$ is the natural ghost current. There is a second BRST operator $\tilde{G}^+_0$ which has a trivial cohomology --- this corresponds to the additional zero mode introduced to the worldsheet theory through the bosonization of the $(\beta,\gamma)$ system of the RNS string. The physical state conditions are given by \cite{Gerigk:2012cq}
\begin{equation}
\label{eq:hybrid_phys_conditions}
    G_0^+\Psi = \tilde{G}^+_0\Psi = \left(J_0 - \frac{1}{2}\right)\Psi = T_0\Psi = 0, \quad \Psi \sim \Psi + G_0^+\tilde{G}^+_0\Lambda.
\end{equation}
These constraints are solved for a compactification-independent vertex operator using the ansatz $\Psi = \Phi e^{2\rho+i\sigma+iH}$ of \cite{Berkovits:1999im}, where $\Phi$ depends only on the $\mathfrak{psu}(1,1|2)_1$ WZW model. This ansatz has picture number $P= -2$, where we define the picture number of a state as in \cite{Naderi:2022bus},
$$P\left(\Phi e^{m\rho+in\sigma+ik_1H^1 + ik_2H^2} \right) = -m.$$
This ansatz provides a nice gauge choice when we consider the constraint $G^+_0\Psi = 0$, which splits up into three separate conditions on $Q$, $\mathcal{T}$ and $G^+_C$ because of their differing ghost dependence. One can show that $(e^{i\sigma}\mathcal{T})_0\Psi = 0$ implies $L_n\Phi = 0$ for all $n \geq 0$ where $L_n$ are the modes of $T^{WZW}$ and similarly $(e^{-\rho}Q)_0\Psi = 0$ implies that $Q_n\Phi = 0$ for all $n\geq 0$ in this picture, recovering \eqref{eq:FFR_phys_state_conditions} from the free field realisation.

\subsection{What is the target space of the $k=1$ string theory?}
\label{sec:target_spaces}

What target space does this worldsheet theory describe an embedding into? We started off with a group manifold description of the $AdS_3\times S^3$ target space. We then introduced the free field realisation of \cite{Eberhardt:2018ouy,Dei:2020zui}, but have perhaps lost sight of what the target space is. In this section we shall review the twistorial interpretation of the free field theory and show that, with a further field redefinition, we can once again recover a conventional spacetime interpretation; this time in terms of the Wakimoto representation \cite{Wakimoto:1986gf}.

\subsubsection{Twistors and ambitwistors}

This section closely follows the $AdS_5$ discussion of \cite{Adamo:2016rtr} but adapted to $AdS_3$.\footnote{
We should highlight the work of \cite{Gaberdiel:2021qbb,Gaberdiel:2021jrv}, where progress has been made in trying to generalise the $AdS_3/CFT_2$ twistor string description to $AdS_5/CFT_4$ in an appropriate limit.
}
We start with the four-dimensional coordinate $X^{\alpha\beta}$ on complexified flat spacetime $\C^4$ where indices are lowered by $\epsilon_{\alpha\beta}$ with $\epsilon^{+-} = -\epsilon_{+-} = +1$. We can parameterise complexified $AdS_3$ by the coordinates\footnote{
We adopt notation similar to \cite{Maldacena:2000kv,deBoer:1998gyt} where an element of Euclidean $AdS_3$ is given by $g= (g_{\alpha\beta}) =e^{\tilde{\gamma} t_+}e^{-2\Phi t_3}e^{-\gamma t_-}$, where
$$
t_+=\left(
\begin{array}{cc}
0 & 1\\
0 & 0
\end{array}
\right),    \qquad
t_-=\left(
\begin{array}{cc}
0 & 0\\
-1 & 0
\end{array}
\right), \qquad
t_3=\frac{1}{2}\left(
\begin{array}{cc}
1 & 0\\
0 & -1
\end{array}
\right),
$$
satisfying the algebra $[t_+,t_-]=-2t_3$, $[t_3,t_{\pm}]=\pm t_{\pm}$.
}
\begin{equation}\label{X}
X^{\alpha\beta}=\left(
\begin{array}{cc}
1 & \quad-\gamma\\
-\tilde{\gamma} & \quad e^{-2\Phi}+\gamma\tilde{\gamma} 
\end{array}
\right),    \qquad  
X_{\alpha\beta}=\left(
\begin{array}{cc}
e^{-2\Phi}+\gamma\tilde{\gamma} & \quad\tilde{\gamma}\\
\gamma & \quad 1
\end{array}
\right).
\end{equation}
We arrive at the group parameterization of $SL(2;\C)$ as 
$$
g^{\alpha\beta}=|X|^{-1}X^{\alpha\beta}\in SL(2;\C),
$$
where $|X|^2=\det(X^{\alpha\beta})=e^{-2\Phi}$. The currents are then given by $J=-g^{-1} \partial g =J^+t_++J^-t_- + 2J^3t_3$. This gives rise to a natural metric in which $\Phi$ is the radial coordinate and $(\gamma,\tilde{\gamma})$ are coordinates on the boundary \cite{Giveon:1998ns},
$$ds^2 = \frac{1}{2} \text{tr} \left[ \left( g^{-1}dg \right)^2 \right]
= d\Phi^2 + e^{2\Phi}d\gamma d\tilde{\gamma}.$$
Note that $(X^2)^{\alpha}{}_{\beta}:= X^{\alpha\gamma}(X^T)_{\gamma\beta}=\delta^{\alpha}{}_{\beta}e^{-2\Phi}$. In practice, it is more helpful to work with the unnormalised $X^{\alpha\beta}$, rather than $g^{\alpha\beta}$, since we more clearly see that as we go to the boundary of $AdS_3$, given by $\Phi\rightarrow \infty$, we have $(X^2)^{\alpha}{}_{\beta} = 0$. We can define the $AdS_3$ twistors as
$$
Z^I=\left(
\begin{array}{c}
\xi^{\alpha}\\
\eta^{\alpha}
\end{array}
\right).
$$
The incidence relation is then
$$
\xi_{\alpha}=(X^T)_{\alpha\beta}\eta^{\beta},
$$
which defines the relationship between the $AdS_3$ spacetime and its twistor space. At the boundary, we have the condition $X^2=0$ and so
$$
X^{\alpha\beta}\xi_{\beta}=(X^2)^{\alpha}{}_{\beta}\eta^{\beta}=0.
$$
Thus the incidence relation on the boundary becomes $X^{\alpha\beta}\xi_{\beta}=0$, which may be written as
\begin{equation}
\label{eq:twistor_incidence_relation}
    \xi^- + \gamma\xi^+ = 0.
\end{equation}
Defining the twistor on the boundary $\C\P^1$ as $z_{\alpha}=\xi_{\alpha}$, we see that this is an incidence relation for the twistor on the boundary. Moreover, in light of the scaling between $\xi^{\alpha}$ and $\eta^{\alpha}$, it is natural to interpret $w_{\alpha}=\eta_{\alpha}$ as the dual twistor for the boundary theory, giving the identification
$$
{\cal Z}^I=\left(
\begin{array}{c}
z^{\alpha}\\
w^{\alpha}
\end{array}
\right)\in\C\P^1\times\C\P^1.
$$
Thus we see that ${\cal Z}^I$, a twistor of $AdS_3$, is simultaneously the ambitwistor of the boundary \cite{Adamo:2016rtr}. Note that the boundary twistors and dual twistors scale in the appropriate way; $z^{\alpha}\sim tz^{\alpha}$ and $w_{\alpha}\sim t^{-1}w_{\alpha}$.  One would then recover different signatures as different real slices on the corresponding spaces. It is straightforward to include the fermionic variables in a supertwistor space
$$
{\cal Z}^M=\left(
\begin{array}{c}
Z^m\\
W^m
\end{array}
\right)\in\C\P^{1|2}\times\C\P^{1|2},\quad\text{where}   \quad Z^m=\left(
\begin{array}{c}
\xi^{\alpha} \\
\psi^A
\end{array}
\right) ,\quad W^m=\left(
\begin{array}{c}
\eta^{\alpha} \\
\chi^A
\end{array}
\right).
$$
The worldsheet action and stress tensor have simple descriptions in these variables,
$$
S=\int \mathrm{d}^2z \;  W_m\bar{D}Z^m+...=\frac{1}{2}\int \mathrm{d}^2z \;  {\cal Z}_M\bar{D}{\cal Z}^M+...+,	\qquad		T=\frac{1}{2}{\cal Z}_M\p {\cal Z}^M+T_{\sigma}+T_{\rho}+T_C,
$$
where the ellipses denote contributions from the antiholomorphic, ghost and $T^4$ sectors. We raise (lower) supertwistor indices using $\Omega^{MN}$ ($\Omega_{MN}$), where $\Omega^{MN}$ is the natural symplectic form given by the (anti-)commutation relations $\left[{\cal Z}^M,{\cal Z}^N \right]_{\pm}=\Omega^{MN}$.\footnote{The bosonic part of which is
$$
\Omega^{IJ}=\left(\begin{array}{cc}
0 & \epsilon^{\alpha\beta}\\
\epsilon^{\alpha\beta} & 0
\end{array}\right), \qquad  \Omega_{IJ}=\left(\begin{array}{cc}
0 & \epsilon_{\alpha\beta}\\
\epsilon_{\alpha\beta} & 0
\end{array}\right).
$$}

The local symmetry $Q(z)$ we encountered in \S\ref{sec:FFR} can also be interpreted naturally in these coordinates as the pull-back of a holomorphic measure along a $\C\P^{1|2}$ in the boundary twistor space. $Q$ has projective weight zero and so we can introduce a non-local operator given by integrating $Q$ over a $\C\P^{1|2}$ in the target superspace. The action of this non-local operator on a boundary supertwistor wavefunction $\Psi(Z)$ may be written as
$$
-\oint Q\Psi= \frac{1}{2} \oint \D\xi\epsilon_{AB}\chi^A\chi^B\Psi = \frac{1}{2} \oint \D\xi\epsilon^{AB}\frac{\partial^2}{\p\psi^A\p\psi^B}\Psi= \oint \D Z \,\Psi,
$$
where $\D Z=\epsilon_{\alpha\beta}z^{\beta}dz^{\alpha}d^2\psi$ is a projective measure on the boundary supertwistor space\footnote{Similarly,
$$
\oint R\Psi=\oint \D W \,\Psi,
$$
is an integral over the dual boundary super-ambitwistor space.
}.

We see that there is an interpretation of the worldsheet theory as describing the embedding of a string into supertwistor space. Our starting point was complexified spacetime and spacetimes of different signatures are recovered by imposing different reality conditions on the spacetime and twistor spaces. That the natural target space is a complexification from which a required signature can be recovered is the perspective we take more generally throughout this paper. Moreover, traditional twistor constructions describe solutions of the spacetime equations of motion in terms of cohomology representatives on twistor space\footnote{See for example \cite{Ward:1990vs,Adamo:2017qyl}.}. It would then perhaps not be too much of a surprise if the $k=1$ string considered here turned out to be a topological string. This has been conjectured in \cite{Eberhardt:2018ouy,Eberhardt:2021jvj} and we shall see further evidence in \S\ref{sec:global_symmetry}. It is interesting to note that, with the exception of the $Q$ constraint, the theory can be described in terms of the bulk twistors ${\cal Z}^M$. The $Q$ constraint requires additional structure as only half of the bulk twistor components are used. Thus, a choice must be made as to whether the constraint is written in terms of boundary twistors or dual twistors ($Z^m$ and $W^m$ respectively). The extra structure needed is provided by the infinity twistor of the complexified flat spacetime $\C^4$ with coordinates $X^{\alpha\beta}$ \cite{Adamo:2017qyl}.

\subsubsection{Recovering spacetime: The Wakimoto representation}
\label{sec:Wakimoto_rep}

Using the above parameterization (\ref{X}), we can read off the classical left-invariant generators $J=-g^{-1} dg$ as
\begin{equation}
\label{eq:J_currents}
    J^+=-\mathscr{B},	\qquad		J^-=2\gamma d\Phi+d\gamma-\mathscr{B}\gamma^2,	\qquad		J^3=d\Phi-\mathscr{B}\gamma,
\end{equation}
where we have introduced the notation $\mathscr{B}=e^{2\Phi}d\tilde{\gamma} = \beta dz + \tilde{\beta} d\bar{z}$. These currents may be pulled back to the worldsheet to give a set of left-moving worldsheet currents (similarly for the right-moving sector). As worldsheet fields, $\beta(z)$ and $\gamma(z)$ are commuting holomorphic fields of weight 1 and 0 respectively  and $\partial\Phi(z)$ is a free boson of weight 1. In order to reproduce the correct current algebra of $\mathfrak{sl}(2,\mathbb{R})_1$, these fields must have the nontrivial OPEs
\begin{equation}
\label{eq:wakimoto_OPEs}
    \beta(z)\gamma(w) \sim \frac{1}{(z-w)}, \quad \partial\Phi(z)\partial\Phi(w) \sim \frac{1}{2(z-w)^2}.
\end{equation}
There is a normal ordering correction to $J^-$, discussed further below and given by \cite{Eberhardt:2019ywk} 
$$J^- = 2\gamma d\Phi -d\gamma - \mathscr{B}\gamma^2.$$
Such a realisation of the current algebra has been motivated by the Wakimoto representation of $\mathfrak{sl}(2,\mathbb{R})_1$, first studied in \cite{Wakimoto:1986gf}, but has since been applied to string theory on $AdS_3$ in order to provide a semi-classical interpretation to the theory (for example, \cite{Eberhardt:2019ywk,Naderi:2022bus,Bhat:2021dez}). It has also been used to construct the boundary CFT \cite{Eberhardt:2019qcl}. The real utility of this (free field) Wakimoto representation is that the (quasi-)primary fields describing it have natural spacetime interpretations.

How can this free field representation be reconciled with the (interacting) non-linear sigma model on $AdS_3$? The common way this is done in the literature is as follows: we start by considering the WZW model on Euclidean $AdS_3$ (denoted by $H_3^+=SL(2;\C)/SU(2)$) \cite{Giveon:1998ns,deBoer:1998gyt}. The bosonic part of the action in first order form is given by
\begin{equation}
\label{eq:S}
    S = \frac{k}{4\pi}\int \mathrm{d}^2z \; (4\partial\Phi\bar{\partial}\Phi + \tilde{\beta}\partial\tilde{\gamma} + \beta\bar{\partial}\gamma - e^{-2\Phi}\beta\tilde{\beta}),
\end{equation}
where $\Phi \to \infty$ is at the boundary of $H_3^+$ and $(\gamma,\tilde{\gamma})$ are complex coordinates that represent coordinates on $S^2$ when we are at large $\Phi$. The equations of motion fix $\beta = e^{2\Phi}\partial\tilde{\gamma}$ and $\tilde{\beta} = e^{2\Phi}\bar{\partial}\gamma$. We may add a Fradkin-Tseytlin term, coupling to the dilaton $\Phi$ \cite{DiFrancesco:1997nk,Fradkin:1985ys}, see Appendix \ref{sec:linear_dilaton}. This gives the revised action \cite{Giveon:1998ns}
\begin{equation}
\label{eq:S_AdS3}
    S_{AdS_3} = \frac{k}{4\pi}\int \mathrm{d}^2z \; \left(4\partial\Phi\bar{\partial}\Phi + \tilde{\beta}\partial\tilde{\gamma} + \beta\bar{\partial}\gamma - e^{-2\Phi}\beta\tilde{\beta} - \frac{1}{k}R\Phi\right),
\end{equation}
with $R$ the worldsheet Ricci scalar. We should point out, however, that at the level of the path integral these two actions are only equivalent if the path integral is dominated by large $\Phi$ contributions \cite{deBoer:1998gyt}.
The classical WZW currents associated to $S_{AdS_3}$ are given by \cite{Eberhardt:2019ywk}
\begin{equation*}
\label{eq:classical_currents}
    J^+ = k\beta,\qquad
    J^- = -2k\gamma\partial\Phi + k\beta\gamma^2 - k\partial\gamma,\qquad
    J^3 = -k\partial\Phi + k\beta\gamma.
\end{equation*}
Close to the boundary the $e^{-2\Phi}\beta\tilde{\beta}$ term is suppressed and the theory is approximately free with a chiral current realisation of the $AdS_3$ isometries. These currents can be quantized to form the current algebra $\mathfrak{sl}(2,\mathbb{R})_{k+2}$, given by the OPE relations
$$\beta(z)\gamma(w) \sim -\frac{1}{k(z-w)}, \quad \partial\Phi(z)\partial\Phi(w) \sim -\frac{1}{2k(z-w)^2},$$
and again we have a normal ordering correction to $J^-$ as
$$J^- = -2k\gamma\partial\Phi + k\beta\gamma^2 - (k+2)\partial\gamma.$$
The fact that we have level $(k+2)$ rather than level $k$ is a consequence of working in the first order formalism. There is an anomaly collected in the change of variables from the $AdS_3$ non-linear sigma model to include $\beta$ and $\tilde{\beta}$ --- this anomaly shifts the level $k \mapsto k+2$. It is immediately apparent that our formal treatment of the $k=1$ string realises a ``$k=-1$'' version of this RNS Wakimoto representation in \eqref{eq:J_currents}.

We have seen that, from the commonly held perspective described above, it appears that the Wakimoto construction is only formally exact in the large radius limit of the theory. Below, we shall see that for the minimal tension string, the Wakimoto construction plays a more direct (and possibly complete) role. Specifically, we will see that the OPEs for this representation will arise in an exact form, not as an approximation in the large $\Phi$ limit. Given that the $k=1$ string is most naturally described in twistor variables, one might anticipate that the adapted variables of the Wakimoto construction provide a more natural spacetime interpretation in light of \eqref{X}.

\subsubsection{Recovering spacetime: Wakimoto from bosonization}
\label{sec:Wakimoto_from_bosonization}

To make contact with the Wakimoto representation, it is useful to bosonize the free fields as in \cite{Naderi:2022bus}. This gives an explicit way of realising the $k=1$ theory in terms of spacetime physics. For the symplectic bosons, we define
\begin{equation}
\label{eq:bosonization}
    \xi^{-} = -e^{-\phi_1 - i\kappa_1}, \quad \eta^+ = e^{\phi_1}\partial(e^{i\kappa_1}), \quad \xi^{+} = e^{\phi_2 + i\kappa_2}, \quad \eta^- = e^{-\phi_2}\partial(e^{-i\kappa_2}),
\end{equation}
where
$$\phi_i(z)\phi_j(w) \sim -\delta_{ij}\log (z-w), \quad \kappa_i(z)\kappa_j(w) \sim -\delta_{ij}\log (z-w).$$
Similarly, for the complex fermions,
\begin{equation*}
\label{eq:bosonized_fermions}
    \chi^+ = e^{iq_1}, \quad \psi^- = -e^{-iq_1}, \quad \psi^+ = e^{iq_2}, \quad \chi^- = e^{-iq_2},
\end{equation*}
where
$$q_i(z)q_j(w) \sim -\delta_{ij}\log (z-w).$$
We have in some sense extended our theory through this bosonization by introducing the fields $\zeta^+ = e^{\phi_1+i\kappa_1}$ and $\zeta^- = e^{-\phi_2 - i\kappa_2}$. However, as is noted in \cite{Naderi:2022bus}, these fields can be seen as a formal trick that do not add additional physical states to the theory.

 The embedding maps from the worldsheet to $\C\P^1\times \C\P^1$, where we think of $(\xi^+,\xi^-)$ and $(\eta^+,\eta^-)$ as projective coordinates on each $\C\P^1$. Before bosonization, we remove a disc $\mathscr{D}_{\epsilon}$ of radius $\epsilon$ around a point in the target space and work in a patch. In what follows we shall choose to work in the chart where $\xi^+\neq 0$ (so $\mathscr{D}_{\epsilon}$ is a disc about the point at which $\xi^+=0$). In doing so, we have chosen here to identify the $\C\P^1$ with projective coordinates $(\xi^+,\xi^-)$ with the boundary $\C\P^1$ on which the dual conformal field theory is defined. Of course, it would be entirely equivalent to choose the other $\C\P^1$, parameterised by $(\eta^+,\eta^-)$, to be identified with the boundary. This choice exchanges twistors and dual twistors in the boundary ambitwistor space.

Removing the target space disc $\mathscr{D}_{\epsilon}$ requires we remove the preimages (under the embedding map) of this disc on the worldsheet. Let us denote the preimages of the point where $\xi^+=0$ by $\{z^*_a\}$, where $a=1,2,...,M$ for some $M$ and the preimages of the disc $\mathscr{D}_{\epsilon}$ by $\mathscr{D}_a$. Thus, the worldsheet (at genus zero) is $\C\P^1$ with $M$ discs $\mathscr{D}_a$ removed. In the bosonized variables we will need to introduce background charges for $\kappa_2$ at these points (see below).

One immediate observation that we can make is that the field $Q$ in \eqref{eq:Q} can be rewritten as
\begin{align}
\label{eq:Q_bosonized}
\begin{split}
    Q = e^{iq_1-iq_2}e^{\Theta}\partial\Sigma
    &= -e^{iq_1+\phi_2+i\kappa_2}e^{-iq_2+\phi_2+i\kappa_2}\partial(e^{-\Sigma})\\
    &= -S^{+++}S^{+-+}\partial(e^{-\Sigma}),
\end{split}
\end{align}
where
\begin{equation*}
\label{eq:Sigma_Theta}
    \Theta = -\phi_1-i\kappa_1+\phi_2+i\kappa_2,    \qquad  \Sigma = \phi_1+i\kappa_1+\phi_2+i\kappa_2.
\end{equation*}
The field $\Sigma$ is projectively invariant under scaling by $Z$, whilst the field $\Theta$ is not. We will find that \eqref{eq:Q_bosonized} will be a useful expression for $Q$ in later sections. For our purposes, we want to adapt this to the hybrid formalism at $k=1$. Remarkably, this bosonization allows us to make contact with the Wakimoto representation and so gives a spacetime interpretation to the $k=1$ string. Consider the fields \cite{Naderi:2022bus}
\begin{equation}
\label{eq:wakimoto_realisation}
    \begin{split}
        \beta = -e^{\Sigma}\partial(i\kappa_1),\qquad
        \gamma = e^{-\Sigma},\qquad
        \partial\Phi = \frac{1}{2}\partial(\phi_1 + \phi_2 + 2i\kappa_2),
    \end{split}
\end{equation}
which satisfy \eqref{eq:wakimoto_OPEs} and so generate $\mathfrak{sl}(2,\mathbb{R})_1$ via
\begin{equation}\label{Waki}
    J^+ = -\beta,\qquad
    J^- = 2\gamma\partial\Phi - \partial\gamma - (\gamma(\beta\gamma)),\qquad
    J^3 = \partial\Phi -(\beta\gamma).
\end{equation}
We have been explicit here with how the normal ordering is taken in the $\beta\gamma^2$ term. Note that zeroes of $\xi^+$ are necessarily zeroes of $\beta$. In a correlation function, we can associate $\gamma\sim-\xi^-/\xi^+$ in the patch where $\xi^+\neq 0$. As such, we see that $\beta$ has zeroes and $\gamma$ has poles at the removed points.

\subsection*{A comment on background charges}

In order to interpret $\xi^-/\xi^+$ as a projective coordinate, we must work in a patch in which $\xi^+\neq 0$. As discussed above, we do this by removing the point corresponding to $\xi^+=0$ from the target space. We therefore also need to excise the pre-images from  the worldsheet. We call the preimages $z^*_a$ ($a=1,...,M$) and so $\xi^+(z^*_a)=0$. The key point we want to make is that, in the bosonized coordinates, the points $z^*_a$ must carry a background charge for the fields $\kappa_2$. We shall argue below that placing a $\kappa_2$-charged state at $z^*_a$ does two things; firstly, it ensures that $\xi^+(z)\rightarrow 0$ as $z\rightarrow z^*_a$ and secondly, it gives rise to the conformal anomaly of the radial Wakimoto coordinate $\Phi$ (see \S\ref{sec:global_symmetry}).

The stress tensor for the bosonized fields is \cite{Naderi:2022bus}
$$
T=-\frac{1}{2}(\partial\phi_1)^2-\frac{1}{2}(\partial\phi_2)^2-\frac{1}{2}(\partial\kappa_1)^2+\frac{i}{2}\partial^2\kappa_1-\frac{1}{2}(\partial\kappa_2)^2-\frac{i}{2}\partial^2\kappa_2.
$$ 
We see from this that $\phi_1$ and $\phi_2$ have no background charge and $\kappa_1$ and $\kappa_2$ have equal and opposite background charges (see Appendix \ref{sec:linear_dilaton}). As we shall explain below, we need only focus on $\kappa_2$. This background charge acts as a source for the field $\kappa_2$ as may be seen in the action
$$
S=\frac{1}{2\pi}\int d^2z\,\left(\frac{1}{2}\partial \kappa_2\bar{\partial} \kappa_2-\frac{i}{4}R\kappa_2 \right),
$$
with equation of motion $\Box\kappa_2+\frac{i}{4}R=0$. The background charge means there is a non-trivial coupling to worldsheet gravity, signifying a conformal anomaly. Following \cite{Friedan:1985ge}, we place a $\kappa_2$-charged state
$$
|q\rangle=:e^{i\kappa_2(z_a^*)}:|0\rangle
$$
at each of the $z^*_a$. It then follows that $i\kappa_2$ has the behaviour
$\lim_{z\rightarrow z_a^*}i\kappa_2(z)\sim \ln(z-z_a^*)$. This behaviour means that, as claimed, $\lim_{z\rightarrow z_a^*}\xi^+(z)=0$ as follows from the bosonization of $\xi^+$ (\ref{eq:bosonization}). We can explicitly include such charged states in the path integral by inserting the non-local operators
\begin{equation}\label{eq:k2_charge}
    \oint dz\;e^{i\kappa_2(z)},
\end{equation}
with the contour being taken around $z_a^*.$ We will comment briefly on this further in \S\ref{sec:discussion}. To all intents, we can think of the points $z_a^*$ as additional punctures in the worldsheet where a state with one unit of $\kappa_2$ charge is inserted.

Despite also carrying a background charge, we do not need to add insertions for $\kappa_1$ in order to construct correlation functions in the Wakimoto representation. Strictly, to realise the Wakimoto construction, we only need to bosonize the $(\xi^+,\eta^-)$ system and thus $\kappa_1$ need not be introduced. The stated realisation of the Wakimoto fields does not live in the ``large Hilbert space'' that includes the zero mode of $e^{i\kappa_1}$, whilst they do depend on the zero mode of $e^{-i\kappa_2}$. This is why only the background charge for $\kappa_2$ affects the correlation functions of Wakimoto fields.\footnote{We could have chosen to work in the opposite chart of $\xi^-\neq 0$, in which case the roles of $\kappa_1$ and $\kappa_2$ would be exchanged. In particular, we would insert $\kappa_1$ charges at the points where $\xi^-=0$. Moreover, if were to work in the fully bosonized theory (i.e. not using the Wakimoto representation as our defining variables but the $\phi_i$ and $\kappa_i$ bosons) then the theory would live most naturally on a cylinder, with charges inserted for both $\kappa_1$ and $\kappa_2$.}
$\Phi$ is charged under $\kappa_2$ and therefore has a conformal anomaly, so we do need to take care of the $\kappa_2$ background charge. Put another way, we identify the $\kappa_2$ conformal anomaly as the source of the radial conformal anomaly in the Wakimoto representation and so we shall place an associated background charge at $z_a^*$.

Given the behaviour of $\kappa_2$ at $z^*_a$, it is not hard to see that
$$
\lim_{z\rightarrow z_a^*}\partial\Phi(z)\sim \frac{1}{z-z_a^*},
$$
whilst $\beta(z)$ has a simple zero and $\gamma(z)$ a simple pole as $z\to z^*_a$. We shall argue for this behaviour of $\partial\Phi(z)$ once  again in \S\ref{sec:wakimoto_radial_coord} but from an alternative perspective, relating $\p\Phi(z)$ to operators that are conformal tensors. Nonetheless, the bosonization gives a clear way to see why zeroes in $\beta$ are associated with poles in $\gamma$ and $\partial\Phi$.

\subsection{Connections to the $AdS_3$ sigma model}\label{sec:connections_to_AdS_3}

It is perhaps not surprising that the most natural spacetime interpretation for the theory is in terms of Wakimoto coordinates. Firstly, we have seen that the twistor geometry has a clean connection with spacetime in Wakimoto coordinates and the theory is a twistor string theory of a novel kind. Secondly, the $Q$ constraint commutes with the boundary coordinate $\gamma(z)$ but not with the radial coordinate $\Phi(z)$, also suggesting the theory naturally treats the radial coordinate differently from other directions. 
Thus far, no approximations or limits have been taken in the construction, yet the OPEs are that of a free theory. This raises the question of how we should interpret the spacetime theory. In what sense are we able to make contact with the anticipated non-linear sigma model on $AdS_3$? We explore two possible explanations for this apparent discrepancy.
\\

\subsubsection*{Large Radius}

The first is the perspective presented in \S\ref{sec:Wakimoto_rep}, where it is assumed that the semi-classical solutions \cite{Eberhardt:2019ywk,Bhat:2021dez} are exact and the worldsheet is in some way pinned to the boundary. In this picture, the meaningful dynamics are in the boundary directions since $\Phi\rightarrow\infty$. In this limit, the Wakimoto action \eqref{eq:S} is replaced by
\begin{equation}\label{eq:FreeWaki}
    S = -\frac{1}{4\pi} \int \mathrm{d}^2z \; \left( 4\partial\Phi \bar{\partial} \Phi + \tilde{\beta}\partial\tilde{\gamma} + \beta \bar{\partial}\gamma \right),
\end{equation}
where the interaction term between $\Phi$ and $\beta$ has negligible effect\footnote{
The central charge of this free theory is $c=-3$, the same as $\mathfrak{sl}(2,\mathbb{R})_1$. We also note that the decoupling of the conjugate $\beta$ and $\gamma$ fields is reminiscent of a similar phenomenon in ambitwistor string constructions \cite{Mason:2013sva}.
}. In this picture, the worldsheet can only probe spacetime in the vicinity of the boundary and one concludes that it is only in this region in which physical perturbations occur (although there have been suggestions that in the limit of large spectral flow the interior can be probed \cite{Eberhardt:2021jvj,Knighton:2022ipy}). The large radius description is the perspective that has been largely adopted; however, there is another possible description of the physics which we describe below.

\subsubsection*{Free Fields with Operator Insertions}

An alternative possibility, and the one that we will advocate for, follows the construction of \cite{Gerasimov:1990fi} (see also chapter 4 of \cite{Ketov:1995yd}). The starting point is the  WZW action, the two terms of which may be written in the Wakimoto parameterization with $g^{-1}\p g=J^mt_m$ as
$$
\frac{1}{2}\int_{\Sigma} Tr(g^{-1}d g\wedge *g^{-1}dg)=\int_{\Sigma} d\Phi\wedge * d\Phi+ \mathscr{B}\wedge *d\gamma,	\qquad		\frac{1}{3}\int_{\cal V} Tr\Big((g^{-1}d g)^3\Big)=\int_{\Sigma} \mathscr{B}\wedge d\gamma
$$
where as in \S\ref{sec:Wakimoto_rep}, $\mathscr{B}=\beta dz+\tilde{\beta}d\bar{z}$ and $\partial {\cal V}=\Sigma$. The WZW action becomes
\begin{equation}\label{eq:S_W}
    S_W =-\frac{1}{4\pi}\int d^2z\,\Big( 2\p\Phi\bar{\p}\Phi+ \beta\bar{\p}\gamma\Big),
\end{equation}
and, in contrast to \eqref{eq:FreeWaki}, $\beta$ now appears as an independent field, rather than as a Lagrange multiplier. Notice that $\tilde{\beta}$ drops out of the classical action \eqref{eq:S_W} yet it is still contained in the path integral measure (so potentially operator insertions can depend on it). Whether we have $\beta$ or $\tilde{\beta}$ appearing in the classical action just depends on which orientation we take for the WZW term. Correlation functions are naively given by
$$
\langle {\cal O}(z_1)...{\cal O}(z_n)\rangle \sim \int {\cal D}\Phi {\cal D}\beta {\cal D}\tilde{\beta}  {\cal D}\gamma\;e^{-S_W}\; {\cal O}(z_1)...{\cal O}(z_n).
$$
What has this free theory got to do with $AdS_3$? The explicit $AdS_3$ non-linear sigma model is recovered by imposing the condition that $e^{-2\Phi}\mathscr{B}$ is single-valued, i.e.
\begin{equation}\label{eq:Insertion}
{\cal S}:=\oint_{C}e^{-2\Phi}\mathscr{B}=0,
\end{equation}
for all closed paths $C$ on the worldsheet. These constraints imply that there exists some scalar $\tilde{\gamma}$, such that $e^{-2\Phi}\mathscr{B} = d\tilde{\gamma}$. We shall assume that this constraint is trivial everywhere on the worldsheet, except when $C$ is a boundary of one of the $M$ removed discs $\mathscr{D}_a$ - we will denote such constraints by $\mathcal{S}_a$.\footnote{
It is clearly sufficient to apply these constraints to the non-contractible cycles. We have not explicitly computed correlation functions so cannot comment on the higher genus case precisely. However, one finds $\mathcal{S} = 0$ around the insertion points of (spectrally flowed) highest weight states, by computing the OPE with \eqref{eq:explicit_V}. The only remaining non-contractible cycles (at genus zero) are about $\{z_a^*\}$, which are removed from the worldsheet to study the theory in the given chart $\xi^+ \neq 0$. As in \cite{Frenkel:2005ku}, to return to the full worldsheet and not merely a chart, we must introduce fictitious vertex operators at these points --- the ``secret representations'' below.} The condition is non-trivial at these points since (focusing on the holomorphic sector) they correspond to where $\beta=0$ whilst $e^{-2\Phi}$ simultaneously diverges at these points.

If we were to impose the constraint $e^{-2\Phi}\mathscr{B} = d\tilde{\gamma}$, the action becomes the familiar non-linear sigma model
$$
S_{AdS_3}=-\frac{1}{4\pi}\int d^2z\,\Big( 2\p\Phi\bar{\p}\Phi+ e^{2\Phi}\p\tilde{\gamma}\bar{\p}\gamma\Big).
$$
The equivalence to the standard WZW model demonstrates that the theory is non-chiral throughout. More precisely, in passing from the NLSM to the WZW model the change of variables from $\tilde{\gamma}$ to $\mathscr{B}$ requires a Jacobian $\det(e^{2\Phi}d)$ in the functional integral, which can be incorporated into the action \eqref{eq:S_W} as an anomaly term
\begin{equation}\label{eq:S_W_hat}
    \widehat{S}_W=-\frac{1}{4\pi}\int d^2z\,\Big( 4\p\Phi\bar{\p}\Phi+ \beta\bar{\p}\gamma+\Phi R\Big),
\end{equation}
where $R$ is the worldsheet Ricci scalar. The change of variables only makes sense on those points where $\mathscr{B}\neq 0$ and so the Jacobian is only defined away from those points (which have been removed).

The delta functions that constrain $\mathscr{B}$ are
$$
\delta^2\Big(\oint e^{-2\Phi}\mathscr{B}\Big)=\delta\Big(\oint e^{-2\Phi}\beta\Big) \delta\Big(\oint e^{-2\Phi}\tilde{\beta}\Big).
$$
Following \cite{Gerasimov:1990fi}, we claim then that the Wakimoto theory requires the constraint \eqref{eq:Insertion} and correlation functions are calculated by\footnote{One can show that
$$
J^+(z)\beta(w)e^{-2\Phi(w)}\sim 0,	\qquad		J^3(z)\beta(w)e^{-2\Phi(w)}\sim 0,	\qquad J^-(z)\beta(w)e^{-2\Phi(w)}\sim \frac{\p}{\p w}\left(\frac{e^{-2\Phi(w)}}{z-w}\right),
$$
such that ${\cal S}$ is invariant under $J^a(z)$.
}
\begin{equation}\label{eq:Gerasimov_correlator}
\int {\cal D}\Phi{\cal D}\beta {\cal D}\tilde{\beta}{\cal D}\gamma\;e^{-\widehat{S}_W[\beta,\gamma,\Phi]}\,\prod_{a=1}^N\delta\Big(\oint_{C_a} e^{-2\Phi}\beta\Big) \delta\Big(\oint_{C_a} e^{-2\Phi}\tilde{\beta}\Big)\;\prod_{i=1}^n{\cal O}_i(\Phi,\beta,\tilde{\beta},\gamma)
\end{equation}
where the contours $C_a$ surround the points $z^*_a$. Prescriptions for how to deal with such delta-function insertions for general $k$ may be found in \cite{Gerasimov:1990fi}. We shall comment on how this might simplify for $k=1$ in \S\ref{sec:discussion}. We expect that these insertions play the role of the ``secret representations'' of \cite{Eberhardt:2019ywk,Hikida:2020kil} (see also \cite{Hikida:2007tq,Hikida:2008pe} and \cite{Frenkel:2005ku} for related ideas). In \cite{Hikida:2020kil}, additional operators are inserted into the path integral which ensure $\beta(z)=-J^+(z)$ has zero modes when $z=z_a^*$. The delta function insertions in \eqref{eq:Gerasimov_correlator} play a similar role and the discussion in \cite{Hikida:2020kil} has some parallels with our discussion in \S\ref{sec:global_symmetry}. An important difference is that the insertions in \eqref{eq:Gerasimov_correlator} are $SL(2)$ invariant, whereas the insertions in \cite{Hikida:2020kil} transform non-trivially under $J^-$. We note also that the $\delta({\cal S})$ insertions fulfill the role of background charges discussed around equation (\ref{eq:k2_charge}) above.

The fact that this description is not restricted to a semi-classical limit will be an important point and will allow us to explain the somewhat unreasonable effectiveness of the Wakimoto representation in the literature. We should emphasise that, whilst the action $\widehat{S}_W$ may appear to describe a chiral theory, the insertions $\delta^2(\mathcal{S}_a)$ encode information about the antiholomorphic sector. They also prevent $\gamma$ from being a globally holomorphic function on the worldsheet, as we will see in \S\ref{sec:global_symmetry}.

\section{The spectrum}
\label{sec:spectrum}

In this section, we will discuss the spectrum of minimal tension strings on $AdS_3 \times S^3 \times T^4$ at the level of detail required for the current work. This section will draw heavily on the existing literature and will not contain novel results\footnote{A detailed discussion of the spectrum is given in \cite{Eberhardt:2018ouy}, including an explanation of the spectral flow automorphism of the current algebra, denoted by $\sigma$. This must be included in any unitary theory of strings on $AdS_3$ \cite{Maldacena:2000hw}. A discussion of the theory's DDF operators can be found in \cite{Naderi:2022bus}.}. There is, however, a slight difference in the interpretation of the $b$-ghost insertions in \S\ref{sec:physical_correlators} compared to the literature, where we view them as screening operators for $U_0$.

\subsection{Physical states}
\label{sec:physical_states}

Usually, a theory of strings on $AdS_3$ (for generic $k$) would include both discrete and continuous representations \cite{Maldacena:2000hw,Eberhardt:2018ouy}. Classically, these correspond to bound states and scattering states, respectively. They are each parameterised by the $\mathfrak{sl}(2,\mathbb{R})$-spin $j$, which is real and positive for the discrete representations, whilst $j = \frac{1}{2} + is$ for real $s\geq 0$ for the continuous representations. It is the continuous representations that are of interest for the current work and we denote these by $\mathscr{C}^j_{\lambda}$, where $\lambda$ is the fractional part of the $J_0^3$ eigenvalue.

These $\mathfrak{sl}(2,\mathbb{R})$ representations should of course sit in a larger multiplet for $\mathfrak{psu}(1,1|2)$. However, in the minimal tension limit of $k=1$, there is a shortening of the spectrum such that the only unitary representations of $\mathfrak{psu}(1,1|2)$ that are allowed for the affine Lie superalgebra $\mathfrak{psu}(1,1|2)_1$ take the form \cite{Eberhardt:2018ouy}
\begin{equation}\label{eq:short_multiplet}
\begin{gathered}
    (\mathscr{C}^{\frac{1}{2}}_{\lambda}, \mathbf{2}) \\
    (\mathscr{C}^1_{\lambda+\frac{1}{2}},\mathbf{1}) \qquad (\mathscr{C}^0_{\lambda + \frac{1}{2}}, \mathbf{1})
\end{gathered}
\end{equation}
where we denote $n$-dimensional representations of $\mathfrak{su}(2)$ by $\mathbf{n}$ and the supercharges $S^{\alpha A \pm}_0$ move between the representations. Since the top $\mathfrak{sl}(2,\mathbb{R})$ representation has $j=\frac{1}{2}$, we refer to this multiplet as the ``$j=\frac{1}{2}$ solution'', which lies at the bottom of the continuum of states.

Other than the vacuum representation, this provides the only highest weight representation of the $k=1$ string. Focusing on the top representation, its highest weight states $|m_1,m_2\rangle$ come from the R-sector of the theory and are labelled by quantum numbers $m_1, m_2$ given by
\begin{align*}
    J_0^3|m_1,m_2\rangle = (m_1+m_2)|m_1,m_2\rangle,\qquad
    U_0|m_1,m_2\rangle = (m_1-m_2-1/2)|m_1,m_2\rangle.
\end{align*}
The spin is defined by $j = m_1 - m_2$, so $m_1 - m_2 = \frac{1}{2}$ will form a physical state condition.\footnote{
In \S\ref{sec:physical_correlators}, we will choose to work in a picture where $Y_0 = 0$, such that upon setting $Z_0=0$, we have $U_0=V_0=0$.
}
In terms of the free fields, such states satisfy
$$\xi^{\pm}_r|m_1,m_2\rangle = \eta^{\pm}_r|m_1,m_2\rangle = 0,$$
for all $r \geq 1$, whilst the action of the zero modes is given by
\begin{align*}
    \xi^+_0|m_1,m_2\rangle &= |m_1,m_2 + 1/2\rangle, \quad &\eta^+_0|m_1,m_2\rangle = 2m_1|m_1 + 1/2,m_2\rangle,\\
    \xi^-_0|m_1,m_2\rangle &= -|m_1-1/2,m_2\rangle, \quad &\eta^-_0|m_1,m_2\rangle = -2m_2|m_1,m_2-1/2\rangle.
\end{align*}
Likewise, for the fermionic modes, we have that
$$
    F^+_r|m_1,m_2\rangle = 0 \quad\text{ for all } r\geq 0,\qquad
    F^-_r|m_1,m_2\rangle = 0 \quad\text{ for all } r\geq 1,
$$
where $F$ is either $\psi$ or $\chi$. This means that $\psi^-_0$ and $\chi^-_0$ are creation modes.
\\

The physical spectrum also includes the spectrally flowed representations of \eqref{eq:short_multiplet}. We can define the action of this spectral flow automorphism $\sigma$ by \cite{Dei:2020zui}
$$\sigma(\xi^{\pm}_r) = \xi^{\pm}_{r\mp 1/2}, \quad \sigma(\eta^{\pm}_r) = \eta^{\pm}_{r\mp 1/2}, \quad \sigma(\psi^{\pm}_r) = \psi^{\pm}_{r\pm 1/2}, \quad \sigma(\chi^{\pm}_r) = \chi^{\pm}_{r\pm 1/2}.$$
States in the $w$-spectrally flowed representation are denoted by $[|m_1,m_2\rangle]^{\sigma^w}$, on which an operator $A_r$ acts as
$$A_r[|m_1,m_2\rangle]^{\sigma^w} \equiv [\sigma^w(A_r) |m_1,m_2\rangle]^{\sigma^w}.$$
Because we are working with the free field realisation, which realises the full $\mathfrak{u}(1,1|2)_1$ algebra before the quotient in \eqref{eq:psu_iso}, there exists a second spectral flow automorphism, denoted by $\hat{\sigma}$. Whilst $\sigma$ acts non-trivially only on $\mathfrak{psu}(1,1|2)_1 \subset \mathfrak{u}(1,1|2)_1$ which is the physical part of the theory we care about, the automorphism $\hat{\sigma}$ acts non-trivially on the $U(1)$ charges $U$ and $V$. To be explicit,
\begin{equation}
\label{eq:sigma_hat}
\hat{\sigma}(\xi^{\pm}_r) = \xi^{\pm}_{r+1/2}, \quad
\hat{\sigma}(\eta^{\pm}_r) = \eta^{\pm}_{r-1/2}, \quad
\hat{\sigma}(\psi^{\pm}_r) = \psi^{\pm}_{r-1/2}, \quad
\hat{\sigma}(\chi^{\pm}_r) = \chi^{\pm}_{r+1/2}.
\end{equation}

Spectrally flowed representations are generically not highest weight, yet it turns out that the $\hat{\sigma}^2$-spectrally flowed representation is the vacuum representation with respect to $\mathfrak{psu}(1,1|2)_1$ \cite{Dei:2020zui}. The vacuum state $|0\rangle^{(1)} = [\psi^+_{-3/2}\psi^-_{-3/2}\psi^+_{-1/2}\psi^-_{-1/2}|0\rangle]^{\hat{\sigma}^2}$ has a $Y_0$ eigenvalue of $2$, in contrast to the vacuum of the unflowed representation which has a $Y_0$ eigenvalue of $0$. This is because of the non-trivial action of $\hat{\sigma}$ on the two $U(1)$ charges. We denote the vertex operator corresponding to the $\hat{\sigma}^2$ vacuum by $W(z)$ and this will be useful later in \S\ref{sec:physical_correlators} for fixing the picture number of $Y$ in physical correlation functions.
\\

A natural way to generate the full spectrum of physical states is to begin with the states in the string theory that correspond to the ground states in the dual CFT. In particular, the $w$-spectrally flowed sector of the string theory gives rise to the $w$-twisted sector of the CFT \cite{Dei:2020zui}. For odd $w$, the $w$-twisted ground states correspond to $[|\Phi_w\rangle]^{\sigma^w}$ \footnote{We are working in picture $P=-2$, so the full physical state is really $[|\Phi_w\rangle]^{\sigma^w} e^{2\rho+i\sigma+iH}$.}, where
\begin{equation}
\label{eq:odd_w}
    |\Phi_w\rangle = \chi^-_{-\frac{w-1}{2}}\psi^-_{-\frac{w-1}{2}}\cdots \chi^-_{-1}\psi^-_{-1}\chi^-_0\psi^-_0 |m_1,m_2\rangle,
\end{equation}
and
\begin{equation}
\label{eq:m1_m2_odd_constraint}
    m_1 + m_2 = -\frac{w^2+1}{4w}, \quad m_1 - m_2 = \frac{1}{2}.
\end{equation}
For even $w$, the ground state is degenerate and is comprised of two states that transform as an $\mathfrak{su}(2)$ doublet. They are $[|\Phi_w^{\pm}\rangle]^{\sigma^w}$, where
\begin{equation}
\label{eq:even_w}
\begin{split}
        |\Phi_w^+\rangle &= \chi^-_{-\frac{w}{2}+1}\psi^-_{-\frac{w}{2}+1}\cdots \chi^-_{-1}\psi^-_{-1}\chi^-_0\psi^-_0 |m_1,m_2\rangle,\\
    |\Phi_w^-\rangle &= \chi^-_{-\frac{w}{2}}\psi^-_{-\frac{w}{2}}\cdots \chi^-_{-1}\psi^-_{-1}\chi^-_0\psi^-_0 |m_1,m_2\rangle,
\end{split}
\end{equation}
and
\begin{equation}
\label{eq:m1_m2_even_constraint}
    m_1 + m_2 = -\frac{w}{4}, \quad m_1 - m_2 = \frac{1}{2}.
\end{equation}

As explained in \cite{Naderi:2022bus}, the full single particle CFT spectrum can then be generated through applications of DDF operators (the spectrum generating algebra) to the states $[|\Phi_w\rangle]^{\sigma^w}$ and $[|\Phi_w^{\pm}\rangle]^{\sigma^w}$. We will assume in this work that the DDF operators of \cite{Naderi:2022bus} also generate the entire physical Hilbert space of the worldsheet CFT (formally, it has only been shown that they provide a lower bound on the spectrum). The conventions for the generators of the $\mathcal{N}=4$ topological algebra of \cite{Naderi:2022bus} differ to our conventions in \eqref{eq:generators} and \eqref{eq:compact_generators} by a similarity transformations. Therefore, the DDF operators in our conventions are a similarity transformation of
\begin{align}
\label{eq:DDF_operators}
    \begin{split}
        \partial\bar{\mathcal{X}}^j_n &= \oint \mathrm{d}z \; \partial\bar{X}^j e^{-n\Sigma},\qquad \partial\mathcal{X}^j_n = \oint \mathrm{d}z \; e^{-n\Sigma}\left[ \partial X^j + ne^{\rho - \Theta + iH^j}\psi^+\psi^- \right],\\
        \Psi^{A,j}_r &= \oint \mathrm{d}z \; \psi^A e^{\phi_1 + i\kappa_1} e^{-\left(r+\frac{1}{2}\right)\Sigma } e^{\rho + iH^j},
    \end{split}
\end{align}
where $A \in \{\pm\}$ and $j \in \{1,2\}$. These are the DDF operators of the free bosons and free fermions of $T^4$, where $\partial X^j$, $\partial \bar{X}^j$ and $H^j$ refer to the compact $T^4$ variables as in \eqref{eq:compact_generators}. The DDF operators for the barred bosons have been given in picture $P=0$ and the free fermions in picture $P = -1$, whilst the unbarred bosons have no definite picture.

\subsection{Vertex operators}

We introduce the vertex operators $\hat{V}^{w}_{m_1,m_2}(x,z)$ where $x$ and $z$ are boundary and worldsheet coordinates respectively. We would like to find the vertex operators that correspond to the ground states $[|\Phi_w\rangle]^{\sigma^w}$ and $[|\Phi_w^{\pm}\rangle]^{\sigma^w}$ i.e. vertex operators that satisfy
\begin{equation}
\label{eq:V_hat}
\begin{split}
    \hat{V}^w_{m_1,m_2}(0;0)|0\rangle &= [|\Phi_w\rangle]^{\sigma^w} \quad \text{for odd } w,\\
    \tensor*{\hat{V}}{*^w_{m_1, m_2}^{\pm}} (0;0)|0\rangle &= [|\Phi^{\pm}_w\rangle]^{\sigma^w} \quad \text{for even } w.
\end{split}
\end{equation}
We add $x$- and $z$-dependence through conjugation, e.g. for odd $w$,
$$\hat{V}^{w}_{m_1,m_2}(x;z) = e^{zL_{-1}}e^{xJ_0^+}\hat{V}^{w}_{m_1,m_2}(0;0)e^{-xJ_0^+}e^{-zL_{-1}},$$
since $L_{-1}$ and $J_0^+$ act as translation operators on the worldsheet and boundary, respectively. In later sections, we will often suppress the $\{\pm\}$ labels for the even $w$ case, so that $\hat{V}^w_{m_1,m_2}$ refers to a generic vertex operator associated to a twisted sector ground state.\footnote{Since $J^+$ and $J^3$ do not commute, $\hat{V}^w_{m_1,m_2}(x;z)$ has definite values for $x$ and $j=m_1-m_2$ but is not a state of definite $m=m_1+m_2$.}

We can deduce the OPE structure of these vertex operators with the free fields using the action of their modes on the ground states. For example,
\begin{equation}
\label{eq:xi_V_OPE}
    \xi^{\pm}(z)[|\Phi_w\rangle]^{\sigma^w} = \sum\limits_{r \leq \pm w/2} [\xi^{\pm}_{r\mp w/2} |\Phi_w\rangle]^{\sigma^w} z^{-r-1/2},
\end{equation}
which implies that $\xi^{\pm}(z)\hat{V}^w_{m_1,m_2}(x_i;z_i) = \mathcal{O} \left((z-z_i)^{\frac{\mp w - 1}{2}} \right)$. Similarly, one finds that $\xi^{\pm}(z) \tensor*{\hat{V}}{*^w_{m_1, m_2}^{A}}(x_i;z_i) = \mathcal{O} \left((z-z_i)^{\frac{\mp w - 1}{2}} \right)$ for even $w$ and $A \in \{\pm\}$.

It was observed in \cite{Naderi:2022bus} that the vertex operator
\begin{equation}
\label{eq:explicit_V}
    V^w_{m_1,m_2}(0;z) = \exp \left[ f^w_{m_1,m_2}(z) \right],
\end{equation}
where
\begin{equation*}
    f^w_{m_1,m_2}(z) = \left(2m_1 + \frac{w-1}{2}\right)\phi_1 + 2m_1 i\kappa_1 + \left( 2m_2 + \frac{w+1}{2}\right)\phi_2 + 2m_2 i\kappa_2,
\end{equation*}
correctly reproduces this OPE structure with the symplectic bosons, where $\phi_i$ and $\kappa_i$ are the bosonized fields given in (\ref{eq:bosonization}). All that remains is to consider the OPE structure with the fermionic free fields $F^{\pm}$.\footnote{We thank Kiarash Naderi for pointing this out to us.}

For the odd $w$ case, we take $r \in \mathbb{Z} + 1/2$ and the state $[|\Phi_w\rangle]^{\sigma^w}$ satisfies
$$F^{\pm}_r[|\Phi_w\rangle]^{\sigma^w} = [F^{\pm}_{r\pm w/2} |\Phi_w\rangle]^{\sigma^w} = 0, \quad \text{for all } r \geq 1/2.$$
This implies the OPE $F^{\pm}(z)\hat{V}^w_{m_1,m_2}(x_i;z_i) = \mathcal{O}(1)$. This is precisely the behaviour of $F^{\pm}$ with the vertex operator $V^w_{m_1,m_2}$ of \eqref{eq:explicit_V}. Therefore, in the case of odd $w$, we have the relationship between states and operators as
\begin{equation}
\label{eq:odd_V}
    V^w_{m_1,m_2}(x;z) \leftrightarrow [|\Phi_w\rangle]^{\sigma^w},
\end{equation}
where $m_1$ and $m_2$ satisfy the constraints in \eqref{eq:m1_m2_odd_constraint}. Unfortunately, the even $w$ case is not quite as simple. Consider $r \in \mathbb{Z}$, such that the states $[|\Phi_w^{\pm}\rangle]^{\sigma^w}$ satisfy
\begin{align*}
    F^+_r[|\Phi_w^+\rangle]^{\sigma^w} &= [F^+_{r+w/2} |\Phi_w^+\rangle]^{\sigma^w} = 0, \quad \text{for all } r \geq 0,\\
    F^-_r[|\Phi_w^+\rangle]^{\sigma^w} &= [F^-_{r-w/2} |\Phi_w^+\rangle]^{\sigma^w} = 0, \quad \text{for all } r \geq 1,\\
    F^+_r[|\Phi_w^-\rangle]^{\sigma^w} &= [F^+_{r+w/2} |\Phi_w^-\rangle]^{\sigma^w} = 0, \quad \text{for all } r \geq 1,\\
    F^-_r[|\Phi_w^-\rangle]^{\sigma^w} &= [F^-_{r-w/2} |\Phi_w^-\rangle]^{\sigma^w} = 0, \quad \text{for all } r \geq 0.
\end{align*}
This OPE behaviour corresponds to  \cite{Naderi:2022bus}
\begin{equation}
\label{eq:even_V}
    e^{\pm\frac{1}{2}(iq_1+iq_2)}V^w_{m_1,m_2}(x;z) \leftrightarrow [|\Phi_w^{\pm}\rangle]^{\sigma^w},
\end{equation}
where this time $m_1$ and $m_2$ satisfy \eqref{eq:m1_m2_even_constraint}.

\subsection{Physical correlation functions}
\label{sec:physical_correlators}

At present, a full discussion of the structure of physical correlation functions of the $k=1$ string requires studying the theory as an $\mathcal{N}=4$ topological string \cite{Berkovits:1999im,Berkovits:1994vy}. We highlight in this section the key results of this approach.\footnote{ Further details may be found in Appendix \ref{sec:hybrid_correlators}.} We will focus on the insertion of twisted sector ground states in picture $P=-2$,
\begin{equation}
\label{eq:P=-2_phys_state}
    \Psi_i = \hat{V}^{w_i}_{m_1^i,m_2^i} e^{2\rho + i\sigma + iH},
\end{equation}
for $i = 1, \dots, n$ where we take $w_i$ to be either odd or even. We expect that it is possible to motivate the form of these correlators in purely twistorial terms, without reference to the hybrid formalism and we hope to return to this elsewhere.
\\

The first novel feature of hybrid correlators as compared with the usual RNS string is the presence of two candidate $b$-ghosts, given by $G^-$ and $\tilde{G}^-$ in \eqref{eq:generators}. This is of course a consequence of the two BRST operators in the $\mathcal{N}=4$ topological string, where
$$G_0^+G^- = \tilde{G}^+_0 \tilde{G}^- = T.$$
Physical correlation functions require an appropriate combination of these $b$-ghosts combined with Beltrami differentials $\{\mu_I\}$ for $I = 1, \dots, \dim(\mathcal{M}_{g,n})$ to construct a measure over $\mathcal{M}_{g,n}$, the moduli space of $n$-punctured, genus $g$ Riemann surfaces \cite{Nakahara:2003nw}. $G^-=e^{-i\sigma} + G^-_C$ contains the usual $b$-ghost of the RNS string, whilst
$$\tilde{G}^- = e^{-2\rho-i\sigma-iH}Q - e^{-\rho-iH}\mathcal{T} - e^{-\rho-i\sigma}\tilde{G}^-_C,$$
appears to complicate the measure on $\mathcal{M}_{g,n}$. Whenever
\begin{equation}\label{b}
\widetilde{\mathbf{b}}(\mu_I) := \int_{\Sigma} \mathrm{d}^2z \; \tilde{G}^-(z) \mu_I(z),
\end{equation}
is inserted inside a physical correlation function, the only term that contributes is the piece involving the field $Q(z)$. The analogous statement is also true for $G^-(z)$, where only the term containing $b(z)$ contributes. We will take a slight liberty in notation to define
\begin{equation*}
    \mathbf{Q}(\mu_I) := \int_{\Sigma} \mathrm{d}^2z \; e^{-2\rho(z)-i\sigma(z)-iH(z)} Q(z) \mu_I(z).
\end{equation*}
The number of insertions of each $b$-ghost can be determined from the ghost charge. The result is that $(g-1)$ copies of $\mathbf{b}(\mu_I)$ and $(n+2g-2)$ copies of $\mathbf{Q}(\mu_I)$ should be inserted, as explained in Appendix \ref{sec:hybrid_correlators}. We will implicitly assume this is the case in what follows.\footnote{Whilst our focus will be on the ground states here, we note that the generalisation to include the DDF operators of \cite{Naderi:2022bus} is not straightforward as they have non-trivial ghost dependence. We comment further on this in Appendix \ref{sec:hybrid_correlators}.}

The states $\hat{V}^w_{m_1,m_2}$ are specified by the free field realisation of $\mathfrak{psu}(1,1|2)_1$. In performing the quotient of \eqref{eq:psu_iso}, we explicitly gauged away $Z$ but found that a second $U(1)$ current $Y$ decoupled from the theory. We treat $Y_0$ as a picture and only consider physical correlators that have an overall vanishing $Y_0$ eigenvalue. By requiring that $j = \frac{1}{2}$ for \emph{all} of the physical state insertions, they satisfy $Y_0 = 0$ already. However, the $\mathbf{Q}(\mu_I)$ insertions are charged under $Y_0$ since
$$
    [U_0, \mathbf{Q}(\mu_I)] = -\mathbf{Q}(\mu_I),\qquad
    [V_0, \mathbf{Q}(\mu_I)] = \mathbf{Q}(\mu_I),
$$
leading to a $Y_0$ charge of $-2$ for each $Q$ insertion. We must cancel this by inserting $(n+2g-2)$ copies of the $\hat{\sigma}^2$-spectrally flowed vacuum state $W(u_{\alpha})$ \cite{Dei:2020zui}. We note in passing that it is natural to pair up one $W(u_{\alpha})$ with each of the $(n+2g-2)$ $\mathbf{Q}(\mu_I)$ and think of them as part of the measure. This means that a generic physical correlation function for the twisted sector ground states in the free field realisation is given by
\begin{equation*}
\label{eq:correlator_expression}
    \left\langle \prod_{\alpha = 1}^{n-2+2g} W(u_{\alpha}) \prod_{i=1}^n \hat{V}^{w_i}_{m_1^i,m_2^i}(x_i;z_i) \right\rangle.
\end{equation*}
where an integral over $\mathcal{M}_{g,n}$, with appropriate measure, is implicit.

The spins of the physical states here all satisfy $j_i = \frac{1}{2}$, as expected in the minimal tension limit where only the bottom of the continuum survives. In \cite{Eberhardt:2019ywk,Eberhardt:2020akk}, it was observed for the $\mathfrak{sl}(2,\mathbb{R})_{k+2}$ WZW model in the RNS formalism that the spins must satisfy the constraint
\begin{equation*}
\label{eq:RNS_constraint}
    \sum_{i=1}^n j_i = \frac{k+2}{2}(n+2g-2) - (n+3g-3),
\end{equation*}
in order for a localising solution to covering maps to exist. At level $k+2 = 3$, this was indeed satisfied by $j_i = \half$ for all $i$. However, as discussed above, this approach is not well-defined in the $k=1$ limit and we should instead apply the hybrid formalism which contains a $\mathfrak{sl}(2,\mathbb{R})_1 \subset \mathfrak{psu}(1,1|2)_1$ current algebra. Therefore, we need to impose the above constraint at level $k+2 = 1$ \cite{Dei:2020zui}
\begin{equation}
\label{eq:spin_constraint}
    \sum_{i=1}^n j_i = \frac{n}{2} - (n-2 + 2g),
\end{equation}
for non-vanishing correlation functions. This appears to be at odds with $j_i = \half$ for all $i$, however one can interpret each $\mathbf{Q}(\mu_I)$ insertion as carrying a spin of $j=-1$. This is because, when acting upon the vertex operators $\hat{V}^{w_i}_{m_1^i,m_2^i}$, the operator $U_0$ determines $m_1 - m_2$, and $\mathbf{Q}(\mu_I)$ carries a $U_0$ charge of $-1$. This means that the constraint \eqref{eq:spin_constraint} is indeed satisfied if we extend it to include the spins of the $\mathcal{N} = 4$ generators (in essence, we view the $\mathbf{Q}(\mu_I)$ as screening operators for $U_0$). It is the presence of an additional $b$-ghost in the hybrid formalism's $\mathcal{N}=4$ topological algebra that makes this possible.

\section{$\mathbf{Q=0}$ and covering map localisation}
\label{sec:localisation}

We provide in this section a discussion of the localisation of physical correlation functions to points in moduli space where a covering map exists. In particular, we explain in \S\ref{sec:Q=0} how the gauge constraint associated to $Q(z)$ implies an efficient method for deriving the localisation, which is equivalent to the incidence relation proposed in \cite{Dei:2020zui,Knighton:2020kuh}. We construct the proof of localisation in \S\ref{sec:proof} and this technical section can be skipped on a first reading. We also provide classical intuition for the localisation in \S\ref{sec:classical_covering_map} by taking advantage of the restriction of the physical spectrum at $k=1$.

\subsection{Covering maps}
\label{sec:covering_map}

The correlation functions of the symmetric product orbifold Sym$^N(T^4)$ may be described by covering maps \cite{Lunin:2000yv,Lunin:2001pw} and hence, if the AdS/CFT duality is to hold true, we should expect the dual string theory correlation functions to also be described in this way.

We define a branched covering map $\Gamma : \Sigma_g \to \mathbb{CP}^1$ where $\Sigma_g$ is a genus $g$ Riemann surface in the following way \cite{Eberhardt:2020akk}. For $i = 1,\dots, n$, let $z_i$ be coordinates on $\Sigma_g$, $x_i$ be coordinates on $\mathbb{CP}^1$ and $w_i \in \mathbb{N}$ be ramification indices. Then $\Gamma$ is a holomorphic map that satisfies:
\begin{enumerate}
    \item $\Gamma(z_i) = x_i$ for all $i$.
    \item $\Gamma(z) = x_i + a_i^{\Gamma}(z-z_i)^{w_i} + \mathcal{O}((z-z_i)^{w_i+1})$ as $z \to z_i$ for some constant $a_i^{\Gamma}$, such that $z_i$ is a ramification point of order $w_i$.
    \item $\Gamma(z)$ has no other critical points.
\end{enumerate}
The degree $N_{\Gamma}$ of $\Gamma$ is defined as the number of preimages of each non-ramification point $x \in \mathbb{CP}^1$ i.e. the number of times the worldsheet covers the boundary. It is given by the Riemann-Hurwitz formula
\begin{equation}
\label{eq:Riemann_Hurwitz}
    N_{\Gamma} = 1 - g + \frac{1}{2}\sum\limits_{i=1}^n (w_i - 1).
\end{equation}

Before we consider the AdS/CFT duality, the CFT knows nothing about the string theory a priori. Hence, $\Sigma_g$ is just some Riemann surface without physical interpretation from the perspective of the CFT. It is known as the covering surface, since the map \linebreak $\Gamma: \Sigma_g \to \mathbb{CP}^1$ covers the boundary on which the CFT lives.

We would now like to take the alternative perspective and consider the correlation functions of the string theory. As will be proven in \S\ref{sec:proof}, these correlation functions are also described through covering maps, since they are localised to points in moduli space where a covering map exists. Interestingly, the Riemann surface $\Sigma_g$ in this context will now be interpreted as the worldsheet of the string theory, with vertex operators inserted on the worldsheet \cite{Pakman:2009zz}. Combining the two perspectives, the covering map provides a manifest construction to relate string theory observables inserted on the worldsheet to CFT observables inserted on the boundary.

This identification of the Riemann surface $\Sigma_g$ as the worldsheet is what makes a covering map localisation possible. The definition of the covering map given above, based on the data $\{x_i,w_i,z_i\}$, is an overconstrained system \cite{Dei:2020zui,Eberhardt:2020akk}. The complex dimension of the space of possible covering maps for the $n$-punctured genus $g$ Riemann surface $\Sigma_{g,n}$ based off the data $\{x_i,w_i,z_i\}$ is $3-3g-n$. This is generically negative and so a covering map need not exist. However, we notice that this is $-\dim (\mathcal{M}_{g,n})$ and so one would expect that an integral over moduli space will increase this dimension back to zero, giving rise to a discrete set of covering maps. This makes intuitive sense in the context of string theory correlators: consider the genus $g=0$ case, whilst $\{x_i,w_i:i=1,\dots, n\}$ and $\{z_i: i = 1, 2, 3\}$ are physical, the remaining $z_i$ are not physical as they break diffeomorphism invariance. The significance of this is that our localisation in \S\ref{sec:proof} will include a sum over the discrete set of possible covering maps.

\subsection{The $Q=0$ constraint}
\label{sec:Q=0}

Consider a ground state in the $w$-twisted sector for odd $w$, $[|\Phi_w\rangle]^{\sigma^w}$. We saw in \eqref{eq:FFR_phys_state_conditions} that such a state, if physical, must satisfy $Q_n[|\Phi_w\rangle]^{\sigma^w} = 0$ for all $n\geq 0$. $Q$ is a weight 3 gauge field, so in vertex operator language, this translates to
$$Q(z)V^w_{m_1,m_2}(x_i;z_i) = \mathcal{O}((z-z_i)^{-2}),$$
where we recall the relationship between $[|\Phi_w\rangle]^{\sigma^w}$ and $V^w_{m_1,m_2}$ \eqref{eq:odd_V}. We can, however, determine precisely what the OPE of any of the free fields with this vertex operator is purely from the representation theory \cite{Dei:2020zui}, as we did with $\xi^{\alpha}(z)$ in \eqref{eq:xi_V_OPE}. Strictly speaking, we do not need to work at the level of the free field realisation and only need the OPE of the supercurrents with $V^w_{m_1,m_2}$, the spectrally flowed highest weight states. To be explicit for our purposes here,
\begin{align*}
    \xi^+(z)[|\Phi_w\rangle]^{\sigma^w} &= \sum\limits_{r \in \mathbb{Z} + 1/2} z^{-r-1/2}\xi^+_r [|\Phi_w\rangle]^{\sigma^w} = z^{-\frac{w+1}{2}} \left[ |\tilde{\Phi}_w\rangle \right]^{\sigma^w} + \mathcal{O}(z^{-\frac{w-1}{2}}),\\
    \chi^{\pm}(z)[|\Phi_w\rangle]^{\sigma^w} &= \sum\limits_{r \in \mathbb{Z} + 1/2} z^{-r-1/2}\chi^{\pm}_r [|\Phi_w\rangle]^{\sigma^w} = \left[ \chi^{\pm}_{-\frac{1}{2} \pm \frac{w}{2}} |\Phi_w\rangle \right]^{\sigma^w} + \mathcal{O}(z),
\end{align*}
where
$$|\tilde{\Phi}_w\rangle = \chi^-_{-\frac{w-1}{2}}\psi^-_{-\frac{w-1}{2}}\cdots \chi^-_{-1}\psi^-_{-1}\chi^-_0\psi^-_0|m_1,m_2 + 1/2 \rangle.$$
Therefore, recalling \eqref{eq:supercurrents}, we deduce that
\begin{equation}
\label{eq:SS_V_OPE}
    S^{+++}S^{+-+}(z) V^w_{m_1,m_2}(x_i;z_i) \sim \mathcal{O}((z-z_i)^{-w-1}).
\end{equation}
It is not hard to adapt this result for the even $w$ case. Using \eqref{eq:even_V}, one can simply check that
$$\chi^+\chi^-(z) [|\Phi_w^{\pm}\rangle]^{\sigma^w} \sim \mathcal{O}(1),$$
whilst the OPE with $\xi^+(z)$ is once again of order $\mathcal{O}(z^{-\frac{w+1}{2}})$. The result analogous to \eqref{eq:SS_V_OPE} then follows.
\\

However, we noted in \eqref{eq:Q_bosonized} that $Q = -S^{+++}S^{+-+}\partial(e^{-\Sigma})$ and the three terms in this normal ordered expression all commute with one another. Hence, if $[|\Phi_w\rangle]^{\sigma^w}$ really is to be a physical state, it must be that
$$\partial(e^{-\Sigma})(z) V^w_{m_1,m_2}(x_i;z_i) \sim \mathcal{O}((z-z_i)^{w-1}),$$
such that $Q(z)V^w_{m_1,m_2}(x_i;z_i) \sim \mathcal{O}((z-z_i)^{-2})$ overall. But this is precisely the condition required for the covering map: it may be integrated to give
\begin{equation}
\label{eq:sigma_V_sketch_OPE}
    e^{-\Sigma}(z)V^w_{m_1,m_2}(x_i;z_i) = \text{ constant } + \mathcal{O}((z-z_i)^w).
\end{equation}
In other words, the $Q=0$ constraint suggests that inserting $e^{-\Sigma}$ inside a physical correlation function of ground states will generate a covering map localisation. We will do this explicitly in \S\ref{sec:proof}.

Moreover, an arbitrary physical state is in the linear span of twisted sector ground states that have been acted on by a string of DDF operators, which are a similarity transformation of \eqref{eq:DDF_operators}. Yet, $e^{-\Sigma}$ commutes with all of these DDF operators and as a result, if we act on a physical state with $e^{-\Sigma}$, the OPE with the ground states will reproduce \eqref{eq:sigma_V_sketch_OPE}. The covering map localisation will then extend to all physical correlation functions.

It is fascinating that this covering map localisation can be seen from the $Q=0$ constraint, combined with knowledge of the highest weight states and spectral flow. Importantly, $Q=0$ removes the half of the global supersymmetries which do not commute with $e^{-\Sigma}$. This appears to be necessary for localisation. Moreover, the $Q=0$ constraint provides a precise motivation for the incidence relation of \cite{Dei:2020zui,Knighton:2020kuh}, as we will discuss in the next subsection, and also provides a more efficient method for deriving the localisation directly in spacetime, without appealing to incidence relations in the twistor space.

\subsection{An alternative proof of the localisation}
\label{sec:proof}

We will now explicitly construct our proof of the covering map localisation. This has been proven before using an incidence relation \cite{Dei:2020zui,Knighton:2020kuh}. We will comment later on the equivalence of these two proofs, and will indeed use the equivalence for the final step.\footnote{
The methods developed in \S\ref{sec:radial_profile} suggest that it is possible to include secret representations that would complete our proof using $e^{-\Sigma}$ more efficiently. In particular, it should not be necessary to refer to the incidence relation to check for the number of poles in the covering map, as is done in Appendix \ref{sec:critical_points}.
}
Aside from the final step, this derivation of the localisation will be more succinct than the literature. It also directly writes the covering map formula in terms of $AdS_3$ coordinates, namely $\gamma(z)=e^{-\Sigma(z)}$ from the Wakimoto representation, rather than the coordinates of its twistor space $\xi^{\pm}(z)$. For different reasons, it has been suggested before in \cite{Eberhardt:2019ywk,Gaberdiel:2022oeu} that inserting the Wakimoto coordinate $\gamma(z)$ should lead to the covering map, but this was only verified up to first sub-leading order.
\\

Let's first analyse the OPE structure of $e^{-\Sigma}$. It is a simple exercise to show that
$$e^{-xJ_0^+}e^{-\Sigma}e^{xJ_0^+} = e^{-\Sigma} + x.$$
This allows us to compute the OPE with twisted sector ground states as
\begin{align}
\begin{split}
\label{eq:sigma_V_OPE}
    e^{-\Sigma(z)}\hat{V}_{m_1^i,m_2^i}^{w_i}(x_i;z_i) &= e^{x_iJ_0^+}(e^{-\Sigma(z)} + x_i)\hat{V}_{m_1^i,m_2^i}^{w_i}(0;z_i) e^{-x_iJ_0^+}\\
    &= x_i\hat{V}_{m_1^i,m_2^i}^{w_i}(x_i;z_i)\\
    &\quad + (z-z_i)^{w_i}e^{x_iJ_0^+} \exp \left[-\Sigma(z) + f^{w_i}_{m_1^i,m_2^i}(z_i) \right] e^{-x_iJ_0^+}\\
    &= x_i\hat{V}_{m_1^i,m_2^i}^{w_i}(x_i;z_i) + (z-z_i)^{w_i} \hat{V}_{m_1^i-1/2,m_2^i-1/2}^{w_i}(x_i;z_i)\\
    &\quad + \mathcal{O}((z-z_i)^{w_i+1}),
\end{split}
\end{align}
where we use the formula \eqref{eq:explicit_V} and we see that our intuition for \eqref{eq:sigma_V_sketch_OPE} was correct. Note that the prefactor of $e^{\pm\frac{1}{2}(iq_1 + iq_2)}$ in $\hat{V}^w_{m_1,m_2}$ for even $w$ has a trivial OPE with $e^{-\Sigma}$ so plays no role. This OPE is not specific to any choice of $m_1$ or $m_2$, meaning it works equally well whether we impose \eqref{eq:m1_m2_odd_constraint} or \eqref{eq:m1_m2_even_constraint}.

Hence, if we insert $e^{-\Sigma(z)}$ inside a correlator and take the limit $z \to z_k$,
\begin{equation*}
    \frac{\left\langle e^{-\Sigma(z)} \prod_{\alpha = 1}^{n-2+2g} W(u_{\alpha}) \prod_{i=1}^n \hat{V}^{w_i}_{m_1^i,m_2^i}(x_i;z_i) \right\rangle}{\left\langle \prod_{\alpha = 1}^{n-2+2g} W(u_{\alpha}) \prod_{i=1}^n \hat{V}^{w_i}_{m_1^i,m_2^i}(x_i;z_i) \right\rangle} = x_k + a_k (z-z_k)^{w_k} + \mathcal{O}((z-z_k)^{w_k+1}),
\end{equation*}
where
\begin{equation}
\label{eq:a_k}
  a_k = \frac{\left\langle \prod_{\alpha = 1}^{n-2+2g} W(u_{\alpha}) \prod_{i\neq k} \hat{V}^{w_i}_{m_1^i,m_2^i}(x_i;z_i) \hat{V}_{m_1^k - 1/2,m_2^k - 1/2}^{w_k}(x_k;z_k) \right\rangle}{\left\langle \prod_{\alpha = 1}^{n-2+2g} W(u_{\alpha}) \prod_{i=1}^n \hat{V}^{w_i}_{m_1^i,m_2^i}(x_i;z_i) \right\rangle}.  
\end{equation}
The shift in the values of $m_1^k$ and $m_2^k$ is such that the constraint \eqref{eq:spin_constraint} is preserved, meaning that the numerator is generically non-zero. In other words, whenever the physical correlation function of twisted sector ground states $\left\langle \prod_{\alpha = 1}^{n-2+2g} W(u_{\alpha}) \prod_{i=1}^n \hat{V}^{w_i}_{m_1^i,m_2^i}(x_i;z_i) \right\rangle \neq 0$, we may define
\begin{equation}
\label{eq:gamma_hat}
    \hat{\Gamma}(z) = \frac{\left\langle e^{-\Sigma(z)} \prod_{\alpha = 1}^{n-2+2g} W(u_{\alpha}) \prod_{i=1}^n \hat{V}^{w_i}_{m_1^i,m_2^i}(x_i;z_i) \right\rangle}{\left\langle \prod_{\alpha = 1}^{n-2+2g} W(u_{\alpha}) \prod_{i=1}^n \hat{V}^{w_i}_{m_1^i,m_2^i}(x_i;z_i) \right\rangle},
\end{equation}
which is a map $\hat{\Gamma} : \Sigma_g \to \C\P^1$. This is because $z$ is a worldsheet coordinate and $x_k$ a coordinate on the Riemann sphere. Moreover, $\hat{\Gamma}$ has the correct behaviour near the insertion points $z_i$ for it to define a covering map. All that remains to be proven is that it has no further critical points, which we will assume for now but is shown explicitly in Appendix \ref{sec:critical_points}. Using the known OPE structure of $e^{-\Sigma(z)}$, we could in principle find expressions for all higher order coefficients in the Taylor series of $\hat{\Gamma}$ as it approaches any of the insertion points.

Reversing the logic above, if a covering map does not exist, then we must have that $\left\langle \prod_{\alpha = 1}^{n-2+2g} W(u_{\alpha}) \prod_{i=1}^n \hat{V}^{w_i}_{m_1^i,m_2^i}(x_i;z_i) \right\rangle = 0$, implying a localisation to points in moduli space where a covering map exists. It is worth noting that we have not strictly proven the converse of this statement: if we are at a point in moduli space such that a covering map exists, it is not guaranteed that the correlation function will be non-zero, even if we expect it to be for non-trivial physics. We will highlight implications of this caveat below.\\

The correlation function is therefore localised to the set of points on ${\cal M}_{g,n}$ where the covering map $\hat{\Gamma}$ of \eqref{eq:gamma_hat} exists. To be concise, we will denote correlation functions (before the integral over moduli space) by
$$
D(\{x_i,w_i,z_i\}) = \left\langle \prod_{\alpha = 1}^{n-2+2g} W(u_{\alpha}) \prod_{i=1}^n \hat{V}^{w_i}_{m_1^i,m_2^i}(x_i;z_i) \right\rangle,
$$
which has support only on the points in moduli space where a covering map exists. Moreover, we must have that 
\begin{equation}\label{eq:corr_localisation}
    \int_{\mathcal{M}_{g,n}} D(\{x_i,w_i,z_i\}) = \sum_{\Gamma} C_{\Gamma},
\end{equation}
where $C_{\Gamma}$ are ordinary functions of the corresponding fixed points in moduli space and $\{x_i,w_i,m_1^i,m_2^i\}$, since the integrand can only be non-zero at the discrete points in moduli space where a covering map $\Gamma$ exists. The $C_{\Gamma}$ do not depend on the insertion points $\{u_{\alpha}\}$, since these correspond to the ($\hat{\sigma}^2$-spectrally flowed) vacuum \cite{Dei:2020zui}.

Under the assumption that the $C_{\Gamma}$ in \eqref{eq:corr_localisation} are non-zero, the above properties are sufficient to show that this distribution precisely satisfies the sampling property of the delta function, by integrating against a test function that is a continuous function of the moduli. Such localization in ${\cal M}_{g,n}$ is common in twistor string theory \cite{Witten:2003nn,Mason:2013sva}. We deduce that
\begin{equation}
\label{eq:localised_correlator}
    \left\langle \prod_{\alpha = 1}^{n-2+2g} W(u_{\alpha}) \prod_{i=1}^n \hat{V}^{w_i}_{m_1^i,m_2^i}(x_i;z_i) \right\rangle = \sum_{\Gamma}\widetilde{C}_{\Gamma} \prod_{I=1}^{n-3+3g}\delta \Big(f_I^{\Gamma} \Big), 
\end{equation}
where the $\{f_I^{\Gamma}\}$ are the constraints that pick out each of the covering maps in moduli space \cite{Eberhardt:2020akk,Knighton:2020kuh} and we acknowledge the possibility of non-trivial Jacobians from the delta functions by changing from $C_{\Gamma}$ to $\widetilde{C}_{\Gamma}$. It is understood that the proposed equality \eqref{eq:localised_correlator} holds in the context of the moduli space integral of \eqref{eq:corr_localisation}. Of course, whilst the case $C_{\Gamma} = 0$ does not give rise to the sampling property, one expects $C_{\Gamma}\neq 0$ for non-trivial physics --- we note that this has not been proven here or elsewhere in the literature to our knowledge. Finally, to maintain the full generality of the solution \eqref{eq:localised_correlator}, we should consider the solution as an equivalence class of distributions, where two distributions are equivalent if they differ only on a set of measure zero and by a finite amount at these points. This is why the possibility of $D(\{x_i,w_i,z_i\})$ being finite but non-zero at points where a covering map exists is contained in \eqref{eq:localised_correlator}.\\

Our insertion of the operator $e^{-\Sigma}$ even allows for a quick method to derive the recursion relation of \cite{Dei:2020zui,Knighton:2020kuh}, which provides a constraint on the dependence of $C_{\Gamma}$ on the parameters $\{m_1^i,m_2^i\}$. In the Taylor expansion of $\hat{\Gamma}(z)$ near an insertion point $z_k$, we learnt in equation \eqref{eq:a_k} that
\begin{equation*}
  a_k = \frac{\left\langle \prod_{\alpha = 1}^{n-2+2g} W(u_{\alpha}) \prod_{i\neq k} \hat{V}^{w_i}_{m_1^i,m_2^i}(x_i;z_i) \hat{V}_{m_1^k - 1/2,m_2^k - 1/2}^{w_k}(x_k;z_k) \right\rangle}{\left\langle \prod_{\alpha = 1}^{n-2+2g} W(u_{\alpha}) \prod_{i=1}^n \hat{V}^{w_i}_{m_1^i,m_2^i}(x_i;z_i) \right\rangle},
\end{equation*}
is the coefficient of the first non-vanishing term after the constant. This coefficient is a property of the covering map, so is determined by the data $\{x_i,w_i,z_i\}$ and is therefore independent of $\{m_1^i,m_2^i\}$. This means we can apply this formula recursively to deduce
\begin{align*}
    \Bigg\langle \prod_{\alpha = 1}^{n-2+2g} W(u_{\alpha}) &\prod_{i=1}^n \hat{V}^{w_i}_{m_1^i,m_2^i}(x_i;z_i) \Bigg\rangle\\
    &= \frac{1}{\left(a_k \right)^s}\left\langle \prod_{\alpha = 1}^{n-2+2g} W(u_{\alpha}) \prod_{i\neq k} \hat{V}^{w_i}_{m_1^i,m_2^i}(x_i;z_i) \hat{V}^{w_k}_{m_1^k - s/2, m_2^k - s/2} \right\rangle,
\end{align*}
for any integer $s$. Defining $h_i = m_1^i + m_2^i + w_i/2$, this leads to our final result
\begin{equation}
\label{eq:corr}
    \left\langle \prod_{\alpha = 1}^{n-2+2g} W(u_{\alpha}) \prod_{i=1}^n \hat{V}^{w_i}_{m_1^i,m_2^i}(x_i;z_i) \right\rangle = \sum\limits_{\Gamma} W_{\Gamma}\prod_{i=1}^n \left( a_i^{\Gamma} \right)^{-h_i} \prod_{I=1}^{n-3+3g} \delta(f_I^{\Gamma}),
\end{equation}
where $W_{\Gamma}$ are undetermined ordinary functions of the fixed points in moduli space for the covering map $\Gamma$ and $\{x_i,w_i\}$.\footnote{There is no dependence on $\{j_i\}$ since these all take the value of $j_i = \frac{1}{2}$.} The constants $a_i^{\Gamma}$ are precisely those given by \eqref{eq:a_k}, except that we also acknowledge precisely where we are in moduli space by labelling them with the corresponding covering map.\\

As was mentioned earlier, this localisation to covering maps \eqref{eq:corr} was previously argued using the following incidence relation: whenever a covering map exists, it is shown in \cite{Dei:2020zui} that
\begin{equation}
\label{eq:incidence_relation}
    \left\langle (\xi^-(z) + \tilde{\Gamma}(z)\xi^+(z)) \prod_{\alpha = 1}^{n-2} W(u_{\alpha}) \prod_{i=1}^n \hat{V}^{w_i}_{m_1^i,m_2^i}(x_i;z_i) \right\rangle = 0.
\end{equation}
In other words, provided the physical correlator is non-vanishing, they define a covering map via
\begin{equation}
\label{eq:gamma_tilde}
    \tilde{\Gamma}(z) = -\frac{\left\langle \xi^-(z) \prod_{\alpha = 1}^{n-2} W(u_{\alpha}) \prod_{i=1}^n \hat{V}^{w_i}_{m_1^i,m_2^i}(x_i;z_i) \right\rangle}{\left\langle \xi^+(z) \prod_{\alpha = 1}^{n-2} W(u_{\alpha}) \prod_{i=1}^n \hat{V}^{w_i}_{m_1^i,m_2^i}(x_i;z_i) \right\rangle}.
\end{equation}
The data $\{x_i,w_i\}$ that define the two covering maps $\tilde{\Gamma}$ and $\hat{\Gamma}$ are the same and so, if the $\{z_i\}$ are such that a covering map exists, we anticipate an equivalence of definitions between \eqref{eq:gamma_hat} and \eqref{eq:gamma_tilde} such that $\tilde{\Gamma} = \hat{\Gamma}$ in the discrete set of covering maps. As such, we will often drop the labels and refer to the covering map simply as $\Gamma$ unless we want to emphasise the definition. This equivalence can be seen from considering the bosonization of the free fields \eqref{eq:bosonization}, where $\xi^- = -e^{-\phi_1 - i\kappa_1}$ and we interpret $\zeta^- = e^{-\phi_2 - i\kappa_2}$ as a formal expression that inverts $\xi^+ = e^{\phi_2 + i\kappa_2}$ \cite{Naderi:2022bus},
\begin{equation}
\label{eq:zeta_OPE}
    \xi^+\zeta^- = 1.
\end{equation}
It then becomes clear that $\tilde{\Gamma}(z) = \hat{\Gamma}(z)$ from $e^{-\Sigma} = -\xi^-\zeta^-$. We therefore deduce that our proof using $e^{-\Sigma}$ is morally equivalent to that given in the literature, but in practice, could be considered more efficient.\footnote{It might be possible to explicitly derive the equivalence of these two covering map definitions using the techniques of \S\ref{sec:wakimoto_radial_coord}. To do so, one could realise $\mathfrak{sl}(2,\mathbb{R})_1 \oplus \mathfrak{u}(1)$ using the fields $(\beta,\gamma,\xi^+,\tilde{\eta}^-)$, where $\tilde{\eta}^- = -e^{-\phi_2-i\kappa_2}\partial(\phi_1+\frac{1}{2}\phi_2 + \frac{3}{2}i\kappa_2)$, before deriving $F_{\gamma\xi^+} = F_{\gamma}F_{\xi^+}$. The equivalence of definitions then immediately follows from $\gamma\xi^+ = -\xi^-$. We are yet to verify the factorisation of the path integral, explicitly checking the Jacobian from the change of basis and how the vertex operators factorise into pieces that depend independently on $\gamma$ and $\xi^+$.
}

The equivalence of $\hat{\Gamma}$ and $\tilde{\Gamma}$ reveals something deeper about the pole structure of the covering map, since we know that it must contain $N_{\Gamma} = \deg \Gamma$ poles. Without loss of generality, we can take all of the $\Gamma(z_i) = x_i$ to be finite. Moreover, it is shown in Appendix \ref{sec:critical_points} that the covering map is also finite at $\{u_{\alpha}\}$. If we define
\begin{equation}
\label{eq:omega_+-}
    \omega^{\pm}(z) = \left\langle \xi^{\pm}(z) \prod_{\alpha = 1}^{n-2+2g} W(u_{\alpha}) \prod_{i=1}^n \hat{V}^{w_i}_{m_1^i,m_2^i}(x_i;z_i) \right\rangle,
\end{equation}
then we know that the poles in $\omega^{\pm}(z)$ can only exist at insertion points (as a consequence of Wick's theorem) but these are not poles in $\Gamma$. This means that the poles of $\tilde{\Gamma}$ can only exist at the zeroes of $\omega^+(z)$ for $z \in \Sigma_g/\{z_i, u_{\alpha}:i=1,2,\dots , n; \; \alpha= 1, 2, \dots n-2+2g\}$. Returning to our new definition $\hat{\Gamma}$, this seems a little strange: the correlation function has poles at non-insertion points, a feature which requires explanation. In fact, these poles are a consequence of having extended the worldsheet theory to be described by the bosonization \eqref{eq:bosonization}, which introduces background charges at the preimages of the point at infinity on the boundary - see \S\ref{sec:Wakimoto_from_bosonization}. These represent spurious singularities as we will discuss in \S\ref{sec:global_symmetry}, and can be explained by the presence of the insertions \eqref{eq:Insertion}. We will also discuss a spacetime interpretation for these poles in \S\ref{sec:classical_hidden_poles}.
\\

Before proceeding, it is worth noting that $e^{-\Sigma}$ was found to be the boundary coordinate $\gamma(z)$ of the Wakimoto representation in \eqref{eq:wakimoto_realisation}. This boundary coordinate was found to act as an operator version of the covering map in \cite{Eberhardt:2019ywk,Gaberdiel:2022oeu}:
\begin{equation}
\label{eq:semi_classical_gamma}
    \left\langle \gamma(z) \prod_{\alpha = 1}^{n-2} W(u_{\alpha}) \prod_{i=1}^n \hat{V}^{w_i}_{m_1^i,m_2^i}(x_i;z_i) \right\rangle = \Gamma(z) \left\langle \prod_{\alpha = 1}^{n-2} W(u_{\alpha}) \prod_{i=1}^n \hat{V}^{w_i}_{m_1^i,m_2^i}(x_i;z_i) \right\rangle,
\end{equation}
giving the semi-classical solution $\gamma(z) \approx \Gamma(z)$. It is worth considering this statement in the context the two perspectives summarised in \S\ref{sec:connections_to_AdS_3}. In the large radius perspective of \cite{Eberhardt:2019ywk,Bhat:2021dez}, \eqref{eq:semi_classical_gamma} was verified up to first sub-leading order. The alternative perspective, of a free theory with $\delta({\cal S})$ insertions from \eqref{eq:Insertion}, does not require any such large radius assumption and the statement \eqref{eq:semi_classical_gamma}, with the appropriate factors of $\delta({\cal S})$ included, can be treated as an exact statement.

Finally, it is worth noting that \eqref{eq:incidence_relation} is very reminiscent of \eqref{eq:twistor_incidence_relation}, where we identify $\Gamma(z)$ with the boundary coordinate $\gamma(z)$ via \eqref{eq:wakimoto_realisation}. This dynamical constraint for a covering map localisation is related to an incidence relation for twistor space. As a consequence of the trivial OPE $\xi^+(z)e^{-\Sigma(z)} = -\xi^-(z)$, note that (\ref{eq:sigma_V_OPE}) implies
$$
\lim_{z\rightarrow z_i}\Big(\xi^-(z)+x_i\xi^+(z)\Big)\hat{V}_{m_1^i,m_2^i}^{w_i}(x_i;z_i)=0,
$$
for each individual vertex operator. Thus the usual interpretation that the vertex operator is inserted at the boundary point $x_i$ is indeed consistent with the boundary twistor incidence relation.

\subsection{Classical intuition for the covering map}
\label{sec:classical_covering_map}

We can see a hint of the special nature of the $k=1$ theory by considering the classical solutions to the model. As mentioned in \S\ref{sec:physical_states}, string theory on $AdS_3$ at generic $k$ contains two types of classical solutions: bound states corresponding to discrete representations of the universal cover of $SL(2,\mathbb{R})$ and long string solutions from the continuous representations \cite{Maldacena:2000hw}. Both the discrete and continuous representations include a label from the $\mathfrak{sl}(2,\mathbb{R})$-spin $j$. The spin is real and $j > 0$ for the discrete representations, whilst $j = \frac{1}{2} + is$ for real $s \geq 0$ for the continuous representations. We will focus only on the long string solutions here.

\begin{figure}[h]
\centering
\begin{minipage}{0.45\textwidth}
    \centering
    \includegraphics[scale = 0.4]{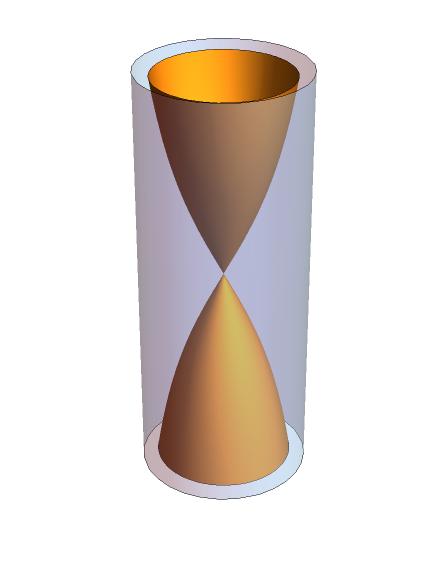}
    \caption{The long string classical solution. The edge of the cylinder is $\partial AdS_3$.}
    \label{fig:Long_string_solution}
\end{minipage}
\hspace{1cm}
\begin{minipage}{0.45\textwidth}
    \centering
    \includegraphics[scale=0.4]{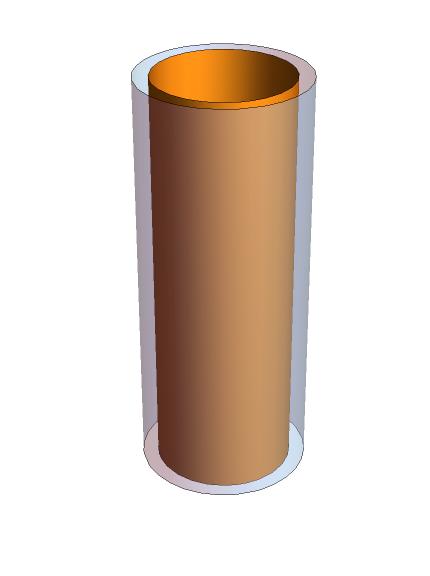}
    \caption{The bottom of the continuum where there is no radial momentum.}
    \label{fig:Static_solution}
\end{minipage}
\end{figure}

In global coordinates the metric of $H_3^+$ (Euclidean $AdS_3$) is given by
\begin{equation}
\label{eq:global_coords}
    ds^2 = \cosh^2\rho dt^2 + d\rho^2 + \sinh^2\rho d\theta^2,
\end{equation}
where $t \in \mathbb{R}$, $\rho \geq 0$ and $\theta \sim \theta + 2\pi$. Then the classical long string solutions are given by
$$t = w\tau, \quad \rho e^{i\theta} = \alpha \tau e^{iw\sigma},$$
where $(\tau,\sigma)$ are worldsheet coordinates, $w \in \mathbb{N}$ is the amount of spectral flow and $\alpha$ is a constant - see Figure \ref{fig:Long_string_solution}. We can also act on this solution with the $SL(2,\mathbb{R}) \times SL(2,\mathbb{R})$ isometry of the WZW model which introduces some fluctuations in the radial profile and gives the string angular momentum.

For these solutions, $\alpha$ is identified with $s$ and determines the radial momentum of the string to give a continuum of long string scattering states. There is an interesting solution at the bottom of this continuum with no radial momentum, where the string sits at constant radius for all time
\begin{equation}
\label{eq:static_solution}
    t = w\tau, \quad \rho = \rho_0, \quad \theta = w\sigma,
\end{equation}
see Figure \ref{fig:Static_solution}. This is the $j=\frac{1}{2}$ solution.

However, this generic classification of the spectrum is greatly simplified in the minimal tension limit at $k=1$, with the $j=\half$ solution forming the only unitary representation of the $\mathfrak{psu}(1,1|2)_1$ WZW model \cite{Eberhardt:2018ouy}. This means that only the constant radius solution is present at $k=1$\footnote{
The discrete representations correspond to spectrally flowed vacuum representations at $k=1$ so do not contribute. More generally, we can focus on the continuous representations since the discrete representations form subrepresentations. Correlation functions involving states in discrete representations can be extracted from residues in the correlation functions of states in continuous representations as their conformal weights are varied \cite{Dei:2022pkr}.}
and we therefore expect the whole worldsheet to be localised near the boundary of $AdS_3$. In fact, as we will now explain, this shortening of the worldsheet spectrum immediately implies that the worldsheet should classically be interpreted as a covering space for the boundary.
\\

\begin{figure}[h]
\centering
\begin{minipage}{0.45\textwidth}
    \centering
    \includegraphics[scale = 0.4]{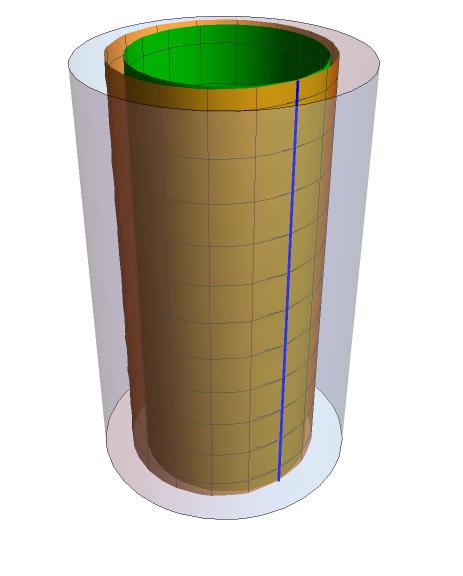}
    \caption{The Riemann surface for the 2-point function.}
    \label{fig:2_point_function}
\end{minipage}
\hspace{1cm}
\begin{minipage}{0.45\textwidth}
    \centering
    \includegraphics[scale=0.4]{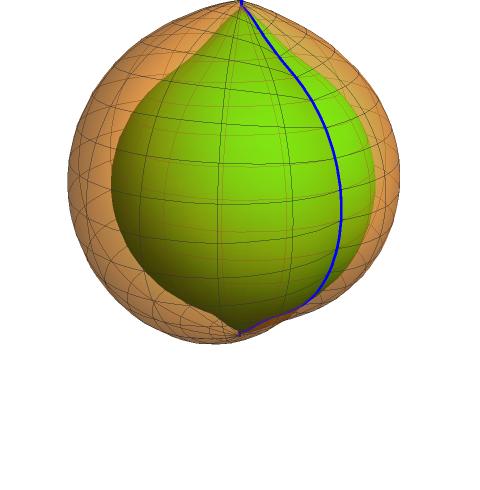}
    \caption{A compactified version of the 2-point function.}
    \label{fig:compactified}
\end{minipage}
\end{figure}

For simplicity, let's focus on the $j=\frac{1}{2}$ and $w = 2$ solution. In the quantum theory, we can think of this as a 2-point function and the worldsheet covers the boundary twice, see Figure \ref{fig:2_point_function}. Each sheet has been given a different colour and the line of intersection is shown in blue. Compactifying the timelike direction, the boundary becomes a sphere (which we have chosen to omit in Figure \ref{fig:compactified}) with the infinite future and past at the north and south poles, respectively.

In the compactified diagram of Figure \ref{fig:compactified}, one notices that there are branch points at the two poles, implying that insertion points of the $k=1$ string should generically correspond to ramification points of the worldsheet. If the insertion corresponds to a state with $w$-spectral flow, the ramification point should involve $w$ sheets all coinciding at a point.

Consider now a generic $n$-point function in the quantum theory. The corresponding classical configuration for the worldsheet is a Riemann surface built of multiple sheets and has ramification points of order $w_i$ at each insertion point $z_i$. This is precisely the definition of a covering surface for the boundary $S^2$ and moreover, the Riemann-Hurwitz formula guarantees that the number of sheets is given by
$$N_{\Gamma} = 1-g + \frac{1}{2}\sum_{i=1}^n (w_i -1).$$
In other words, the restriction to the bottom of the continuum of states at $k=1$ imposes that the only possible configurations for the worldsheet are covering surfaces (i.e. points for which a covering map exists).

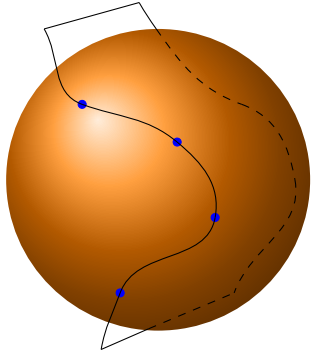
\begin{figure}[h]
    \centering
    \begin{tikzpicture}[scale=0.5]
        \shade[ball color=orange] (0,0) circle (4cm); 
        \filldraw [blue] (-2,2) circle (3pt);
        \filldraw [blue] (0.5,1) circle (3pt);
        \filldraw [blue] (1.5,-1) circle (3pt);
        \filldraw [blue] (-1,-3) circle (3pt);
        \draw (-3,4) to[out=300,in=160] [sloped,above] (-2,2);
        \draw (-2,2) to[out=340,in=140] [sloped,above] (0.5,1);
        \draw (0.5,1) to[out=320,in=80] [sloped,above] (1.5,-1);
        \draw (1.5,-1) to[out=260,in=70] [sloped,above] (-1,-3);
        \draw (-1,-3) to[out=250,in=80] [sloped,above] (-1.5,-4.5);
        \draw (-3,4) to (-0.5,4.7);
        \draw (-0.5,4.7) to[out=300,in=125] [sloped,above] (-0.05,4);
        \draw [dashed] (-0.05,4) to[out=305,in=160] [sloped,above] (2.2,2);
        \draw [dashed] (2.2,2) to[out=340,in=100] [sloped,above] (3.6,0);
        \draw [dashed] (3.6,0) to[out=280,in=70] [sloped,above] (2,-3);
        \draw (-1.5,-4.5) to (-0.25,-3.95);
        \draw [dashed] (-0.25,-3.95) to (2,-3);
    \end{tikzpicture}
    \caption{A slice through the boundary sphere that passes through all insertion points (the blue dots).}
    \label{fig:slice}
\end{figure}

To elaborate a little further, the possible non-zero contributions to correlation functions come from the distinct topological possibilities corresponding to how the sheets are ramified. We can represent this through what we call ``ramification diagrams''. Consider an $n$-point function with ramification indices $w_i$ at the insertion points $z_i$. We can take a slice through the boundary sphere which passes through each of these insertion points --- see Figure \ref{fig:slice} for an example with a 4-point function.

We then build ramification diagrams by considering the projection of the Riemann surface onto this slice. At each ramification point, there are $w_i$ coinciding sheets, which we sketch in Figure \ref{fig:ramification_point}. The distinct topological possibilities come from the different ways to glue these ramification points together globally --- an example of such a diagram is shown in Figure \ref{fig:ramification_diagram}. We can compactify the time direction in the diagram by connecting each sheet on the left hand side to the sheet on the right hand side of the same height. Each such diagram corresponds to a specific choice of covering map and we should sum over all distinct configurations in the correlation function. We anticipate that these ramification diagrams can be used to directly recreate the Feynman rules of the dual symmetric product orbifold \cite{Pakman:2009zz}.
\begin{figure}[h]
    \centering
    \begin{minipage}{0.45\textwidth}
        \centering
        \begin{tikzpicture}[scale = 0.65]
            \draw (-2,1) to (-1,1)
            (-1,1) to[out = 0, in = 110] (0,0)
            (0,0) to[out = 290, in = 180] (1,-1)
            (1,-1) to (2,-1);
            \draw (-2,-1) to (-1,-1)
            (-1,-1) to[out = 0, in = 250] (0,0)
            (0,0) to[out = 70, in = 180] (1,1)
            (1,1) to (2,1);
            \draw (-2,0) to (2,0);
            \filldraw [blue] (0,0) circle (3pt);
        \end{tikzpicture}
        \caption{The projection of a $w_i=3$ ramification point.}
        \label{fig:ramification_point}
    \end{minipage}
    \hspace{1cm}
    \begin{minipage}{0.45\textwidth}
        \centering
        \begin{tikzpicture}[scale = 0.65]
            \draw (-4,2) to (-3,2)
            (-3,2) to[out = 0, in = 110] (-2,1)
            (-2,1) to[out = 290, in = 180] (-1,0)
            (-1,0) to (0,0);
            \draw (-4,0) to (-3,0)
            (-3,0) to[out = 0, in = 250] (-2,1)
            (-2,1) to[out = 70, in = 180] (-1,2)
            (-1,2) to (0,2);
            \draw (-4,1) to (0,1);
            \draw (-4,-1) to (0,-1);
            \filldraw [blue] (-2,1) circle (3pt);
            
            \draw (0,2) to[out = 0, in = 110] (1,1.5)
            (1,1.5) to[out = 290, in = 180] (2,1)
            (2,1) to (3,1);
            \draw (0,1) to[out = 0, in = 250] (1,1.5)
            (1,1.5) to[out = 70, in = 180] (2,2)
            (2,2) to (3,2);
            \filldraw [blue] (1,1.5) circle (3pt);
            
            \draw (0,0) to[out = 0, in = 110] (1,-0.5)
            (1,-0.5) to[out = 290, in = 180] (2,-1)
            (2,-1) to (3,-1);
            \draw (0,-1) to[out = 0, in = 250] (1,-0.5)
            (1,-0.5) to[out = 70, in = 180] (2,0)
            (2,0) to (3,0);
            \filldraw [blue] (1,-0.5) circle (3pt);
            
            \draw (3,1) to[out = 0, in = 110] (4,0)
            (4,0) to[out = 290, in = 180] (5,-1)
            (5,-1) to (6,-1);
            \draw (3,-1) to[out = 0, in = 250] (4,0)
            (4,0) to[out = 70, in = 180] (5,1)
            (5,1) to (6,1);
            \draw (3,0) to (6,0);
            \draw (3,2) to (6,2);
            \filldraw [blue] (4,0) circle (3pt);
    \end{tikzpicture}
    \caption{One possible ramification diagram for the 4-point function with ramification indices $\underline{w} = (3,2,2,3)$.}
    \label{fig:ramification_diagram}
    \end{minipage}
\end{figure}

\section{The radial profile of the worldsheet}
\label{sec:radial_profile}

Building on the conjecture that the free field action \eqref{eq:S_W_hat} is exact at $k=1$, we shall demonstrate that correlation functions of ground states \eqref{eq:V_hat} in the $k=1$ theory have a global symmetry $\Phi(z)\rightarrow \Phi(z)+\alpha$, generated by the charge
$$
{\cal K}=2 \oint \mathrm{d}z \; \p\Phi(z),
$$
where $\Phi(z)$ is defined in \eqref{eq:wakimoto_realisation}. The proof of this will be given in \S\ref{sec:global_symmetry}. We first investigate the pole structure of correlation functions with ${\cal K}$ insertions in \S\ref{sec:wakimoto_radial_coord} and the results of this investigation will prove useful in \S\ref{sec:global_symmetry}. We shall gain useful insight into the relationship of the $k=1$ string and the Wakimoto formulation along the way. The significance of this symmetry is that $\Phi$ plays a role in the $k=1$ theory analogous to the field in the Wakimoto formulation, i.e. it encodes information about the radial direction in the target space.

\subsection{The Wakimoto radial coordinate}
\label{sec:wakimoto_radial_coord}

This is a somewhat technical section, the key purpose of which is to show that
\begin{eqnarray}\label{me}
\oint_{C_0} \mathrm{d}z \; 
      \frac{\left\langle \partial\Phi(z)  \prod_{\alpha = 1}^{n-2+2g} W(u_{\alpha}) \prod_{i=1}^n \hat{V}^{w_i}_{m_1^i,m_2^i}(x_i;z_i) \right\rangle}{\left\langle \prod_{\alpha = 1}^{n-2+2g} W(u_{\alpha}) \prod_{i=1}^n \hat{V}^{w_i}_{m_1^i,m_2^i}(x_i;z_i) \right\rangle}
      = \left( \sum_{i=1}^n\frac{w_i+1}{2}-N_{\Gamma} \right),
\end{eqnarray}
where $C_0$ is a contour around $z=0$ traversed anticlockwise.
\\

Our operator for the boundary coordinate of the Wakimoto representation $\gamma = e^{-\Sigma}$ gave a realisation of the semi-classical solution \eqref{eq:semi_classical_gamma} in the quantum theory. There is a similar ground state semi-classical solution for the radial coordinate \cite{Eberhardt:2019ywk}
\begin{equation}
\label{eq:semi_classical_phi}
    \left\langle \partial\Phi(z) \prod_{\alpha = 1}^{n-2} W(u_{\alpha}) \prod_{i=1}^n \hat{V}^{w_i}_{m_1^i,m_2^i}(x_i;z_i) \right\rangle = -\frac{\partial^2\Gamma(z)}{2\partial\Gamma (z)} \left\langle \prod_{\alpha = 1}^{n-2} W(u_{\alpha}) \prod_{i=1}^n \hat{V}^{w_i}_{m_1^i,m_2^i}(x_i;z_i) \right\rangle.
\end{equation}
One might hope to recreate this semi-classical solution in the quantum theory by inserting
\begin{equation}
\label{eq:phi_operator}
    \partial\Phi = \frac{1}{2}\partial(\phi_1 + \phi_2 + 2i\kappa_2),
\end{equation}
and using \eqref{eq:gamma_hat} as our definition of the covering map. We will see that this is almost, but not quite, the case. First define
\begin{equation}
\label{eq:phi}
    \partial\varphi(z) := \frac{1}{2}\frac{\left\langle \partial(\phi_1+\phi_2 + 2i\kappa_2)(z) \prod_{\alpha = 1}^{n-2+2g} W(u_{\alpha}) \prod_{i=1}^n \hat{V}^{w_i}_{m_1^i,m_2^i}(x_i;z_i) \right\rangle}{\left\langle \prod_{\alpha = 1}^{n-2+2g} W(u_{\alpha}) \prod_{i=1}^n \hat{V}^{w_i}_{m_1^i,m_2^i}(x_i;z_i) \right\rangle},
\end{equation}
as our Wakimoto radial coordinate in the quantum theory. And so, to make contact with \eqref{eq:semi_classical_phi}, we seek to understand the singularity structure of $\p\varphi(z)$. As usual, we shall see there are poles near insertion points $z_i$ but that there are also poles due to the fact that $\Phi(z)$ is not a conformal tensor.

\subsubsection*{Poles of $\p\varphi(z)$ as $z\rightarrow z_i$}

The behaviour of $\partial\varphi(z)$ near insertion points is determined by the OPE of $\partial\Phi(z)$ with $\hat{V}^{w_i}_{m_1^i,m_2^i}(x_i;z_i)$. We can deal with odd and even $w_i$ simultaneously since there is no dependence on the free fermions in \eqref{eq:phi_operator}. Moreover, the OPE of $J^+(z)$ with $\partial\Phi(z)$ is trivial, so there is no $x_i$-dependence in the OPE with $\hat{V}^{w_i}_{m_1^i,m_2^i}(x_i;z_i)$. Therefore, using \eqref{eq:explicit_V} we deduce
\begin{align*}
    \partial\Phi(z) \hat{V}^{w_i}_{m_1^i,m_2^i}(x_i;z_i) &= \frac{1}{2}\frac{1}{z-z_i}\left( 2m^i_2 - 2m^i_1 - w_i  \right) \hat{V}^{w_i}_{m_1^i,m_2^i}(x_i;z_i) + \dots \\
    &= -\frac{w_i+1}{2}\frac{1}{z-z_i} \hat{V}^{w_i}_{m_1^i,m_2^i}(x_i;z_i) + \dots,
\end{align*}
where we have used $m_1 - m_2 = \frac{1}{2}$ from \eqref{eq:m1_m2_odd_constraint}\footnote{
A key step in this calculation is that in \S\ref{sec:physical_correlators} we chose to set $j_i = \frac{1}{2}$ for all insertions $\hat{V}^{w_i}_{m_1^i,m_2^i}$ and interpret the insertions $\mathbf{Q}(\mu_I)$ as carrying a spin of $j = -1$.}.
This suggests that, as $z \to z_i$,
$$\partial\varphi(z) = -\frac{w_i + 1}{2}\frac{1}{z-z_i} + \dots,$$
which is slightly different to \eqref{eq:semi_classical_phi}, where
$$-\frac{\partial^2 \Gamma}{2\partial\Gamma} = -\frac{w_i-1}{2}\frac{1}{z-z_i} + \dots$$
This is necessary to ensure that all vertex operators are indeed inserted at the boundary of $AdS_3$: as $z \to z_i$, we have that
$$\varphi(z) \sim -\frac{w_i+1}{2}\log (z-z_i) \to +\infty$$
for all $w_i \in \mathbb{N}$. By contrast, any $w_i = 1$ insertions need not be at the boundary if we were to simply apply \eqref{eq:semi_classical_phi}.

\subsubsection*{$\p\varphi(z)$ is regular at $z=u_{\alpha}$}

To uncover the full pole structure of \eqref{eq:phi}, we should also consider $z \to u_{\alpha}$. Recall that the field $W(u_{\alpha})$ corresponds to the state $|0\rangle^{(1)} = [\psi^+_{-3/2}\psi^-_{-3/2}\psi^+_{-1/2}\psi^-_{-1/2}|0\rangle]^{\hat{\sigma}^2}$, such that
\begin{equation}
\label{eq:W_phi_OPE}
    \partial\Phi(z) |0\rangle^{(1)} = \left[ \hat{\sigma}^2\left( \partial\Phi \right)(z) \psi^+_{-3/2}\psi^-_{-3/2}\psi^+_{-1/2}\psi^-_{-1/2} |0\rangle \right]^{\hat{\sigma}^2}.
\end{equation}
To perform this spectral flow, first note that
$$\hat{\sigma}^2 (\xi^{\alpha})(z) = z\xi^{\alpha}(z), \quad \hat{\sigma}^2 (\eta^{\alpha})(z) = \frac{1}{z}\eta^{\alpha}(z),$$
from \eqref{eq:sigma_hat}. This can be translated into the bosonized variables \eqref{eq:bosonization} via
$$\hat{\sigma}^2(\phi_1)(z) = \phi_1(z) - \log z, \quad \hat{\sigma}^2(\phi_2)(z) = \phi_2(z) + \log z, \quad \hat{\sigma}^2(\kappa_i)(z) = \kappa_i(z),$$
from which we deduce that $\hat{\sigma}^2\left( \partial\Phi \right)(z) = \partial\Phi(z)$ is invariant. Since the OPE of $\p\Phi(z)$ with $\psi^{\pm}(u_{\alpha})$ is trivial, we conclude that \eqref{eq:W_phi_OPE} must be regular as $z \to 0$ and the OPE of $\partial\Phi(z)$ with $W(u_{\alpha})$ is trivial. We conclude that there are no poles at $z=u_{\alpha}$.

\subsubsection*{Hidden poles of $\p\varphi(z)$}

Is it possible for there to be other poles so far not accounted for? If there are any other poles in \eqref{eq:phi}, they must come from non-insertion points, which is a consequence of working with the bosonized theory as we have seen before. This might seem strange but we must note that $\Phi(z)$ does not transform as a conformal tensor --- it has an anomaly as we'll see in \S\ref{sec:global_symmetry}. In this section, we will give general consistency arguments to motivate the existence of these poles but, as we shall see later, the insertion of the delta functions in \eqref{eq:Gerasimov_correlator} provide a concrete realisation of these general considerations.

To check for such poles, it will be helpful to have an analogous statement to \eqref{eq:zeta_OPE}. This comes from noticing that $\partial\Phi$ has a trivial OPE with $J^+ = \xi^+\eta^+ = e^{\Sigma}\partial(i\kappa_1)$. The normal ordered product is then given by
\begin{equation}
\label{eq:phi_J_OPE}
    \begin{split}
        \partial\Phi J^+ &= \frac{1}{2}\partial(\phi_1 + \phi_2 + 2i\kappa_2) \partial(i\kappa_1) e^{\Sigma}\\
        &= \partial(\phi_2 + i\kappa_2) \partial(i\kappa_1) e^{\Sigma} + \frac{1}{2}\partial(\phi_1 - \phi_2) \partial(i\kappa_1)e^{\Sigma}\\
        &= \eta^+ \partial \xi^+ - U J^+,
    \end{split}
\end{equation}
which is contained in the free field realisation. This implies that, if we insert this normal ordered expression inside a physical correlator, it will be finite away from all insertion points. We therefore conjecture that poles in $\partial\varphi(z)$ at non-insertion points can only exist at the zeroes of
\begin{equation*}
\label{eq:J_correlator}
    \left\langle J^+(z) \prod_{\alpha = 1}^{n-2+2g} W(u_{\alpha}) \prod_{i=1}^n \hat{V}^{w_i}_{m_1^i,m_2^i}(x_i;z_i) \right\rangle.
\end{equation*}
Stated differently, $\partial\Phi J^+$ is a conformal tensor and the zeroes of $J^+$ ensure that the places where $\p\Phi$ behaves badly under conformal transformations do not ruin this fact.

Let us introduce some useful notation. We will denote a correlation function weighted average of a field $\Psi(z)$ as
\begin{equation}\label{eq:F_Psi}
    F_{\Psi}(z):=\frac{\left\langle \Psi(z) \prod_{\alpha = 1}^{n-2+2g} W(u_{\alpha}) \prod_{i=1}^n \hat{V}^{w_i}_{m_1^i,m_2^i}(x_i;z_i) \right\rangle}{\left\langle \prod_{\alpha = 1}^{n-2+2g} W(u_{\alpha}) \prod_{i=1}^n \hat{V}^{w_i}_{m_1^i,m_2^i}(x_i;z_i) \right\rangle}.
\end{equation}
We show in Appendix \ref{sec:expectations} that there is an exact factorisation $F_{\p\Phi J^+}(z) = F_{\p\Phi}(z) F_{J^+}(z)$, which follows from the fact that $\p\Phi(z)$ and $J^+(z)$ commute. Since $\p\varphi(z) = F_{\p\Phi}(z)$, this factorisation along with \eqref{eq:phi_J_OPE} confirms that poles in $\p\varphi(z)$ away from insertion points can only occur at the zeroes of $F_{J^+}(z)$.

Recalling $J^+ = \xi^+\eta^+$, we will assume that $J^+(z) \to 0$ when either $\xi^+(z) \to 0$ or $\eta^+(z) \to 0$. This assumption is supported by our discussion of background charges in \S\ref{sec:Wakimoto_from_bosonization}. The locations of the zeroes of $\xi^+(z)$ are given by $\{z_a^*\}$ (we know from Appendix \ref{sec:critical_points} that $a = 1, \dots N_{\Gamma}$ such that $M = N_{\Gamma}$) and we denote the zeroes of $\eta^+(z)$ by $\{t_k\}$ for some label $k$. These provide the possible locations for the ``hidden poles'' of $\p\varphi(z)$. However, as we will now show, the $\{t_k\}$ do not in fact give rise to poles in $\p\varphi(z)$.

We learn from \eqref{eq:phi_J_OPE} that
$$F_{\p\Phi J^+}(z) = F_{\eta^+\p\xi^+}(z) - F_{UJ^+}(z).$$
The fields $U$ and $J^+$ can be viewed independently in the path integral, so we can use the same intuition as above to justify $F_{UJ^+}(t_k) = F_{UJ^+}(z_a^*) = 0$, since these points correspond to zeroes of $F_{J^+}(z)$. Note that $U$ is a conformal tensor contained in the free field realisation, so it will not have any poles away from insertion points to counteract these zeroes of $F_{J^+}(z)$.\footnote{In fact, the arguments of Appendix \ref{sec:expectations} can simply be applied to this case to show an analogous factorisation $F_{UJ^+}(z) = F_U(z)F_{J^+}(z)$ from which the zeroes at $z\in \{z_a^*,t_k\}$ trivially follow.} Moreover, we can deduce that $F_{\eta^+\p\xi^+}(z)$ has a zero whenever $\eta^+ \to 0$, such that $F_{\eta^+\p\xi^+}(t_k) = 0$ for all $k$. Overall, we find that $F_{\p\Phi J^+}(t_k) = F_{\p\Phi}(t_k)F_{J^+}(t_k) = 0$, such that there is no pole in $\p\varphi(z)$ at the points $t_k$.

By contrast, whilst $F_{UJ^+}(z_a^*) = 0$, this is not the case for $F_{\eta^+\p\xi^+}(z_a^*)$, leading to hidden poles of $\p\varphi(z)$. Applying the product rule and the linearity in $\Psi$ in the definition \eqref{eq:F_Psi} of $F_{\Psi}(z)$,
$$F_{\eta^+\p\xi^+}(z) = \p F_{J^+}(z) - F_{\p\eta^+ \xi^+}(z).$$
Again, the second term will vanish at $z_a^*$, leaving\footnote{If $z_a^*$ is a higher order zero of $F_{J^+}(z)$, then of course $\p F_{J^+}(z_a^*) = 0$. However, in the following we could account for this by simply expanding in a Taylor series as $z \to z_a^*$ to find the first non-zero contribution. The final result \eqref{eq:hidden_pole} will persist to hold after Taylor expanding the numerator, except that the residue of the simple pole will be given by the degree of the zero. We do not expect higher order zeroes to appear based off our background charge arguments and we therefore make the assumption that $\p F_{J^+}(z_a^*) \neq 0$.}
$$F_{\p\Phi J^+}(z_a^*) = F_{\eta^+\p\xi^+}(z_a^*) = \p F_{J^+}(z_a^*).$$
It then follows that, as $z \to z_a^*$,
\begin{align*}
    F_{\p\Phi J^+}(z) &= F_{\p\Phi}(z)F_{J^+}(z)\\
    &= F_{\p\Phi}(z) \left[(z-z^*_a)\p F_{J^+}(z^*_a) + \mathcal{O}((z-z_a^*)^2) \right]\\
    &= F_{\p\Phi}(z) \left[(z-z^*_a)F_{\p\Phi J^+}(z^*_a) + \mathcal{O}((z-z_a^*)^2) \right].
\end{align*}
Rearranging, we have that
\begin{equation}\label{eq:hidden_pole}
    F_{\p\Phi}(z)=\frac{1}{z-z^*_a}\frac{F_{\p\Phi J^+}(z)}{F_{\p\Phi J^+}(z^*_a)}+...=\frac{1}{z-z^*_a}+...
\end{equation}
in the limit as $z \to z_a^*$. This is precisely the behaviour expected from \eqref{eq:semi_classical_phi}, where we note that $\{z_a^*\}$ are the simple poles of $\Gamma(z)$ by \eqref{eq:gamma_tilde}. Note that an alternative derivation of this statement followed from a consideration of the $\kappa_2$ background charge in \S\ref{sec:Wakimoto_from_bosonization}.
\\

In summary, we find that 
$$
\p\varphi(z) = F_{\p\Phi}(z)=\frac{\left\langle \p\Phi(z) \prod_{\alpha = 1}^{n-2+2g} W(u_{\alpha}) \prod_{i=1}^n \hat{V}^{w_i}_{m_1^i,m_2^i}(x_i;z_i) \right\rangle}{\left\langle \prod_{\alpha = 1}^{n-2+2g} W(u_{\alpha}) \prod_{i=1}^n \hat{V}^{w_i}_{m_1^i,m_2^i}(x_i;z_i) \right\rangle}=-\frac{1}{2}\frac{w_j+1}{z-z_j}+...,
$$
as $z \to z_j$, where $z_j$ is the location of one of the punctures and 
$$
\p\varphi(z) = F_{\p\Phi}(z)=\frac{\left\langle \p\Phi(z) \prod_{\alpha = 1}^{n-2+2g} W(u_{\alpha}) \prod_{i=1}^n \hat{V}^{w_i}_{m_1^i,m_2^i}(x_i;z_i) \right\rangle}{\left\langle \prod_{\alpha = 1}^{n-2+2g} W(u_{\alpha}) \prod_{i=1}^n \hat{V}^{w_i}_{m_1^i,m_2^i}(x_i;z_i) \right\rangle}=\frac{1}{z-z_a^*}+...,
$$
as $z\rightarrow z_a^*$, where $z_a^*$ is a zero of $\xi^+(z)$ (or equivalently a zero of $\omega^+(z)$ from \eqref{eq:omega_+-}). By contrast, $\p\varphi(z)$ is regular near the $W(u_{\alpha})$ insertions. Integrating $\p\varphi(z)$ around a contour at $z=0$ which excludes the vertex operators and zeroes of $\xi^+(z)$ gives the expression (\ref{me}).

\subsection{An unexpected global symmetry}
\label{sec:global_symmetry}

We noted earlier in \S\ref{sec:connections_to_AdS_3} that the Wakimoto fields \eqref{eq:wakimoto_realisation} have OPEs generated by the action\footnote{Whilst we choose to work with the perspective of free fields with operator insertions via \eqref{eq:S_W_hat}, the following considerations would equally hold for \eqref{eq:FreeWaki} in a large $\Phi$ limit. Of course, to what extent the global radial symmetry holds under this assumption is unclear and hence, our perspective appears more natural.}
\begin{equation*}
\label{eq:OPE_action}
    S = -\frac{1}{4\pi} \int \mathrm{d}^2z \; \left( 4\partial\Phi \bar{\partial} \Phi  + \beta \bar{\partial}\gamma \right).
\end{equation*}
Naively, it appears that this action contains a global symmetry associated to radial translations, $\Phi \mapsto \Phi + \alpha$ for constant $\alpha$, generated by the charge
$${\cal K} = 2 \oint \mathrm{d}z \; \partial\Phi.$$
As we may see by adding a Fradkin-Tseytlin term
\begin{equation}
\label{eq:quantum_wakimoto_action}
    S = -\frac{1}{4\pi} \int \mathrm{d}^2z \; \left( 4\partial\Phi \bar{\partial} \Phi +  \beta \bar{\partial}\gamma + qR\Phi \right),
\end{equation}
on a curved worldsheet, this symmetry is potentially anomalous. In particular, the action changes as \cite{DiFrancesco:1997nk}
$$\delta_{{\cal K}} S = -\frac{q\alpha}{4\pi} \int \mathrm{d}^2z \; \sqrt{h}R(h) = -2q\alpha(1-g).$$
We found that the background charge was $q=1$ in \eqref{eq:S_W_hat}, by performing a coordinate transformation from the standard Euclidean $AdS_3$ action. Alternatively, we can pin down the background charge by studying how $\Phi(z)$ transforms under conformal transformations. The general story is reviewed in Appendix \ref{sec:linear_dilaton} and, for the case at hand, the  stress tensor associated to \eqref{eq:quantum_wakimoto_action} is \cite{Naderi:2022bus}
$$T = (\partial\Phi)^2 - \partial^2\Phi + \beta\partial\gamma,$$
from which we read off $q=1$. The field $\partial\Phi$ is quasiprimary with OPE
$$T(z)\partial\Phi(w) = \frac{1}{(z-w)^3} + \frac{\partial\Phi}{(z-w)^2} + \frac{\partial^2\Phi}{z-w} + \dots$$
This implies that $\Phi(z)$ is not a conformal tensor.\footnote{
The finite transformation law is \cite{Eberhardt:2019ywk}
$$\partial\Phi(z) \mapsto \partial\Phi (z) = \left[\partial f \left(f^{-1}(z)\right) \right]^{-1} \partial\Phi(f^{-1}(z)) - \frac{\partial^2 f\left( f^{-1}(z) \right)}{2\left[\partial f \left( f^{-1}(z) \right) \right]^2},$$
under a conformal transformation $z \mapsto f(z)$.
}
In particular, under the inversion transformation $z \mapsto -\frac{1}{z}$,
$$\partial\Phi (z) \mapsto \frac{1}{z^2} \partial\Phi \left(-\frac{1}{z} \right) - \frac{1}{z},$$
such that $(\partial\Phi)_0^{\dagger} = -(\partial\Phi)_0 - 1$.
This anomalous transformation law is precisely why we found additional poles away from insertion points in $\partial\varphi(z)$, which came with a residue of 1. These poles should be interpreted as the location of background charges for $\Phi$ (or equivalently for $i\kappa_2$) which are spurious singularities. From the boundary perspective, these charges should sit at $x = \infty$, which are the poles of the covering map. Recall that our choice of chart in \S\ref{sec:target_spaces} covered the region where $\xi^+\neq 0$ and so the point $x=\infty$ corresponds to these omitted points via the incidence relation $x=\gamma=-\frac{\xi^-}{\xi^+}$. This means we have precisely $N_{\Gamma}$ charges sitting at $\{z_a^*\}$ (the preimages of $x=\infty$) as argued earlier.

To better understand what is going on, we consider the behaviour of a correlation function of ground states \eqref{eq:V_hat} under this transformation generated by ${\cal K}$. After the integral over moduli space, correlation functions depend on the following data: the location $\{x_i\}$ on the boundary where the vertex operators are inserted and the ramification $\{w_i\}$ at that insertion. For brevity we denote this data by two $n$-vectors $\underline{x}=(x_1,...,x_n)$ and $\underline{w}=(w_1,...,w_n)$ respectively. A correlation function ${\cal C}_g(\underline{x},\underline{w})$ may be written as a path integral\footnote{Since the worldsheet fields may have NS or R boundary conditions, there is also a GSO projection from the sum over worldsheet spin structures. Moreover, to implement the gauging of the $U(1)$ current $Z$ of the free field realisation, we must integrate over the moduli space Jac$(\Sigma_g)$ of flat $U(1)$-bundles over the Riemann surface \cite{Eberhardt:2021jvj}. Finally, if we take our alternative interpretation for the Wakimoto free fields, we should insert the delta functions of \eqref{eq:Gerasimov_correlator}.}
\begin{equation}
\label{eq:C_g}
\begin{split}
    \mathcal{C}_g(\underline{x},\underline{w}) &= \left\langle \prod_{\alpha = 1}^{n-2+2g} W(u_{\alpha}) \prod_{i=1}^n \hat{V}^{w_i}_{m_1^i,m_2^i}(x_i;z_i) \right\rangle \\
    &= \int_{{\cal M}_{g,n}} d\mu\; \int{\cal D}\Psi\, e^{-S[\Psi]}\,\prod_{\alpha = 1}^{n-2+2g} W(u_{\alpha}) \prod_{i=1}^n \hat{V}^{w_i}_{m_1^i,m_2^i}(x_i;z_i),
\end{split}
\end{equation}
where $d\mu$ is an appropriate measure on the moduli space ${\cal M}_{g,n}$ of $n$-punctured, genus $g$ Riemann surfaces, $\Psi$ is a generic label denoting all of the fields in the theory and $S[\Psi]$ is the action \eqref{eq:S_W_hat} which includes the Fradkin-Tseytlin term for $\Phi$. We will say more about $d\mu$ below but for now, we take it to include $3g-3$ $b$-ghost insertions
$$
\mathbf{b}(\mu_I)=\int_{\Sigma}d^2z\,G^-(z)\mu_I(z) =\int_{\Sigma}d^2z\,b(z)\mu_I(z) + \dots, 
$$
where the ellipsis denotes the compact part of $G^-$, which we know from Appendix \ref{sec:hybrid_correlators} does not contribute to physical correlation functions. Moreover, $\{\mu_I\}$ denote a basis of Beltrami differentials so that
$$
d\mu=\prod_{I=1}^{n+3g-3}\mathbf{b}(\mu_I)d\tau_I,
$$
where $\{\tau_I\}$ are coordinates on ${\cal M}_{g,n}$. Assuming $d\mu$ is invariant under the transformation generated by ${\cal K}$, the variation of the correlation function can be found using the commutator $\delta_{\cal K}(\cdot) = \left[ \alpha{\cal K}, \cdot \right]$,
\begin{align*}
    \delta_{{\cal K}} {\cal C}_g(\underline{x},\underline{w}) &= -2\alpha\oint_{C_0} \mathrm{d}z \; \left\langle \partial\Phi(z) \prod_{\alpha = 1}^{n-2+2g} W(u_{\alpha}) \prod_{i=1}^n \hat{V}^{w_i}_{m_1^i,m_2^i}(x_i;z_i) \right\rangle\\
    &\quad + 2\alpha(1-g)\left\langle \prod_{\alpha = 1}^{n-2+2g} W(u_{\alpha}) \prod_{i=1}^n \hat{V}^{w_i}_{m_1^i,m_2^i}(x_i;z_i) \right\rangle
\end{align*}
where the second term comes from the variation of the Fradkin-Tseytlin term. The minus sign in front of the first term comes from viewing the contour $C_0$ to be around the complement of the disc around $z=0$ in $\Sigma_g$, rather than $z=0$. As in (\ref{me}), we rewrite the first term as
\begin{align*}
-\oint_{C_0} \mathrm{d}z \; \p\varphi(z){\cal C}_g(\underline{x},\underline{w})& = -\oint_{C_0} \mathrm{d}z \; \left( -\frac{1}{2}\sum_{i=1}^n\frac{w_i+1}{z-z_i}+\sum_{a=1}^{N_{\Gamma}}\frac{1}{z-z^*_a}\right){\cal C}_g(\underline{x},\underline{w})\\
&=\left(-\frac{1}{2}\sum_{i=1}^n\oint_{C_i} \mathrm{d}z \;  \frac{w_i+1}{z-z_i} + \sum_{a=1}^{N_{\Gamma}}\oint_{C_a} \mathrm{d}z \; \frac{1}{z-z^*_a}\right){\cal C}_g(\underline{x},\underline{w})\\
&=-\left(\sum_{i=1}^n\frac{w_i+1}{2}-N_{\Gamma}\right){\cal C}_g(\underline{x},\underline{w}),
\end{align*}
where $C_i$ and $C_a$ are contours around the insertion points $z_i$ and hidden poles $z_a^*$, respectively. We may then rewrite the variation of a correlation function under ${\cal K}$ as
\begin{align*}
 \delta_{{\cal K}} {\cal C}_g(\underline{x},\underline{w}) &= 2\alpha \left[ -\oint \mathrm{d}z \; \partial\varphi(z) + (1-g) \right]
      \left\langle \prod_{\alpha = 1}^{n-2+2g} W(u_{\alpha}) \prod_{i=1}^n \hat{V}^{w_i}_{m_1^i,m_2^i}(x_i;z_i) \right\rangle\\
    & =2\alpha \left[  \;  (1-g) - \left( \sum_{i=1}^n\frac{w_i+1}{2}-N_{\Gamma} \right) \right]
      \left\langle \prod_{\alpha = 1}^{n-2+2g} W(u_{\alpha}) \prod_{i=1}^n \hat{V}^{w_i}_{m_1^i,m_2^i}(x_i;z_i) \right\rangle.
\end{align*}

Thus far, the symmetry appears to be anomalous; however, we have assumed that the measure $d\mu$ is invariant under the transformation which is not in fact the case\footnote{We still assume that the \emph{functional} measure ${\cal D}\Psi$ in \eqref{eq:C_g} is invariant.}. As we shall see, the replacement of $n+2g-2$ of the $\mathbf{b}(\mu_I)$ with
$$
\mathbf{b}(\mu_I)\rightarrow \widetilde{\mathbf{b}}(\mu_I):=\int_{\Sigma}d^2z\,\widetilde{G}^-(z)\mu_I(z),	\qquad		d\mu\rightarrow \prod_{I=1}^{n+2g-2}\widetilde{\mathbf{b}}(\mu_I)d\tau_I\,\prod_{J=n+2g-1}^{n+3g-3}\mathbf{b}(\mu_J)d\tau_J,
$$
where $\widetilde{G}^-(z)$ is defined by \eqref{eq:generators}, gives a measure that transforms in a way such that the correlation function is invariant under global radial transformations. Indeed, a measure constructed of $g-1$ insertions of $\mathbf{b}(\mu_I)$ and $n+2g-2$ insertions of $\widetilde{\mathbf{b}}(\mu_I)$ is precisely what the hybrid formalism requires \cite{Dei:2020zui}. We can think of the $\widetilde{\mathbf{b}}(\mu_I)$ as screening operators for this symmetry \cite{DiFrancesco:1997nk}. The non-trivial contribution from $\widetilde{\mathbf{b}}(\mu_I)$ in physical correlators comes from the presence of the weight $(3,0)$ field $Q(z)$ in the definition of $\widetilde{G}^-(z)$.  The non-trivial OPE is
\begin{equation*}
\label{eq:phi_Q_OPE}
    \partial\Phi(z) Q(w) = \frac{Q(w)}{z-w} + \dots
\end{equation*}
implying that $\delta_{{\cal K}} \widetilde{\mathbf{b}}(\mu_I) = 2\alpha\widetilde{\mathbf{b}}(\mu_I)$. Taking account of all $n+2g-2$ insertions, we have $\delta_{{\cal K}}d\mu=2\alpha(n+2g-2) d\mu$. Therefore, using this measure, we see that
\begin{equation}\label{RH}
 \delta_{{\cal K}} {\cal C}_g(\underline{x},\underline{w}) =2\alpha \left[ (n+2g-2)+ (1-g) - \; \left( \sum_{i=1}^n\frac{w_i+1}{2}-N_{\Gamma} \right) \right]
     {\cal C}_g(\underline{x},\underline{w}),
\end{equation}
which vanishes as a result of the Riemann-Hurwitz formula \eqref{eq:Riemann_Hurwitz}. The correlation function of ground states is therefore invariant under global radial translations.\footnote{We have been able to show that this result generalises to states that have been acted on by the $\partial\bar{\mathcal{X}}^j_n$ and $\Psi^{A,j}_r$ DDF operators of \eqref{eq:DDF_operators}, but have not verified the radial symmetry when there are insertions containing $\partial \mathcal{X}^j_n$. We expect the result to fully generalise and hope to return to this elsewhere.}

It is interesting that we needed exactly $n+2g-2$ of the $\widetilde{\mathbf{b}}(\mu_I)$ insertions for a non-anomalous radial translation. We mentioned in \S\ref{sec:physical_correlators} that we should include this number of $\widetilde{\mathbf{b}}(\mu_I)$ insertions to cancel the anomalies of the ghost currents of the hybrid formalism (this is shown in detail in Appendix \ref{sec:hybrid_correlators}). It appears that the background charge of $\Phi$ is therefore related to that of the $\rho$ ghost.
\\

To reiterate, the global radial symmetry has been derived here using consistency arguments for the existence of spurious poles in the correlation function. To be more precise as to how these poles can actually occur, notice that the insertions $\delta({\cal S})$ \eqref{eq:Insertion} have the required properties to account for these spurious poles. In particular, correlation functions \eqref{eq:Gerasimov_correlator} of highest weight states are of the form
$$
    \mathcal{C}_g(\underline{x},\underline{w}) = \left\langle \prod_{\alpha = 1}^{n-2+2g} W(u_{\alpha}) \prod_{i=1}^n \hat{V}^{w_i}_{m_1^i,m_2^i}(x_i;z_i)\prod_{a=1}^{M}\delta\Big({\cal S}_a\Big) \right\rangle.
$$
We therefore need to consider how the $\delta({\cal S})$ change under ${\cal K}$. Let the contour $C$ in \eqref{eq:Insertion} be a small contour around the origin. The infinitesimal change in ${\cal S}$ is
\begin{eqnarray}
[\alpha{\cal K},{\cal S}]&=&2\alpha\oint_{z=0}\mathrm{d}z\oint_{w=z}\mathrm{d}w\; \p\Phi(w)e^{-2\Phi}\beta (z)=-2\alpha{\cal S}.
\end{eqnarray}
The delta-function transforms with the opposite sign\footnote{One can see this by noting that the commutator $[{\cal K},{\cal S}]=-2{\cal S}$ can be realised by the Euler operator ${\cal K}=-2{\cal S} \frac{d}{d{\cal S}}$. Since $\delta(t{\cal S})=t^{-1}\delta({\cal S})$, we see that $[{\cal K},\delta({\cal S})]=2\delta({\cal S})$.} giving
$$
[\alpha{\cal K},\delta({\cal S})]=2\alpha\delta({\cal S}).
$$
Thus, the insertions $\delta({\cal S})$ are exactly what is required to produce the spurious pole contribution. Under an infinitesimal variation, (\ref{RH}) is naturally satisfied if we identify $M$ with $N_{\Gamma}$. This makes sense: we are working on the patch where $\xi^+\neq 0$ and so have removed a point in the target space. Since no vertex operator is inserted there, if the worldsheet is a covering map, the removal of the point in the target space corresponds to the removal of the $N_{\Gamma}$ preimages of that point on the worldsheet. It is then consistent to conclude that the locations of the spurious poles $z^*_a$ are the $N_{\Gamma}$ preimages of the excluded point in the target space. Around these points are $N_{\Gamma}$ excluded discs, centred on these points and the contours $C_a$ are the $N_{\Gamma}$ boundaries of the worldsheet about these excluded discs.

It's interesting to note that a very similar story has been found for topological strings in the large volume limit \cite{Frenkel:2005ku}. In that case, a non-linear sigma model can be described by a free theory with additional insertions. These additional insertions can be understood as nonlinear deformations of the free worldsheet theory.

\subsection{Classical intuition for the ``hidden'' poles}
\label{sec:classical_hidden_poles}

One way to think about the hidden poles in the genus zero case is the following. We can remove a point on the boundary and work in a chart that is a copy of $\C$. As discussed previously, a natural choice is to remove the point where $\xi^+=0$, corresponding to the point at infinity in the coordinate $\gamma=-\xi^+/\xi^-$. Given the worldsheet localises on a covering map, we remove the corresponding points $z^*_a$ for which $\Gamma(z^*_a)=\infty$. For a generic point on the worldsheet (i.e. not a ramification point), by the Riemann-Hurwitz formula \eqref{eq:Riemann_Hurwitz}, there will be $N_{\Gamma}$ preimages of this point and so $a=1,...,N_{\Gamma}$ as found above. With the point at infinity on the boundary removed, we can choose to describe the theory by a flat metric and the coordinates $\Phi$, $\gamma$ and $\tilde{\gamma}$ are well-defined. This can be made quite explicit for the classical $j=\frac{1}{2}$ solution of \S\ref{sec:classical_covering_map} with the divergence of $\phi$ at the removed point becoming particularly apparent.

We can translate the global coordinates of \eqref{eq:global_coords} to Wakimoto-adapted coordinates \cite{Giveon:1998ns} by first defining $r = \sinh \rho \geq 0$ and
\begin{equation}
\label{eq:coords}
    \phi = t + \frac{1}{2}\log (1+r^2), \quad \gamma = \frac{r}{\sqrt{1+r^2}}e^{-t+i\theta}, \quad \tilde{\gamma} = \frac{r}{\sqrt{1+r^2}}e^{-t-i\theta}.
\end{equation}
In global coordinates, the boundary of $AdS_3$ is given by $\rho \to \infty$, whilst in the Wakimoto-adapted coordinates it is not quite as simple. For finite or large $t$, the boundary of $AdS_3$ is at $\phi \to \infty$, whilst $(\gamma,\tilde{\gamma})$ become $S^2$ coordinates for the boundary with $(\theta,t)$ determining the azimuthal and polar angles (assuming we have compactified the time coordinate). However, the value of $\phi$ is not well-defined at the south pole of the $(r,\theta,t)$ sphere since both $t \to -\infty$ and $r \to +\infty$ there. This means that $\phi$ tends to a different value depending on the direction in which you approach the south pole.

Nevertheless, since $\phi \to \infty$ at all other points on the boundary of $AdS_3$, it makes sense to use $\phi$ as a radial coordinate. We define the manifestly positive radial coordinate
\begin{equation*}
\label{eq:r_tilde}
    \tilde{r} = e^{\phi} = e^{t}\sqrt{1+r^2},
\end{equation*}
and we compactify the coordinates such that the spacetime looks like a sphere again. The ambiguity in the value of $\phi$ as we tended to the south pole of the $(r,\theta,t)$ sphere is now replaced by a straight line from the origin to the south pole of the $(\tilde{r},\theta,t)$ sphere, because both $r$ and $t$ are infinite on this line. This line is therefore not contained in the bulk spacetime but forms part of the boundary.
To be explicit, the boundary of $AdS_3$ in these coordinates is given by the boundary of the sphere (where $\phi \to \infty$ and $t \in \mathbb{R}\cup\{+\infty\}$) and the line between the origin and south pole (where $\phi \in \mathbb{R}\cup\{\pm\infty\}$ and $t \to -\infty$).

\begin{figure}[h]
    \centering
    \includegraphics[scale=0.4]{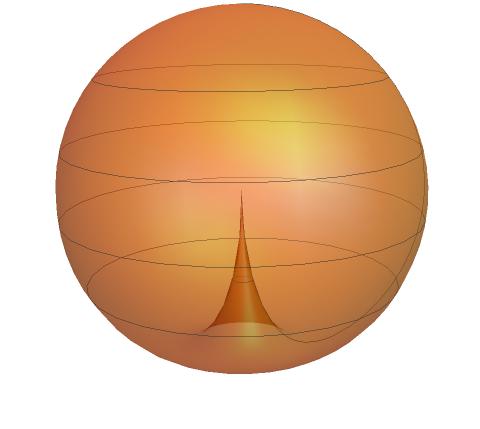}
    \caption{The classical solution \eqref{eq:static_solution} with no radial momentum sketched in the coordinates \eqref{eq:coords} to give a classical picture for the worldsheet. Some strings are sketched to show how it propagates from the infinite past in the centre to the future at the north pole.}
    \label{fig:classical_worldsheet}
\end{figure}

What does the solution \eqref{eq:static_solution} look like in these coordinates? It is given by
\begin{equation}
\label{eq:probing_solution}
    \tilde{r} = e^{w\tau}\sqrt{1 + r_0^2}, \quad \gamma = \frac{r_0}{\sqrt{1+r_0^2}}e^{-w(\tau-i\sigma)}, \quad \tilde{\gamma} = \frac{r_0}{\sqrt{1+r_0^2}}e^{-w(\tau+i\sigma)},
\end{equation}
for constant $r = r_0$, such that the radial coordinate $\tilde{r}$ is not constant. Instead, the infinite past of the string always probes finite $\phi$, whilst $\gamma \to \infty$. A sketch of \eqref{eq:probing_solution} is shown in Figure \ref{fig:classical_worldsheet} where we assume that $r_0 \gg 1$ (meaning that $\gamma$ and $\tilde{\gamma}$ are $S^2$ coordinates).

If we combine this picture with the covering map intuition of \S\ref{sec:classical_covering_map}, then provided $t\to -\infty$ is not an insertion point, we expect to have $N_{\Gamma}$ points on $\Sigma_g$ that probe finite $\phi$ with $\gamma \to \infty$. However, we know that $e^{-\Sigma}$ can be identified with the boundary coordinate $\gamma$, so these cusps in the worldsheet should precisely be the points where the covering map \eqref{eq:gamma_hat} diverges. We therefore see that the poles in $\hat{\Gamma}$ do not represent physical observables but are a consequence of covering the boundary $S^2$ with only one coordinate chart, including $\{\infty\}$ as a point in spacetime. They represent the infinite past of the string.

These cusps may have implications for the dependence of the theory on the background bulk geometry. It was proposed in \cite{Eberhardt:2021jvj} that a theory of minimal tension strings on $AdS_3 \times S^3 \times T^4$ is in fact independent of the bulk geometry of the spacetime, up to some equivalence classes for the asymptotics of the geometry. Nevertheless, it is thought that a bulk geometry does emerge in the limit of large spectral flow and this geometry contains a conical deficit \cite{Eberhardt:2021jvj,Knighton:2022ipy}. This fits nicely with the picture given in Figure \ref{fig:classical_worldsheet}, since the worldsheet is localised near the boundary of the $(\tilde{r},\theta,t)$ sphere at every point except near the infinite past, where a conical deficit appears to emerge. This conical deficit grows as the spectral flow $w \in \mathbb{N}$ is increased. In the limit that $w \to \infty$ (with $r_0$ held constant) this deficit angle will tend towards $\pi$, with $\tilde{r}$ being suppressed by $e^{w\tau}$ at any point south of the equator where $\tau<0$. This is precisely the same range of possible deficit angles observed in \cite{Eberhardt:2021jvj}.

Significantly, we see that a theory of minimal tension strings on a background containing $AdS_3 \times S^3$ (for the bosonization of the free fields) always comes with a global symmetry $\Phi \mapsto \Phi + \alpha$, which we interpret as a shift in the radial coordinate. This implies that the worldsheet can be taken to live arbitrarily close to the points on the boundary of $AdS_3$ where $\phi \to \infty$ in the coordinates of \eqref{eq:coords}, except at the infinite past where $\phi \to -\infty$ (clearly this cannot be removed through a shift by a constant). This means that the worldsheet can always be taken to be of the form of Figure \ref{fig:classical_worldsheet}. There may still be dynamics in the radial direction; the symmetry is global, not local, so only the zero mode is removed.

In this context, the charges at $\{z_a^*\}$ can be thought of as singularities in the radial coordinate of the embedding into spacetime. At genus 0, the Riemann surface is conformally equivalent to a sphere, so we may take a chart that covers the whole Riemann surface except the infinite past. The restriction of the worldsheet to covering map configurations and the $\Phi$ symmetry in this chart can be used to place the worldsheet at the boundary and we are left with a cusp at the infinite past, where $\Phi$ diverges as we approach the image of $z^*_a$.

\section{Discussion}
\label{sec:discussion}

In this paper, we have attempted to gain some insight into the physics that underpins the minimal tension string proposal of \cite{Gaberdiel:2018rqv,Eberhardt:2018ouy}. Interpreting $G^+(z)$ in (\ref{eq:generators}) as a BRST current, the presence of the stress tensor $T(z)$ and its associated ghost $\sigma(z)$ is easy to understand. Given that the internal CFT on the $T^4$ is a topologically twisted version of the ${\cal N}=(1,1)$ RNS theory, it is as expected that the twisted supercurrent $G^+_C(z)$ also appears. As shown in Appendix \ref{sec:Qappendix}, $P(z)$ vanishes in the $k=1$ string so what remains is to understand the inclusion of the $e^{-\rho(z)}Q(z)$ term. We have shown that $Q(z)$ generates a local symmetry of the theory and $\rho(z)$ can be thought of as the ghost associated with it. One of our aims in this paper has been to show that $Q(z)$ plays a key role in some of the more distinctive features of the $k=1$ string. In particular, the $Q=0$ constraint is an efficient method to see that correlation functions of physical states are non-trivial only when the worldsheet is a covering map of the boundary. We have also seen that this symmetry via the insertion of $Q$-weighted modes into the correlation function ensures that the rigid translation symmetry in the radial $AdS_3$ direction is not anomalous. Furthermore, we wanted to clarify the target space structure of the theory: we have seen that, in the first instance, the minimal tension string embeds into the supertwistor space of $AdS_3$ and we have presented an alternative perspective on the connection with the $AdS_3$ geometry, which we hope sheds some light on the nature of the secret representations of \cite{Eberhardt:2019ywk}.

There is much work still to be done before we can confidently conclude what the principles underlying this theory are. Of particular interest would be a complete construction of the $k=1$ string without recourse to the hybrid formalism and fully motivated by the geometry of the twistor space. A more detailed understanding of the theory as a twistor string, including the form of generic vertex operators as cohomology representatives in twistor space is clearly desirable and, we believe, within reach. In terms of the results in this paper, we shall end by briefly commenting and speculating on three particular issues.

\subsection*{Consequences of the global symmetry}

In \S\ref{sec:global_symmetry}, we argued that there is an exact global symmetry of the minimal tension string, corresponding to a shift in the zero mode of the radial coordinate of $AdS_3$. The radial translation global symmetry and the localisation to covering maps suggests that the theory has some level of background independence and is perhaps topological \cite{Eberhardt:2018ouy,Eberhardt:2021jvj}. Furthermore, the fact that the zero mode of $\Phi$ is not physical has consequences for the interpretation of the theory. The rigid symmetry of the worldsheet translates into a local (scaling) symmetry of the target space theory. In constructing a string field theory, the conjugates to the zero modes of the embedding fields take on the role of the target space coordinates\footnote{For a string in flat spacetime, it is the $x^{\mu}$, conjugate to the zero mode $\alpha^{\mu}_0+\tilde{\alpha}^{\mu}_0$ of $\p X^{\mu}(z)+\bar{\p}X^{\mu}(\bar{z})$ that provides a coordinate for spacetime. Similarly, on a torus, conjugates of both momentum and winding zero modes provide coordinates on the doubled target space geometry. See, for example, \cite{Hull:2009mi} for details.}. As such, if we were to construct a minimal tension string field theory in the usual way, then we would expect to find the target space of the theory to be constructed in terms of the boundary coordinates only, as there is no observable radial zero mode. The non-zero modes of $\Phi$ still play a role in the physics but the interpretation would seem to be in terms of a theory defined directly on the boundary. It would be interesting to construct the string field theory explicitly and to see if it gave a direct description of the boundary theory.

\subsection*{Target spaces for the $k=1$ string}

We have highlighted in \S\ref{sec:connections_to_AdS_3} two possible interpretations of the Wakimoto free field description \cite{Wakimoto:1986gf} of the theory at $k=1$; either as a theory in which the worldsheet is pinned to the boundary, or as a free field realisation along the lines of that proposed in \cite{Gerasimov:1990fi}. There are pros and cons to each perspective. In the $\Phi\rightarrow \infty$ limit, it is unclear to what extent this theory really describes the full physics on $AdS_3$ as the interaction term will  be absent. Indeed, a connection to the full $AdS_3$ sigma model requires the inclusion of this interaction term, which renders $\beta$ and $\tilde{\beta}$ as Lagrange multipliers rather than dynamical fields. Conversely, whilst the free field approach akin to \cite{Gerasimov:1990fi} certainly describes the full $AdS_3$ sigma model, the details of how the theory simultaneously encodes the degrees of freedom of the holomorphic and antiholomorphic sectors is unclear. For example, it is currently not understood how the vertex operators in \eqref{eq:Gerasimov_correlator} depend on $\tilde{\beta}$. Moreover, whilst it is simple to realise the left-moving current $J = g^{-1}\p g$ in the coordinates of \S\ref{sec:connections_to_AdS_3}, it does not appear to be possible to write the right-moving current $\tilde{J} = \bar{\p}g g^{-1}$ in these coordinates locally. In essence, this is because we cannot express $g$ locally in these coordinates. We hope to explore these issues in the future to provide a clearer connection between these two perspectives.

This raises the conceptual question of to what extent are these physically distinct proposals? At the classical level, the theory described by \eqref{eq:S} and the approach of \cite{Gerasimov:1990fi} are both equivalent to the $AdS_3$ NLSM for all finite $\Phi$, as explained in \S\ref{sec:k=1_strings}. Moreover, the equivalence of \eqref{eq:S} to the NLSM at the quantum level has been discussed previously in \cite{deBoer:1998gyt}, where it is stated that the equivalence can be trusted when the path integral is dominated by large $\Phi$ contributions. We are not aware of any substantive work relating the approach of \cite{Gerasimov:1990fi} to the NLSM at the quantum level. However, if a formal limit of $\Phi \to \infty$ is taken and the interaction term of \eqref{eq:S} is dropped, then this theory appears to be qualitatively different --- it is not clear how such a theory would relate to the NLSM on $AdS_3$.

A curious feature of the $k=1$ string that we have tried to emphasise is that it describes a worldsheet embedding into the supertwistor space of $AdS_3$, with the $S^3$ being encoded in the fermionic components of the supertwistor. Furthermore, it has been known for some time  that the twistor space of $AdS_d$ can be identified as the ambitwistor space of its boundary. It is therefore satisfying that both of the above descriptions of the theory give a direct relationship between the bulk and boundary theories, as required by the twistor construction, either requiring the dynamics to be pinned to the boundary or by the shift symmetry in the radial zero mode. As such, this raises a hope that the origin of the AdS/CFT correspondence at $k=1$ can be understood from the twistor description. A key question then is how this is modified for $k>1$, where any such connection to twistor space is less well understood.

As a final comment on the free field interpretation for the Wakimoto representation, we note that the insertions
$$
\delta({\cal S})=\delta\left(\oint e^{-2\Phi}\beta\right)
$$
 have many qualities in common with the secret representations of \cite{Eberhardt:2019ywk}. Another candidate is the related non-local operator
$$
\widetilde{\cal S}:=\oint e^{i\kappa_2},
$$
discussed around (\ref{eq:k2_charge}). Like $\delta({\cal S})$, $\widetilde{\cal S}$ is invariant under $SL(2)$ and has the correct behaviour at the spurious poles $[{\cal K},\widetilde{\cal S}]=2\widetilde{\cal S}$. The insertions $\widetilde{\cal S}$ carry precisely the charge needed to balance the conformal anomaly. It would be interesting to explore correlation functions with such insertions further.

\subsection*{$\mathbf{Q}$ and localisation}

We conclude with some speculative comments and discuss two possible mechanisms by which localisation might occur. It has been shown here and previously in \cite{Eberhardt:2019ywk,Eberhardt:2020akk} that non-trivial correlation functions are in one-to-one correspondence with points on the moduli space where the worldsheet is a covering map of the boundary. What we still do not have is an explicit derivation, at the level of the path integral, that the correlation functions localise the integral over moduli space to precisely those points where the covering map exists. How might the insights gleaned in this paper move us forward?

We saw in \S\ref{sec:Q=0} that $Q(z)$ is related to localisation to covering maps for the $k=1$ string. It is unclear what role $Q(z)$ plays for $k>1$. In particular, the constraint algebra for $Q(z)$ at $k>1$ requires the inclusion of the  composite field $P(z)$ in order to close. As we show in Appendix \ref{sec:Qappendix},
$$
Q(z) Q(w) \sim \frac{6 P(w)}{(z-w)^2} + \frac{3\partial P(w)}{z-w},
$$
significantly complicating the interpretation of not just $Q(z)$ but also $\rho(z)$. And so $Q(z)$ is only nilpotent when $P(z)=0$, which is guaranteed when $k=1$. That matters are more complicated for $k>1$ should not be surprising; after all, it is the hoped-for  simplicity of the $k=1$ case that has attracted physicists to study this case for many years and we do not expect the covering map story to generalise to $k>1$.

The fact that $Q(z)$ is nilpotent only for $k=1$ provides some circumstantial evidence that this feature may lie at the heart of the moduli space localisation phenomenon. This is reminiscent of path integral localisation arguments in QFT \cite{Pestun:2016zxk} and is perhaps suggestive that something similar be true in the $k=1$ case, where the Abelian symmetry generated by $Q(z)$ gives a localisation in the moduli space of Riemann surfaces. What is not yet clear is how the action of $Q(z)$ might be extended to the worldsheet moduli and the exact sense in which $Q^2(z)=0$ might lead to a localisation of the moduli space integral. We have no concrete suggestion of how this might occur.

As an alternative localisation mechanism, we note that similar moduli space localisation occurs in twistor string theories, where the vertex operators are distributions with delta-function support \cite{Witten:2003nn,Berkovits:2004hg,Mason:2013sva}. Given the natural twistor interpretation of the theory, it is possible that the origin of the localisation may be most simply understood in terms of the natural twistor wavefunctions of the theory. Investigating such considerations requires a much better understanding of the vertex operators as natural constructions in twistor space than we currently have. We hope to pursue this issue in the future.\\

\begin{center}
    \textbf{Acknowledgments}
\end{center}
This work has been partially supported by STFC consolidated grant ST/T000694/1. NM is funded by an EPSRC studentship. RR is the Thomas and Stephan K\"{o}rner Fellow at Trinity Hall and is grateful to the Avery-Wong Foundation for their continued support of this Fellowship. We would like to thank Matthias Gaberdiel, Rajesh Gopakumar, and David Skinner for helpful conversations and correspondence and especially Kiarash Naderi and Bob Knighton for helpful comments on an early draft.

\appendix

\section{$\mathbf{Q^2=0}$ when $\mathbf{k=1}$}
\label{sec:Qappendix}

In this appendix, we show in the special case of $k=1$, that the current $Q(z)$ is nilpotent, but we shall begin by thinking about $Q(z)$ for general $k$. In the $k=1$ theory, we were able to apply the free field realisation of $\mathfrak{psu}(1,1|2)_1$. This is no longer possible for $k>1$. We should instead return to a WZW description of the theory, either in the RNS or hybrid formalism. We write our $\mathfrak{psu}(1,1|2)_k$ generators as $K^{ab} = -K^{ba}$ and $S^a_{\alpha}$ which are bosonic and fermionic, respectively. The index $a \in \{1,2,3,4\}$ is an $\mathfrak{so}(4)$-index and $\alpha \in \{1,2\}$ labels the supersymmetries. The generators satisfy the commutation relations \cite{Gaberdiel:2022bfk}
\begin{align*}
    \left[ K^{ab}_m, K^{cd}_n \right] &= mk \epsilon^{abcd}\delta_{m+n,0} + \delta^{ac}K^{bd}_{m+n} - \delta^{ad}K^{bc}_{m+n} - \delta^{bc}K^{ad}_{m+n} + \delta^{bd}K^{ac}_{m+n},\\
    \left\{ (S^a_{\alpha})_m, (S^b_{\beta})_n \right\} &= mk\delta^{ab}\epsilon_{\alpha\beta} \delta_{m+n,0} + \frac{1}{2}\epsilon_{\alpha\beta} \epsilon^{abcd}K^{cd}_{m+n},\\
    \left[ K^{ab}_m, (S^c_{\alpha})_n \right] &= \delta^{ac}(S^b_{\alpha})_{m+n} - \delta^{bc}(S^a_{\alpha})_{m+n},
\end{align*}
where the $\epsilon$ tensors are totally antisymmetric with $\epsilon_{12} = \epsilon^{1234} = +1$. The transformations relating these variables to the basis used in \S\ref{sec:FFR} can be found in Appendix A.2 of \cite{Gaberdiel:2022bfk}. In terms of these fields, $Q$ is given by\footnote{Our convention has reversed the roles of the supercharges relative to \cite{Gaberdiel:2022bfk} to follow \cite{Gerigk:2012cq}, but note that there is a relative factor of $i$ between our definition of $K^{ab}$ and the one given in \cite{Gerigk:2012cq}.}
\begin{equation}
\label{eq:Q_WZW}
    Q = \frac{1}{2\sqrt{k}}\left( K^{ab}S^a_2 S^b_2 + 4S^a_2\partial S^a_2 \right).
\end{equation}
The OPE of $Q$ with itself is
$$Q(z) Q(w) \sim \frac{6 (S_2)^4}{(z-w)^2} + \frac{3\partial (S_2)^4}{z-w},$$
where $(S_2)^4 = \frac{1}{24}\epsilon^{abcd} \, S^a_2 \, S^b_2 \, S^c_2 \, S^d_2 = S^1_2 \, S^2_2 \, S^3_2 \, S^4_2$, so generically the simple pole does not vanish. This term $(S_2)^4 \equiv P$ in \eqref{eq:generators}, so we see how the BRST charge $G^+_0$ in the hybrid formalism accommodates for this non-trivial $Q\cdot Q$ OPE at generic $k$ and $P$ then plays an important role in the closure of the constraint algebra generated by $Q$. A glance at the form of the BRST current $G^+(z)$ shows that the simple interpretation of $Q(z)$ as a constraint and $\rho(z)$ as the associated ghost becomes unclear when $P(z)$ is also involved.

There is a simplification when $k=1$. In this special case, we have that $(S_2)^4 = 0$, and so $Q(z)$ is nilpotent. This can be seen by translating back into our usual $\mathfrak{psu}(1,1|2)_1$ basis \cite{Gaberdiel:2022bfk} and then applying the free field realisation \eqref{eq:supercurrents},
$$(S_2)^4 = S^{+++}S^{--+}S^{+-+}S^{-++} = \xi^+\chi^+\xi^-\chi^-\xi^+\chi^-\xi^-\chi^+ = 0,$$
where we apply implicit normal ordering. We arrive at the trivial OPE $Q(z) Q(w) \sim 0$.

\section{Gauge fixing the worldsheet metric in the hybrid formalism}
\label{sec:hybrid_correlators}

We mentioned in \S\ref{sec:physical_correlators} that the correlation functions of the hybrid formalism are constructed in terms of $\mathcal{N}=4$ topological string correlators. The purpose of this section is to elaborate on this construction and identify precisely which parts of the correlation function are non-vanishing when the twisted sector ground states of \eqref{eq:V_hat} are inserted in $P=-2$. We will focus on the holomorphic sector of the theory.

The double cohomology associated to the $\mathcal{N}=4$ topological string \eqref{eq:hybrid_phys_conditions} means we have two BRST charges $G^+_0$ and $\tilde{G}^+_0$, which give rise to two $b$-ghost type variables, $G^-$ and $\tilde{G}^-$. These can be combined with Beltrami differentials $\{\mu_I\}$ to gauge fix the worldsheet metric \cite{Nakahara:2003nw}. The similarities with the $b$-ghost can be seen from
$$G^+_0 G^- = \tilde{G}^+_0 \tilde{G}^- = T,$$
which is precisely the expected behaviour of a $b$-ghost under the action of a BRST charge. This choice of $b$-ghost forms an $SU(2)$ outer automorphism of the theory, which means the most general correlation function is given by a sum over all distinct choices between $G^-$ and $\tilde{G}^-$ \cite{Berkovits:1994vy}. Moreover, to gauge fix the double cohomology in \eqref{eq:hybrid_phys_conditions}, we must insert $J$ once and $\tilde{G}^+$ $(g-1)$ times into our correlation function --- this ensures that deformations of the form $G^+_0\tilde{G}^+_0\Lambda$ vanish inside the correlation function. Overall, the general form for the $n$-point correlation function of the hybrid formalism for genus $g \geq 1$ is \cite{Berkovits:1999im,Dei:2020zui}
\begin{equation}
\label{eq:hybrid_correlator}
    F_{g,n} = \sum_{\{\hat{G}^-\}} \int_{\mathcal{M}_{g,n}} \left\langle \prod\limits_{I = 1}^{n+3g-3} \hat{G}^-(\mu_I) \left[ \int \tilde{G}^+ \right]^{g-1} \int J \prod\limits_{i=1}^{n} \Psi_i \right\rangle,
\end{equation}
where $\hat{G}^-$ refers to either $\mathbf{b}$ or $\widetilde{\mathbf{b}}$. The fields $\Psi_i$ are the physical state insertions, which we take to be in picture $-2$ with ghost contribution $e^{2\rho+i\sigma+iH}$ as in \eqref{eq:P=-2_phys_state}. This correlation function appears to be very complicated, however we will soon see that there is only one non-zero term in the sum by ghost charge conservation. Moreover, \eqref{eq:hybrid_correlator} can be adapted for genus $g=0$, by requiring that $n\geq 3$ to fix the conformal Killing group and also inverting $\tilde{G}^+$ with $e^{-\rho-iH}$ (see \eqref{eq:generators}). It is worth noting that sometimes $n$ of the Beltrami differentials are taken to act on these states to give terms of the form
$$\int_{\Sigma_g} \hat{G}^-_{-1} \Psi_i .$$

The ghost fields $\rho$, $\sigma$, $H^1$ and $H^2$ all form $\epsilon=-1$ linear dilaton type theories in the hybrid formalism (see Appendix \ref{sec:linear_dilaton}) and carry background charges
$$q_{\rho} = 3, \qquad q_{\sigma} = 3i, \qquad q_{H^1}=q_{H^2} = i,$$
as can be read off from the stress tensors in \eqref{eq:generators} and \eqref{eq:compact_generators}. This implies that the overall ghost contribution for a non-vanishing correlator must be \cite{Dei:2020zui}
$$e^{(g-1)(-3\rho-3i\sigma - iH)},$$
to act as an appropriate screening operator. Let's consider a general term in \eqref{eq:hybrid_correlator} with $k$ insertions of $G^-$ and $n+3g-3-k$ insertions of $\tilde{G}^-$. Then the overall ghost contribution is given by
\begin{align*}
    &k \times (\text{ghosts of } G^-) + (n+3g-3-k) \times (\text{ghosts of } \tilde{G}^-) + (g-1) \times (\text{ghosts of } \tilde{G}^+) +\\
    &n \times(\text{ghosts of physical state insertions}).
\end{align*}
According to \eqref{eq:generators} and \eqref{eq:compact_generators}, the possible ghost contributions of $G^-$ are either $-i\sigma$, $-iH^1$ or $-iH^2$, whilst those of $\tilde{G}^-$ are $(-2\rho-i\sigma-iH)$, $(-\rho-iH)$, $(-\rho-i\sigma-iH^1)$ or $(-\rho-i\sigma-iH^2)$ and for $\tilde{G}^+$ they are $(\rho+iH)$, $(\rho+i\sigma+iH^1)$ or $(\rho+i\sigma+iH^2)$. By contrast, the physical states necessarily contribute $n$ copies of $(2\rho + i\sigma + iH)$. One can show that there is a one-parameter family of combinations which provide the correct ghost charge. These are given by
\begin{align*}
    &k(-i\sigma) + (n+g-1+k)(-2\rho - i\sigma-iH) + (g-1-k)(-\rho-i\sigma-iH^1)\\
    &+ (g-1-k)(-\rho-i\sigma-iH^2) + (g-1)(\rho+iH) + n(2\rho+i\sigma+iH)\\
    &= -(g-1)(3\rho+3i\sigma+iH),
\end{align*}
for $0\leq k \leq g-1$, where $G^-$ only provides contributions from $e^{-i\sigma}$ and $\tilde{G}^+$ only from $e^{\rho+iH}$. $\tilde{G}^-$ can either contribute via $e^{-2\rho-i\sigma-iH}Q$ or via the compact pieces $e^{-\rho-i\sigma-iH^1}\partial\bar{X}^2$ and $e^{-\rho-i\sigma-iH^2}\partial\bar{X}^1$.

Suppose we do insert one of these compact pieces into the correlation function, for instance $(e^{-\rho-i\sigma-iH^1}\partial\bar{X}^2)(\mu_I)$. The contour integral around this insertion can then be deformed away from $\mu_I$. However, $e^{-\rho-i\sigma-iH^1}\partial\bar{X}^2$ has trivial OPEs with all of the other insertions (since the physical states of interest are compactification independent) so this correlation function must vanish! The same could be said for the other compact insertion.

We conclude that the only non-vanishing combination with the correct ghost charge is when $k = g-1$ and the non-zero pieces are the $e^{-i\sigma}$ part of $G^-$, the $e^{-2\rho-i\sigma-iH}Q$ part of $\tilde{G}^-$ and the $e^{\rho+iH}$ part of $\tilde{G}^+$. In particular, there are $(g-1)$ copies of $\mathbf{b}(\mu_I)$ and $(n+2g-2)$ copies of $\widetilde{\mathbf{b}}(\mu_I)$ as claimed in \S\ref{sec:physical_correlators}. We will implicitly include these insertions in all physical correlators.
\\

So far, we have only considered correlation functions of the twisted sector ground states, but we could equally insert arbitrary physical states by acting with DDF operators. In \S\ref{sec:physical_states}, we highlighted the work of \cite{Naderi:2022bus} where the following DDF operators were proposed:
\begin{align*}
\label{eq:DDF_operators_appendix}
    \begin{split}
        \partial\bar{\mathcal{X}}^j_n &= \oint \mathrm{d}z \; \partial\bar{X}^j e^{-n\Sigma},\qquad \partial\mathcal{X}^j_n = \oint \mathrm{d}z \; e^{-n\Sigma}\left[ \partial X^j + ne^{\rho - \Theta + iH^j}\psi^+\psi^- \right],\\
        \Psi^{A,j}_r &= \oint \mathrm{d}z \; \psi^A e^{\phi_1 + i\kappa_1} e^{-\left(r+\frac{1}{2}\right)\Sigma } e^{\rho + iH^j},
    \end{split}
\end{align*}
where $A \in \{\pm\}$ and $j \in \{1,2\}$. One should act with a similarity transformation on these operators to translate to our conventions for the $\mathcal{N}=4$ topological algebra generators. However, this similarity transformation will have a trivial effect in terms of identifying which terms in the correlation function are non-vanishing.

Since the $\Psi_r^{A,j}$ and $\p\mathcal{X}^j_n$ have non-trivial ghost contributions, they will change which terms contribute to the measure for non-vanishing correlation functions. We are yet to fully understand the contributions from $\p\mathcal{X}^j_n$, since these have a non-trivial commutator with $\tilde{G}^-_C$. It therefore cannot be deduced so simply that any insertions of $e^{-\rho-i\sigma-iH^1}\partial\bar{X}^2$ or $e^{-\rho-i\sigma-iH^2}\partial\bar{X}^1$ will vanish. Nevertheless, if we restrict to considering correlation functions of physical states that are formed only from application of the operators $\p\bar{\mathcal{X}}^j_n$ and $\Psi_r^{A,j}$, then we can generalise our statements about which terms are non-vanishing in \eqref{eq:hybrid_correlator}. In particular, there will be no non-vanishing contributions from insertions including $\tilde{G}^-_C$ as before.

If we insert $\p\bar{\mathcal{X}}^j_n$, then the argument above is completely unchanged, as this operator contains no additional ghost contribution. By contrast, the operator $\Psi_r^{A,j}$ is in picture $P=-1$, so we need to be careful with charge conservation. There are no new contributions for the $\sigma$-ghost, so the only possible contributing terms persist to be $-i\sigma$ from $G^-$, $(-2\rho-i\sigma-iH)$ from $\tilde{G}^-$, $(\rho+iH)$ from $\tilde{G}^+$ and the ghosts of the physical states. One finds that it is only possible to impose charge conservation if there as many DDF operators $\Psi^{A,1}_r$ as $\Psi^{A,2}_r$. The constraint then becomes
\begin{align*}
    &(g-1-\ell)(-i\sigma) + (n+2g-2+\ell)(-2\rho - i\sigma-iH)\\
    &+ (g-1)(\rho+iH) + n(2\rho+i\sigma+iH) + \ell(\rho+iH^1) + \ell(\rho+iH^2)\\
    &= -(g-1)(3\rho+3i\sigma+iH),
\end{align*}
where $\ell$ is the number of each $\Psi^{A,j}_r$ insertion for $j=1,2$.

It is worth noting how these changes to the measure affect the arguments of \S\ref{sec:radial_profile} for deriving a global radial translation symmetry. As $\ell$ is increased, there are additional insertions of $e^{-2\rho-i\sigma-iH}Q$ which carry a charge under $\mathcal{K}$ of $+2$. This is precisely cancelled by the additional operators $\Psi^{A,1}_r$ and $\Psi^{A,2}_r$ which each carry a $\mathcal{K}$-charge of $-1$. As such, the global symmetry remains.

\section{Linear dilaton systems}
\label{sec:linear_dilaton}

A linear dilaton can be described by the action
$$S = -\frac{\epsilon}{2\pi} \int \mathrm{d}^2 z \; \partial \phi \bar{\partial} \phi,$$
for $\epsilon = \pm 1$, which implies the OPE
$$\phi(z) \phi(w) \sim \epsilon \log(z-w).$$
This OPE is usually taken as the definition of $\epsilon$. We can deform this action to add a background charge $q$ via \cite{DiFrancesco:1997nk}
$$S = -\frac{\epsilon}{2\pi} \int \mathrm{d}^2 z \; \left( \partial \phi \bar{\partial} \phi + \frac{1}{4} qR\phi \right),$$
where $R$ is the Ricci scalar for the worldsheet metric. This gives a family of CFTs with stress tensors
$$T_q(z) = \epsilon \left( \frac{1}{2}(\partial\phi)^2 - \frac{1}{2}q\partial^2\phi \right),$$
and associated central charge
$$c = 1 - 3\epsilon q^2.$$
The operator $\partial\phi$ is quasiprimary, with OPE
$$T_q(z) \partial\phi(w) \sim \frac{q}{(z-w)^3} + \frac{\partial\phi}{(z-w)^2} + \frac{\partial^2\phi}{z-w},$$
whist $e^{m\phi}$ is primary with weight
$$h(e^{m\phi}) = \frac{1}{2}\epsilon m(m+ \epsilon q).$$
In the absence of a background charge, $\partial\phi$ would generate a symmetry $\phi \mapsto \phi + \epsilon\alpha$ for constant $\alpha$ via the Noether charge
$$\tilde{Q} = \oint \mathrm{d}z \; \partial\phi.$$
This symmetry is now anomalous, with a non-trivial variation in the action
$$\delta S = -\frac{\alpha q}{8\pi} \int \mathrm{d}^2z \; \sqrt{h}R(h) = -\alpha q (1-g),$$
where we have been explicit to include the volume form for the worldsheet metric. In physical correlation functions, this anomaly is cancelled by screening operators in the measure with a non-trivial transformation under $\tilde{Q}$ and poles in the correlation function from inserting $\partial\phi(z)$. We include two examples below.

\subsection*{The $\mathbf{(b,c)}$ ghost system:}
We can bosonize the $(b,c)$ ghost system as \cite{Polchinski:1998rq,Polchinski:1998rr}
$$b = e^{-i\sigma}, \quad c = e^{i\sigma},$$
where $\sigma$ is a linear dilaton with $\epsilon = -1$. The stress tensor is
$$T = (\partial b)c - 2\partial(bc) = -\frac{1}{2}(\partial\sigma)^2 + \frac{3i}{2}\partial^2\sigma,$$
from which we read off $q = 3i$. The action is therefore
$$S = \frac{1}{2\pi}\int \mathrm{d}^2z \; \left( \partial\sigma \bar{\partial}\sigma + \frac{3i}{4}R\sigma \right).$$
Under the transformation $i\sigma \mapsto i\sigma + \alpha$, we have that $\delta(e^{-S}) = -3\alpha(1-g)e^{-S}$. We can use the $b$ and $c$ ghosts as screening operators, which respectively carry charges of $-1$ and $+1$ under $i\sigma \mapsto i\sigma + \alpha$. For this reason, the measure contains $(n+3g-3)$ insertions of $b$ and each physical state carries a factor of $c$ to give
$$\# b \text{ ghosts} - \# c \text{ ghosts} = 3(g-1).$$

\subsection*{The Wakimoto radial coordinate:}

The Wakimoto radial coordinate at $k=1$ is given by
$$\partial\Phi = \frac{1}{2}\partial(\phi_1 + \phi_2 + 2i\kappa_2),$$
which satisfies
$$\Phi(z)\Phi(w) \sim \frac{1}{2}\log(z-w).$$
The stress tensor can be found from the bosonization of the free fields \eqref{eq:bosonization} to give
$$T = (\partial\Phi)^2 - \partial^2\Phi.$$
To make contact with the above general construction, we can normalise the field by defining $\tilde{\Phi} = \sqrt{2}\Phi$, which forms an $\epsilon = +1$ linear dilaton system. The stress tensor is then
$$T = \frac{1}{2}(\partial\tilde{\Phi})^2 - \frac{1}{\sqrt{2}}\partial^2\tilde{\Phi},$$
from which we read off $q = \sqrt{2}$. The action is therefore
$$S = -\frac{1}{2\pi}\int \mathrm{d}^2z \; \left( \partial\tilde{\Phi}\bar{\partial}\tilde{\Phi} + \frac{\sqrt{2}}{4}R\tilde{\Phi}  \right) = -\frac{1}{4\pi} \int \mathrm{d}^2z \; \left( 4\partial\Phi\bar{\partial}\Phi + R\Phi \right).$$
Under the transformation $\Phi \mapsto \Phi + \alpha$, we have that $\delta (e^{-S}) = 2\alpha(1-g) e^{-S}$. There are screening operators in the measure from $\widetilde{\mathbf{b}}(\mu_I)$ as outlined in \S\ref{sec:global_symmetry}.

\section{The critical points of $\mathbf{\hat{\Gamma}(z)}$}
\label{sec:critical_points}

In \S\ref{sec:proof}, we claimed that the formula \eqref{eq:gamma_hat} provided a covering map despite only checking the first two conditions in the definition given in \S\ref{sec:covering_map}. We will now verify that $\hat{\Gamma}(z)$ contains no further critical points (away from $\{z_i\}$) such that it really is the covering map that we hoped for. The check will be formulated in the spirit of \cite{Knighton:2020kuh} and is included here for completeness. In particular, we will use the equivalence to the incidence relation from \eqref{eq:zeta_OPE}, explained in the main text, to apply the work of \cite{Knighton:2020kuh}. This of course means that our novel method for proving the localisation to covering maps via $e^{-\Sigma}$ is currently no faster than work already in the literature for this part of the proof. Nevertheless, it appears that a formal treatment of the proposed secret representations in \S\ref{sec:connections_to_AdS_3} would allow for a complete proof of the localisation in terms of Wakimoto variables.
\\

We should first comment on how $e^{-\Sigma}(z)$ interacts with the insertions of $G^-(\mu_I)$, $\tilde{G}^-(\mu_I)$ and $\int \tilde{G}^+$, since we ignored these earlier. Recall from Appendix \ref{sec:hybrid_correlators} that the pieces which contribute from these insertions are $e^{-i\sigma}$, $e^{-2\rho-i\sigma -iH}Q$ and $e^{\rho+iH}$. All of these terms have trivial OPEs with $e^{-\Sigma}$ and so we are safe to ignore them.

Next, we need to check the behaviour of $e^{-\Sigma}(z)$ as $z \to u_{\alpha}$, the insertion points of the fields $W(u_{\alpha})$. This field corresponds to the state $|0\rangle^{(1)} = [\psi^+_{-3/2}\psi^-_{-3/2}\psi^+_{-1/2}\psi^-_{-1/2}|0\rangle]^{\hat{\sigma}^2}$ \cite{Dei:2020zui}, for which
\begin{equation}
\label{eq:W_OPE}
    e^{-\Sigma}(z)|0\rangle^{(1)} = \left[ \hat{\sigma}^2(e^{-\Sigma})(z) \psi^+_{-3/2}\psi^-_{-3/2}\psi^+_{-1/2}\psi^-_{-1/2} |0\rangle \right]^{\hat{\sigma}^2},
\end{equation}
by definition. The modes of the free fields transform as $\hat{\sigma}(\xi^{\pm}_r) = \xi^{\pm}_{r+1/2}$ under this automorphism, implying that $\hat{\sigma}^2(\xi^{\pm})(z) = z\xi^{\pm}(z)$. However, from the identity $\xi^+\zeta^- = 1$, it must be that $\zeta^-$ scales in the opposite direction, $\hat{\sigma}^2(\zeta^-)(z) = \frac{1}{z}\zeta^-(z)$. As such, we find that $e^{-\Sigma}(z)$ is invariant under $\hat{\sigma}^2$.\footnote{
Note that we could have alternatively derived $\hat{\sigma}^2(e^{-\Sigma}) = e^{-\Sigma}$ by using the action of the spectral flow automorphism on the bosonized variables as in \S\ref{sec:wakimoto_radial_coord}. We have chosen to present a perhaps more intuitive explanation here.
}
It also has no interaction with the free fermions, so commutes to act on the vacuum and by the state-operator correspondence, \eqref{eq:W_OPE} must be $\mathcal{O}(1)$ as $z \to 0$. We deduce that the OPE of $e^{-\Sigma}(z)$ with $W(u_{\alpha})$ is trivial and $\hat{\Gamma}(z)$ is finite as $z \to u_{\alpha}$.
\\

All that remains to be done is to demonstrate there are no remaining critical points at generic points $z$ which are away from the insertions. We do this through the application of the Riemann-Hurwitz formula \eqref{eq:Riemann_Hurwitz} and counting the number of preimages of $\infty \in \mathbb{CP}^1$, the poles in $\hat{\Gamma}$. If we choose to count the multiplicity in the poles, then this will tell us the degree $N_{\Gamma}$ of $\hat{\Gamma}$.

Without loss of generality, we will take $\hat{\Gamma}(z_i) = x_i$ to be finite for all $i$. Therefore, as explained in \S\ref{sec:proof}, the equivalence of $\hat{\Gamma}$ and $\tilde{\Gamma}$ implies that all poles in $\hat{\Gamma}(z)$ come from the zeroes of
\begin{equation*}
\label{eq:omega_+}
    \omega^+(z) = \left\langle \xi^+(z) \prod_{\alpha = 1}^{n-2+2g} W(u_{\alpha}) \prod_{i=1}^n \hat{V}^{w_i}_{m_1^i,m_2^i}(x_i;z_i) \right\rangle,
\end{equation*}
where $z \in \Sigma_g/\{z_i, u_{\alpha}:i=1,2,\dots , n; \; \alpha= 1, 2, \dots n-2+2g\}$. $\omega^+(z)$ defines a meromorphic 1/2-form on $\Sigma_g$ \cite{Knighton:2020kuh}, which means it satisfies $Z(\omega^+) - P(\omega^+) = g-1$, where $Z(\omega^+)$ is the number of zeroes and $P(\omega^+)$ the number of poles on the whole of $\Sigma_g$, counting multiplicity. We would like the zeroes away from insertion points, but let's first find $P(\omega^+)$ so that we can use this formula.

Recall from \eqref{eq:xi_V_OPE} that as $z \to z_i$, we have the OPE
$$\xi^+(z)\hat{V}^{w_i}_{m_1^i,m_2^i}(x_i;z_i) = (z-z_i)^{-\frac{w_i + 1}{2}} \hat{V}^{w_i}_{m_1^i,m_2^i + 1/2} + \dots$$
By Wick's theorem, the only other places we could possibly have poles in $\omega^+(z)$ are at the insertion points $\{u_{\alpha}\}$. However, from $\hat{\sigma}^2(\xi^{\pm})(z) = z\xi^{\pm}(z)$, we know that
$$\xi^+(z)W(u_{\alpha}) = \mathcal{O}((z-u_{\alpha})).$$
Therefore, $P(\omega^+) = \sum_{i=1}^n \frac{w_i+1}{2}$, whilst there are $n+2g-2$ simple zeroes coming from the insertion points $\{u_{\alpha}\}$. We deduce that
$Z(\omega^+) = g-1 + \sum_{i=1}^n \frac{w_i+1}{2}.$
The poles in $\hat{\Gamma}(z)$ are given by the zeroes in $\omega^+(z)$ away from the insertions, meaning
$$P(\hat{\Gamma}) \leq Z(\omega^+) - (n+2g-2) = 1 - g + \sum_{i=1}^n \frac{w_i - 1}{2} = N_{\Gamma},$$
by the Riemann-Hurwitz formula \eqref{eq:Riemann_Hurwitz}. The less than or equal to sign is to acknowledge that some of the zeroes in $\omega^+(z)$ could strictly be cancelled by zeroes in
$\omega^-(z)$ --- recall $e^{-\Sigma} = -\xi^-\zeta^-$.

Nevertheless, we can demonstrate this bound is saturated by recalling the definition of the degree of a covering map as the number of preimages of each non-ramification point. Since $P(\hat{\Gamma})$ counts multiplicity in the poles, $P(\hat{\Gamma})$ is precisely the degree of $\hat{\Gamma}(z)$. But the Riemann-Hurwitz formula says that the degree is at least $P(\hat{\Gamma}) \geq N_{\Gamma}$, where $N_{\Gamma}$ only contains the known ramification points at the insertions $\{z_i\}$.

We therefore deduce that $P(\hat{\Gamma}) = N_{\Gamma}$, such that there are no further ramification points (i.e. critical points) in $\hat{\Gamma}(z)$ other than those at the insertions of the $\hat{V}^{w_i}_{m_1^i,m_2^i}(x_i;z_i)$. Hence, our proof of the localisation given in \S\ref{sec:proof} was indeed valid.

\section{Comments on factorisation}
\label{sec:expectations}

The goal of this appendix is to explain the factorisation $F_{\p\Phi J^+}(z) = F_{\p\Phi}(z) F_{J^+}(z)$ used in \S\ref{sec:wakimoto_radial_coord}. More generally, we will discuss conditions under which a factorisation $F_{AB}(z) = F_A(z)F_B(z)$ occurs for operators $A(z)$ and $B(z)$. The conditions we provide will be sufficient for our key example of $A = \p\Phi$ and $B = J^+$, yet it may well be possible to provide less stringent conditions on $A$ and $B$ for the factorisation to hold. For simplicity, in this appendix we will ignore contributions from the fermionic fields $\psi^A$ and $\chi^A$ as well as the ghost fields. Including such contributions does not change the central argument.
\\

First recall the definition \eqref{eq:F_Psi} of the correlation function weighted average of a field $\Psi(z)$,
\begin{equation*}
    F_{\Psi}(z):=\frac{\left\langle \Psi(z) \prod_{\alpha = 1}^{n-2+2g} W(u_{\alpha}) \prod_{i=1}^n \hat{V}^{w_i}_{m_1^i,m_2^i}(x_i;z_i) \right\rangle}{\left\langle \prod_{\alpha = 1}^{n-2+2g} W(u_{\alpha}) \prod_{i=1}^n \hat{V}^{w_i}_{m_1^i,m_2^i}(x_i;z_i) \right\rangle}.
\end{equation*}
If we view $f(x_i;z_i) = e^{-S}\prod_{\alpha = 1}^{n-2+2g} W(u_{\alpha}) \prod_{i=1}^n \hat{V}^{w_i}_{m_1^i,m_2^i}(x_i;z_i)$ as a probability density functional for these correlation functions, we can apply standard results from probability theory. Namely, given independent continuous random variables $A(z)$ and $B(z)$, the expectation value of $AB(z)$ factorises,
\begin{equation}\label{eq:factorisation}
    F_{AB}(z) = F_A(z)F_B(z).
\end{equation}
Of course, since we are working with operators, we need to provide a meaning for ``independence'' in our context and to be careful with normal ordering. We define two operators $A$ and $B$ to be ``independent'' if there exists a basis of the field content built from linear combinations of $\{\phi_i,\kappa_i\}$ that satisfies three constraints. The restriction to linear combinations is such that the Jacobian factor associated to the change of basis will be a constant and so will not complicate our later discussions about the path integral. The constraints are given by:
\begin{enumerate}
    \item The elements of the basis must have trivial OPEs with each other.
    \item If $A$ depends on any basis element, then $B$ does not depend on that element and vice versa.
    \item If $A$ depends on a basis element $e_1$ and $B$ depends on a basis element $e_2$, then $J^+$ cannot depend on both $e_1$ \emph{and} $e_2$.
\end{enumerate}
If such a basis exists, the first and second constraints imply that $A(z)$ and $B(z)$ must have trivial OPEs with each other (such that $AB(z)$ is unambiguous) and the path integral measure will independently integrate over the degrees of freedom associated to $A$ and $B$, such that the measure factorises. The reason for requiring the third constraint is a little more subtle.

The first and second constraints are sufficient to show that $f(x_i;z_i)$ will factorise when all $x_i = 0$, as can be seen from \eqref{eq:explicit_V}. We introduce non-trivial $x$-dependence via
$$V^{w}_{m_1,m_2}(x;z) = e^{xJ_0^+}V^{w}_{m_1,m_2}(0;z)e^{-xJ_0^+},$$
and therefore, we require that the conjugation by $J^+_0$ does not break the factorisation property --- this is imposed by the third constraint. If a basis satisfying all three constraints exists then, without loss of generality, we may assume that $J^+(z)$ does not depend on $A(z)$ such that the vertex operators factorise as
\begin{equation}\label{eq:x_dep_factorisation}
    e^{xJ_0^+}V_A(0;z)V_B(0;z)e^{-xJ_0^+} = V_A(0;z)e^{xJ_0^+}V_B(0;z)e^{-xJ_0^+} = V_A(0;z)V_B(x;z).
\end{equation}
The full probability density functional $f(x_i;z_i)$ will then factorise into a piece containing all of the $A$-dependence and a piece containing all of the $B$-dependence.

Such a basis exists in the case of interest to this paper, when $A = \p\Phi$ and $B = J^+$. It is given by
$$e_1 = \phi_1 + \phi_2 + 2i\kappa_2, \quad e_2 = \phi_1 - \phi_2, \quad e_3 = \phi_1 + \phi_2 + i\kappa_2, \quad e_4 = i\kappa_1,$$
such that $\p\Phi = \frac{1}{2}\p e_1$ and $J^+ = e^{e_3}\p(e^{e_4})$. The vertex operators \eqref{eq:explicit_V} then factorise as $V(0;z)=V_A(0;z)V_B(0;z)$, where we suppress the labels on $V^w_{m_1,m_2}(0;z)$ and
\begin{align*}
    V(0;z) &= \exp\left[ \left( m_2 - m_1 - \frac{w}{2} \right)e_1 + \left(m_1 - m_2 - \frac{1}{2}\right)e_2 + (2m_1 + w)e_3 + 2m_1 e_4 \right]\\
    &= \exp\left[ \left( m_2 - m_1 - \frac{w}{2} \right)e_1 + \left(m_1 - m_2 - \frac{1}{2}\right)e_2 \right] \exp\Big[ (2m_1 + w)e_3 + 2m_1 e_4 \Big]\\
    &= V_A(0;z)V_B(0;z),
\end{align*}
where without loss of generality we have chosen the $e_2$ contribution to be in $V_A(0;z)$, despite being independent of both $A$ and $B$. We may reintroduce non-trivial $x$-dependence via \eqref{eq:x_dep_factorisation} to find that $V(x;z) = V_A(0;z)V_B(x;z)$.

Similarly the classical action, without constraints, splits into the sum of two terms $S[\phi_i,\kappa_i]=S_A[e_1, e_2]+S_B[e_3,e_4]$. Thus, the measure
$$
f:=e^{-S}\prod_{i=1}^nV_i(x_i;z_i)
$$
factorises as $f=f_Af_B$, where $f_A=e^{-S_A}\prod_{i=1}^nV_A^i(0;z_i)$ and $f_B=e^{-S_B}\prod_{i=1}^nV_B^i(x_i;z_i)$.\footnote{This argument can be straightforwardly generalised to include the ghost and fermionic sectors, for which the term $\prod_{\alpha=1}^{n-2+2g}W(u_{\alpha})$ should also be included in the measure.} We shall assume that this measure is normalisable and moreover that
\begin{equation}\label{eq:normalisation_condition}
    \int {\cal D}e_1 {\cal D}e_2 \;f_A\neq 0, \quad \int  {\cal D}e_3 {\cal D}e_4 \;f_B\neq 0.
\end{equation}
In the context of the results in \S\ref{sec:proof}, this implies we must be at a point in moduli space where a covering map exists. We define the expectation as in \eqref{eq:F_Psi}
$$
F_{\p\Phi J^+}(z)=\frac{\int \prod_{j=1}^4{\cal D}e_j \;\p\Phi(z)J^+(z)\,e^{-S}  \prod_{i=1}^nV_i(x_i;z_i)}{\int \prod_{j=1}^4{\cal D}e_j \;e^{-S} \prod_{i=1}^nV_i(x_i;z_i)}.
$$
Under the assumptions of \eqref{eq:normalisation_condition} and that the probability density functional $f(x_i;z_i)$ factorises, this expectation factorises as
$$
F_{\p\Phi J^+}(z)=\frac{\int {\cal D}e_1 {\cal D}e_2 \;\p\Phi(z)\,f_A}{\int {\cal D}e_1 {\cal D}e_2 \;f_A}\times \frac{\int {\cal D}e_3 {\cal D}e_4 \;J^+(z)\,f_B}{\int {\cal D}e_3 {\cal D}e_4 \;f_B}.
$$
We now multiply by\footnote{This is a somewhat formal argument as we are assuming all integrals are finite in some appropriate sense. The important point is that they do not vanish by assumption.}
$$
1=\frac{\int {\cal D}\hat{e}_1 {\cal D}\hat{e}_2  \; \hat{f}_A}{\int {\cal D}\hat{e}_1 {\cal D}\hat{e}_2 \; \hat{f}_A}\times \frac{\int {\cal D}\hat{e}_3 {\cal D}\hat{e}_4  \; \hat{f}_B}{\int {\cal D}\hat{e}_3 {\cal D}\hat{e}_4 \; \hat{f}_B},
$$
where the hatted fields are copies of the un-hatted linear dilaton systems we started with. We have also introduced a shorthand notation $\hat{f}_A = f_A(\hat{e}_1,\hat{e}_2)$  and $\hat{f}_B = f_B(\hat{e}_3,\hat{e}_4)$. We then have
$$
F_{\p\Phi J^+}(z)=\frac{\int {\cal D}e_1{\cal D}e_2 {\cal D}\hat{e}_3{\cal D}\hat{e}_4 \;\p\Phi(z)\,f_A\hat{f}_B}{\int {\cal D}e_1{\cal D}e_2 {\cal D}\hat{e}_3{\cal D}\hat{e}_4 \;f_A\hat{f}_B}
\times \frac{\int  {\cal D}\hat{e}_1{\cal D}\hat{e}_2{\cal D}e_3{\cal D}e_4 \;J^+(z)\,f_B\hat{f}_A}{\int {\cal D}\hat{e}_1{\cal D}\hat{e}_2 {\cal D}e_3{\cal D}e_4 \;f_B\hat{f}_A}.
$$
The distinction between the hatted and un-hatted fields is artificial and so we conclude that, under the assumptions stated,
\begin{equation*}\label{eq:factorisation_appendix}
    F_{\p\Phi J^+}(z)=F_{\p\Phi}(z)F_{J^+}(z),
\end{equation*}
as required.

Before concluding, it is worth commenting on our assumption in \S\ref{sec:wakimoto_radial_coord} that the zeroes of $F_{J^+}(z)$ away from insertion points are given by the points where either $\xi^+(z) \to 0$ or $\eta^+(z) \to 0$. We cannot apply the factorisation techniques of this appendix to $J^+ = \xi^+\eta^+$, since the third condition in our definition of ``independent'' operators is not satisfied by $\xi^+$ and $\eta^+$. The conjugation of vertex operators by $J_0^+$ for non-trivial $x_i$-dependence therefore prevents an exact factorisation of physical correlators. Nevertheless, we do not require the full factorisation to derive our results in \S\ref{sec:wakimoto_radial_coord} --- we only need to know the location of the zeroes. Since $\xi^+$ and $\eta^+$ are conformal tensors, they do not have poles away from insertion points, implying we are safe to assume that $J^+(z) \to 0$ whenever $\xi^+(z) \to 0$ or $\eta^+(z) \to 0$.

\bibliographystyle{JHEP}
\bibliography{covering_maps.bib}

\providecommand{\href}[2]{#2}\begingroup\raggedright\begin{thebibliography}{10}

\bibitem{Maldacena:1997re}
J.~M. Maldacena, {\it {The Large N limit of superconformal field theories and supergravity}},  {\em Adv. Theor. Math. Phys.} {\bf 2} (1998) 231--252, [\href{http://arxiv.org/abs/hep-th/9711200}{{\tt hep-th/9711200}}].

\bibitem{Gaberdiel:2018rqv}
M.~R. Gaberdiel and R.~Gopakumar, {\it {Tensionless string spectra on AdS$_{3}$}},  {\em JHEP} {\bf 05} (2018) 085, [\href{http://arxiv.org/abs/1803.04423}{{\tt arXiv:1803.04423}}].

\bibitem{Eberhardt:2018ouy}
L.~Eberhardt, M.~R. Gaberdiel, and R.~Gopakumar, {\it {The Worldsheet Dual of the Symmetric Product CFT}},  {\em JHEP} {\bf 04} (2019) 103, [\href{http://arxiv.org/abs/1812.01007}{{\tt arXiv:1812.01007}}].

\bibitem{Eberhardt:2019ywk}
L.~Eberhardt, M.~R. Gaberdiel, and R.~Gopakumar, {\it {Deriving the AdS$_{3}$/CFT$_{2}$ correspondence}},  {\em JHEP} {\bf 02} (2020) 136, [\href{http://arxiv.org/abs/1911.00378}{{\tt arXiv:1911.00378}}].

\bibitem{Eberhardt:2019qcl}
L.~Eberhardt and M.~R. Gaberdiel, {\it {String theory on AdS$_3$ and the symmetric orbifold of Liouville theory}},  {\em Nucl. Phys. B} {\bf 948} (2019) 114774, [\href{http://arxiv.org/abs/1903.00421}{{\tt arXiv:1903.00421}}].

\bibitem{Eberhardt:2020akk}
L.~Eberhardt, {\it {AdS$_{3}$/CFT$_{2}$ at higher genus}},  {\em JHEP} {\bf 05} (2020) 150, [\href{http://arxiv.org/abs/2002.11729}{{\tt arXiv:2002.11729}}].

\bibitem{Dei:2020zui}
A.~Dei, M.~R. Gaberdiel, R.~Gopakumar, and B.~Knighton, {\it {Free field world-sheet correlators for ${\rm AdS}_3$}},  {\em JHEP} {\bf 02} (2021) 081, [\href{http://arxiv.org/abs/2009.11306}{{\tt arXiv:2009.11306}}].

\bibitem{Gaberdiel:2020ycd}
M.~R. Gaberdiel, R.~Gopakumar, B.~Knighton, and P.~Maity, {\it {From symmetric product CFTs to AdS$_{3}$}},  {\em JHEP} {\bf 05} (2021) 073, [\href{http://arxiv.org/abs/2011.10038}{{\tt arXiv:2011.10038}}].

\bibitem{Knighton:2020kuh}
B.~Knighton, {\it {Higher genus correlators for tensionless AdS$_{3}$ strings}},  {\em JHEP} {\bf 04} (2021) 211, [\href{http://arxiv.org/abs/2012.01445}{{\tt arXiv:2012.01445}}].

\bibitem{Gaberdiel:2021njm}
M.~R. Gaberdiel and K.~Naderi, {\it {The physical states of the Hybrid Formalism}},  {\em JHEP} {\bf 10} (2021) 168, [\href{http://arxiv.org/abs/2106.06476}{{\tt arXiv:2106.06476}}].

\bibitem{Gaberdiel:2021kkp}
M.~R. Gaberdiel, B.~Knighton, and J.~Vo\v{s}mera, {\it {D-branes in AdS$_{3}$ \texttimes{} S$^{3}$ \texttimes{} \ensuremath{\mathbb{T}}$^{4}$ at k = 1 and their holographic duals}},  {\em JHEP} {\bf 12} (2021) 149, [\href{http://arxiv.org/abs/2110.05509}{{\tt arXiv:2110.05509}}].

\bibitem{Gaberdiel:2022bfk}
M.~R. Gaberdiel, K.~Naderi, and V.~Sriprachyakul, {\it {The free field realisation of the BVW string}},  \href{http://arxiv.org/abs/2202.11392}{{\tt arXiv:2202.11392}}.

\bibitem{Dei:2022pkr}
A.~Dei and L.~Eberhardt, {\it {String correlators on $\text{AdS}_3$: Analytic structure and dual CFT}},  {\em SciPost Phys.} {\bf 13} (2022), no.~3 053, [\href{http://arxiv.org/abs/2203.13264}{{\tt arXiv:2203.13264}}].

\bibitem{Gaberdiel:2022oeu}
M.~R. Gaberdiel and B.~Nairz, {\it {BPS correlators for AdS$_{3}$/CFT$_{2}$}},  {\em JHEP} {\bf 09} (2022) 244, [\href{http://arxiv.org/abs/2207.03956}{{\tt arXiv:2207.03956}}].

\bibitem{Naderi:2022bus}
K.~Naderi, {\it {DDF operators in the Hybrid Formalism}},  \href{http://arxiv.org/abs/2208.01617}{{\tt arXiv:2208.01617}}.

\bibitem{Eberhardt:2019}
L.~Eberhardt, {\em {Strings on AdS3}}.
\newblock PhD thesis, ETH Zurich, 2019.

\bibitem{Giribet:2018ada}
G.~Giribet, C.~Hull, M.~Kleban, M.~Porrati, and E.~Rabinovici, {\it {Superstrings on AdS$_{3}$ at $k =$ 1}},  {\em JHEP} {\bf 08} (2018) 204, [\href{http://arxiv.org/abs/1803.04420}{{\tt arXiv:1803.04420}}].

\bibitem{Maldacena:2000hw}
J.~M. Maldacena and H.~Ooguri, {\it {Strings in AdS(3) and SL(2,R) WZW model 1.: The Spectrum}},  {\em J. Math. Phys.} {\bf 42} (2001) 2929--2960, [\href{http://arxiv.org/abs/hep-th/0001053}{{\tt hep-th/0001053}}].

\bibitem{Maldacena:2000kv}
J.~M. Maldacena, H.~Ooguri, and J.~Son, {\it {Strings in AdS(3) and the SL(2,R) WZW model. Part 2. Euclidean black hole}},  {\em J. Math. Phys.} {\bf 42} (2001) 2961--2977, [\href{http://arxiv.org/abs/hep-th/0005183}{{\tt hep-th/0005183}}].

\bibitem{Maldacena:2001km}
J.~M. Maldacena and H.~Ooguri, {\it {Strings in AdS(3) and the SL(2,R) WZW model. Part 3. Correlation functions}},  {\em Phys. Rev. D} {\bf 65} (2002) 106006, [\href{http://arxiv.org/abs/hep-th/0111180}{{\tt hep-th/0111180}}].

\bibitem{Giveon:1998ns}
A.~Giveon, D.~Kutasov, and N.~Seiberg, {\it {Comments on string theory on AdS(3)}},  {\em Adv. Theor. Math. Phys.} {\bf 2} (1998) 733--782, [\href{http://arxiv.org/abs/hep-th/9806194}{{\tt hep-th/9806194}}].

\bibitem{deBoer:1998gyt}
J.~de~Boer, H.~Ooguri, H.~Robins, and J.~Tannenhauser, {\it {String theory on AdS(3)}},  {\em JHEP} {\bf 12} (1998) 026, [\href{http://arxiv.org/abs/hep-th/9812046}{{\tt hep-th/9812046}}].

\bibitem{Berkovits:1999im}
N.~Berkovits, C.~Vafa, and E.~Witten, {\it {Conformal field theory of AdS background with Ramond-Ramond flux}},  {\em JHEP} {\bf 03} (1999) 018, [\href{http://arxiv.org/abs/hep-th/9902098}{{\tt hep-th/9902098}}].

\bibitem{Eberhardt:2021jvj}
L.~Eberhardt, {\it {Summing over Geometries in String Theory}},  {\em JHEP} {\bf 05} (2021) 233, [\href{http://arxiv.org/abs/2102.12355}{{\tt arXiv:2102.12355}}].

\bibitem{Bhat:2021dez}
F.~Bhat, R.~Gopakumar, P.~Maity, and B.~Radhakrishnan, {\it {Twistor coverings and Feynman diagrams}},  {\em JHEP} {\bf 05} (2022) 150, [\href{http://arxiv.org/abs/2112.05115}{{\tt arXiv:2112.05115}}].

\bibitem{Pakman:2009zz}
A.~Pakman, L.~Rastelli, and S.~S. Razamat, {\it {Diagrams for Symmetric Product Orbifolds}},  {\em JHEP} {\bf 10} (2009) 034, [\href{http://arxiv.org/abs/0905.3448}{{\tt arXiv:0905.3448}}].

\bibitem{Eberhardt:2020bgq}
L.~Eberhardt, {\it {Partition functions of the tensionless string}},  {\em JHEP} {\bf 03} (2021) 176, [\href{http://arxiv.org/abs/2008.07533}{{\tt arXiv:2008.07533}}].

\bibitem{Fiset:2022erp}
M.-A. Fiset, M.~R. Gaberdiel, K.~Naderi, and V.~Sriprachyakul, {\it {Perturbing the symmetric orbifold from the worldsheet}},  {\em JHEP} {\bf 07} (2023) 093, [\href{http://arxiv.org/abs/2212.12342}{{\tt arXiv:2212.12342}}].

\bibitem{Vonk:2005yv}
M.~Vonk, {\it {A Mini-course on topological strings}},  \href{http://arxiv.org/abs/hep-th/0504147}{{\tt hep-th/0504147}}.

\bibitem{Berkovits:1993xq}
N.~Berkovits and C.~Vafa, {\it {On the Uniqueness of string theory}},  {\em Mod. Phys. Lett. A} {\bf 9} (1994) 653--664, [\href{http://arxiv.org/abs/hep-th/9310170}{{\tt hep-th/9310170}}].

\bibitem{Berkovits:1994vy}
N.~Berkovits and C.~Vafa, {\it {N=4 topological strings}},  {\em Nucl. Phys. B} {\bf 433} (1995) 123--180, [\href{http://arxiv.org/abs/hep-th/9407190}{{\tt hep-th/9407190}}].

\bibitem{DiFrancesco:1997nk}
P.~Di~Francesco, P.~Mathieu, and D.~Senechal, {\em {Conformal Field Theory}}.
\newblock Graduate Texts in Contemporary Physics. Springer-Verlag, New York, 1997.

\bibitem{Frenkel:1980rn}
I.~B. Frenkel and V.~G. Kac, {\it {Basic Representations of Affine Lie Algebras and Dual Resonance Models}},  {\em Invent. Math.} {\bf 62} (1980) 23--66.

\bibitem{Segal:1981ap}
G.~Segal, {\it {Unitarity Representations of Some Infinite Dimensional Groups}},  {\em Commun. Math. Phys.} {\bf 80} (1981) 301--342.

\bibitem{Goddard:1987td}
P.~Goddard, D.~I. Olive, and G.~Waterson, {\it {Superalgebras, Symplectic Bosons and the Sugawara Construction}},  {\em Commun. Math. Phys.} {\bf 112} (1987) 591.

\bibitem{Berkovits:2004hg}
N.~Berkovits, {\it {An Alternative string theory in twistor space for N=4 superYang-Mills}},  {\em Phys. Rev. Lett.} {\bf 93} (2004) 011601, [\href{http://arxiv.org/abs/hep-th/0402045}{{\tt hep-th/0402045}}].

\bibitem{Gaiotto:2017euk}
D.~Gaiotto and M.~Rap\v{c}\'ak, {\it {Vertex Algebras at the Corner}},  {\em JHEP} {\bf 01} (2019) 160, [\href{http://arxiv.org/abs/1703.00982}{{\tt arXiv:1703.00982}}].

\bibitem{Knighton:2022ipy}
B.~Knighton, {\it {Classical geometry from the tensionless string}},  \href{http://arxiv.org/abs/2207.01293}{{\tt arXiv:2207.01293}}.

\bibitem{Green:1987sp}
M.~B. Green, J.~H. Schwarz, and E.~Witten, {\em {Superstring Theory. Vol. 1: Introduction}}.
\newblock Cambridge Monographs on Mathematical Physics. Cambridge University Pressknin, 7, 1988.

\bibitem{Gerigk:2012cq}
S.~Gerigk, {\it {String States on AdS$_3 \times \rm{S}^3$ from the Supergroup}},  {\em JHEP} {\bf 10} (2012) 084, [\href{http://arxiv.org/abs/1208.0345}{{\tt arXiv:1208.0345}}].

\bibitem{Wakimoto:1986gf}
M.~Wakimoto, {\it {Fock representations of the affine lie algebra A1(1)}},  {\em Commun. Math. Phys.} {\bf 104} (1986) 605--609.

\bibitem{Adamo:2016rtr}
T.~Adamo, D.~Skinner, and J.~Williams, {\it {Twistor methods for AdS$_{5}$}},  {\em JHEP} {\bf 08} (2016) 167, [\href{http://arxiv.org/abs/1607.03763}{{\tt arXiv:1607.03763}}].

\bibitem{Gaberdiel:2021qbb}
M.~R. Gaberdiel and R.~Gopakumar, {\it {String Dual to Free N=4 Supersymmetric Yang-Mills Theory}},  {\em Phys. Rev. Lett.} {\bf 127} (2021), no.~13 131601, [\href{http://arxiv.org/abs/2104.08263}{{\tt arXiv:2104.08263}}].

\bibitem{Gaberdiel:2021jrv}
M.~R. Gaberdiel and R.~Gopakumar, {\it {The worldsheet dual of free super Yang-Mills in 4D}},  {\em JHEP} {\bf 11} (2021) 129, [\href{http://arxiv.org/abs/2105.10496}{{\tt arXiv:2105.10496}}].

\bibitem{Ward:1990vs}
R.~S. Ward and R.~O. Wells, {\em {Twistor geometry and field theory}}.
\newblock Cambridge Monographs on Mathematical Physics. Cambridge University Press, 8, 1991.

\bibitem{Adamo:2017qyl}
T.~Adamo, {\it {Lectures on twistor theory}},  {\em PoS} {\bf Modave2017} (2018) 003, [\href{http://arxiv.org/abs/1712.02196}{{\tt arXiv:1712.02196}}].

\bibitem{Fradkin:1985ys}
E.~S. Fradkin and A.~A. Tseytlin, {\it {Quantum String Theory Effective Action}},  {\em Nucl. Phys. B} {\bf 261} (1985) 1--27. [Erratum: Nucl.Phys.B 269, 745--745 (1986)].

\bibitem{Friedan:1985ge}
D.~Friedan, E.~J. Martinec, and S.~H. Shenker, {\it {Conformal Invariance, Supersymmetry and String Theory}},  {\em Nucl. Phys. B} {\bf 271} (1986) 93--165.

\bibitem{Mason:2013sva}
L.~Mason and D.~Skinner, {\it {Ambitwistor strings and the scattering equations}},  {\em JHEP} {\bf 07} (2014) 048, [\href{http://arxiv.org/abs/1311.2564}{{\tt arXiv:1311.2564}}].

\bibitem{Gerasimov:1990fi}
A.~Gerasimov, A.~Morozov, M.~Olshanetsky, A.~Marshakov, and S.~L. Shatashvili, {\it {Wess-Zumino-Witten model as a theory of free fields}},  {\em Int. J. Mod. Phys. A} {\bf 5} (1990) 2495--2589.

\bibitem{Ketov:1995yd}
S.~V. Ketov, {\em {Conformal field theory}}.
\newblock World Scientific, 1995.

\bibitem{Frenkel:2005ku}
E.~Frenkel and A.~Losev, {\it {Mirror symmetry in two steps: A-I-B}},  {\em Commun. Math. Phys.} {\bf 269} (2006) 39--86, [\href{http://arxiv.org/abs/hep-th/0505131}{{\tt hep-th/0505131}}].

\bibitem{Hikida:2020kil}
Y.~Hikida and T.~Liu, {\it {Correlation functions of symmetric orbifold from AdS$_{3}$ string theory}},  {\em JHEP} {\bf 09} (2020) 157, [\href{http://arxiv.org/abs/2005.12511}{{\tt arXiv:2005.12511}}].

\bibitem{Hikida:2007tq}
Y.~Hikida and V.~Schomerus, {\it {H+(3) WZNW model from Liouville field theory}},  {\em JHEP} {\bf 10} (2007) 064, [\href{http://arxiv.org/abs/0706.1030}{{\tt arXiv:0706.1030}}].

\bibitem{Hikida:2008pe}
Y.~Hikida and V.~Schomerus, {\it {The FZZ-Duality Conjecture: A Proof}},  {\em JHEP} {\bf 03} (2009) 095, [\href{http://arxiv.org/abs/0805.3931}{{\tt arXiv:0805.3931}}].

\bibitem{Nakahara:2003nw}
M.~Nakahara, {\em {Geometry, topology and physics}}.
\newblock Institute of Physics Publishing, 2003.

\bibitem{Lunin:2000yv}
O.~Lunin and S.~D. Mathur, {\it {Correlation functions for M**N / S(N) orbifolds}},  {\em Commun. Math. Phys.} {\bf 219} (2001) 399--442, [\href{http://arxiv.org/abs/hep-th/0006196}{{\tt hep-th/0006196}}].

\bibitem{Lunin:2001pw}
O.~Lunin and S.~D. Mathur, {\it {Three point functions for M(N) / S(N) orbifolds with N=4 supersymmetry}},  {\em Commun. Math. Phys.} {\bf 227} (2002) 385--419, [\href{http://arxiv.org/abs/hep-th/0103169}{{\tt hep-th/0103169}}].

\bibitem{Witten:2003nn}
E.~Witten, {\it {Perturbative gauge theory as a string theory in twistor space}},  {\em Commun. Math. Phys.} {\bf 252} (2004) 189--258, [\href{http://arxiv.org/abs/hep-th/0312171}{{\tt hep-th/0312171}}].

\bibitem{Hull:2009mi}
C.~Hull and B.~Zwiebach, {\it {Double Field Theory}},  {\em JHEP} {\bf 09} (2009) 099, [\href{http://arxiv.org/abs/0904.4664}{{\tt arXiv:0904.4664}}].

\bibitem{Pestun:2016zxk}
V.~Pestun et~al., {\it {Localization techniques in quantum field theories}},  {\em J. Phys. A} {\bf 50} (2017), no.~44 440301, [\href{http://arxiv.org/abs/1608.02952}{{\tt arXiv:1608.02952}}].

\bibitem{Polchinski:1998rq}
J.~Polchinski, {\em {String theory. Vol. 1: An introduction to the bosonic string}}.
\newblock Cambridge Monographs on Mathematical Physics. Cambridge University Press, 12, 2007.

\bibitem{Polchinski:1998rr}
J.~Polchinski, {\em {String theory. Vol. 2: Superstring theory and beyond}}.
\newblock Cambridge Monographs on Mathematical Physics. Cambridge University Press, 12, 2007.

\end{thebibliography}\endgroup

\end{document}